\documentclass[twocolumn,showpacs,amsmath,floatfix,prb,aps]{revtex4}

\usepackage{color}
\usepackage{amsmath}
\usepackage{pifont}   % Ding symbols
\usepackage{graphicx} % Include figure files
\usepackage{dcolumn}  % Align table columns on decimal point
\usepackage{bm}       % bold math
\usepackage{amsfonts} % Some more fonts
\usepackage{amssymb}  % More symbols
\usepackage{multirow} % Table functions
\usepackage{bbm}
\usepackage{hyperref} % hyperref

\begin{document}
%\DeclareGraphicsRule{*}{png}{*}{}

%%%% User-defined commands %%%%
\newcommand{\ba}{{\bf a}}
\newcommand{\BB}{{\bf b}}
\newcommand{\bd}{{\bf d}}
\newcommand{\br}{{\bf r}}
\newcommand{\bp}{{\bf p}}
\newcommand{\bk}{{\bf k}}
\newcommand{\bg}{{\bf g}}
\newcommand{\bt}{{\bf t}}
\newcommand{\bu}{{\bf u}}
\newcommand{\bq}{{\bf q}}
\newcommand{\bG}{{\bf G}}
\newcommand{\bP}{{\bf P}}
\newcommand{\bJ}{{\bf J}}
\newcommand{\bK}{{\bf K}}
\newcommand{\bL}{{\bf L}}
\newcommand{\bR}{{\bf R}}
\newcommand{\bS}{{\bf S}}
\newcommand{\bT}{{\bf T}}
\newcommand{\bQ}{{\bf Q}}
\newcommand{\bA}{{\bf A}}
\newcommand{\bH}{{\bf H}}
\newcommand{\bX}{{\bf X}}

\newcommand{\bra}[1]{\left\langle #1 \right |}
\newcommand{\ket}[1]{\left| #1 \right\rangle}
\newcommand{\braket}[2]{\left\langle #1 | #2 \right\rangle}
\newcommand{\mel}[3]{\left\langle #1 \left| #2 \right| #3 \right\rangle}

\newcommand{\bdel}{\boldsymbol{\delta}}
\newcommand{\bsig}{\boldsymbol{\sigma}}
\newcommand{\beps}{\boldsymbol{\epsilon}}
\newcommand{\bnu}{\boldsymbol{\nu}}
\newcommand{\bnab}{\boldsymbol{\nabla}}
\newcommand{\bchi}{\boldsymbol{\chi}}
\newcommand{\bGam}{\boldsymbol{\Gamma}}

\newcommand{\bgt}{\tilde{\bf g}}

\newcommand{\brh}{\hat{\bf r}}
\newcommand{\bph}{\hat{\bf p}}

\author{N. Ray$^1$}
\author{F. Rost$^1$}
\author{D. Weckbecker$^1$}
\author{M. Vogl$^1$}
\author{S. Sharma$^2$}
\author{R. Gupta$^{2}$}
\author{O. Pankratov$^1$}
\author{S. Shallcross$^1$}
\email{sam.shallcross@fau.de}
\affiliation{1 Lehrstuhl f\"ur Theoretische Festk\"orperphysik, Staudstr. 7-B2, 91058 Erlangen, Germany,}
\affiliation{2 Max-Planck-Institut f\"ur Mikrostrukturphysik Weinberg 2, D-06120 Halle, Germany.}

\title{Going beyond $\bk.\bp$ theory: a general method for obtaining effective Hamiltonians in both high and low symmetry situations}
\date{\today}

%%%%%%%%%%%%
% ABSTRACT %
%%%%%%%%%%%%

\begin{abstract}

We provide a method for the generation of effective continuum Hamiltonians that goes beyond the well known $\bk.\bp$ method in being equally effective in both high and no symmetry systems, as well as for situations involving both perturbative as well as non-perturbative structural deformations. This, as we demonstrate, provides for the emerging realm of low dimensional materials a method with the wide applicability and usefulness that the $\bk.\bp$ method brings to the study of three dimensional materials. Our approach is based on a exact map of the two-centre tight-binding method onto a compact continuum Hamiltonian, with a precise condition given for the Hermiticity of the latter object. We apply this method to a broad range of low dimensional systems of both high and no symmetry: graphene, graphdiyne, $\gamma$-graphyne, 6,6,12-graphyne, twist bilayer graphene, and partial dislocation networks in Bernal stacked bilayer graphene. For the single layer systems the method yields Hamiltonians for the ideal lattices, as well as a systematic theory for corrections due to deformation. In the case of bilayer graphene we provide a compact expression for an effective field capable of describing \emph{any stacking deformation} of the bilayer; twist bilayer graphene, as well as the partial dislocation network in AB stacked graphene, emerge as special cases of this field. For the latter system we find (i) charge pooling on the mosaic of AB and AC segments near the Dirac point and (ii) localized current carrying states on the partials with the current density characterized by both intralayer and interlayer components. The formalism is equally applicable to any bilayer system, for instance few layer dichalcogenides, and we discuss possibility of extended defects in these materials.

\end{abstract}

%\pacs{73.20.At, 73.21.Ac, 81.05.Uw}

\maketitle

%%%%%%%%%%%%%%%%
% INTRODUCTION %
%%%%%%%%%%%%%%%%

\section{Introduction}

In many situations the motion of electrons in matter is well described by compact effective Hamiltonians treating only quasiparticles near the Fermi surface of the material. Historical examples of such Hamiltonians include the Luttinger-Kohn and Kane models for III-V semiconductors\cite{bir74,voon09,sun10}, and the effective Dirac equation for IV-VI semi-conductors\cite{pan83,pan87}. Perhaps the most famous example of an effective Hamiltonian is the Dirac-Weyl equation which in particle physics describes massless neutrinos, but in condensed matter governs the low energy quasi-particles of graphene, a two dimensional honeycomb lattice of carbon. This description of graphene provides both great insight into the physics of this remarkable material, as well as a framework in which electronic effects from the large length scale deformations that this 2d membrane is subject to, for example flexural rippling, can be efficiently calculated. 

The experimental fabrication of graphene in 2004\cite{nov04} can now be seen a heralding the emergence of a new science of low dimensional materials\cite{geim13}. Such materials include both close cousins of graphene, for example the Bernal stacked\cite{mcc13} or twist bilayer graphenes\cite{lop07,hass08,lop13,bist11,mel12,shall13,shall16}, as well as more distant relatives in the form of complex all carbon allotropes\cite{bau87,nar98,en11,mal12,kim12,liu12,yue13,mi14,mi14a,yang15} or silicene\cite{vog12}. Recently, attention has started to focus on interesting non-carbon low dimensional materials such as the layered transition metal
dichalcogenides\cite{wang12,liu12,shi13,kos13,he14,huang15,chu15}, for example MoS$_2$, MoSe$_2$, and CuCl. It might have been expected that the emergence of this multitude of new low dimensional materials would have been accompanied by the development of a range of effective Hamiltonians, offering comparable power and insight to the Dirac-Weyl Hamiltonian in the case of graphene, however this has not proven to be the case.

A principle reason for this is that the $\bk.\bp$ method, a tool that has proved profoundly useful in the context of high symmetry three dimensional materials\cite{bir74,voon09}, is both technically more difficult to apply in the reduced symmetry situations often found in low dimensional materials as well as, more importantly, being quite incapable of treating the \emph{non-perturbative} structural deformations that such materials are often subject to. As an example of such a deformation consider a twist fault in Bernal stacked graphene bilayer. A small mutual rotation of the layers generates a moir\'e within which all stacking types are present and, therefore, a system that cannot be considered a small structural perturbation of the Bernal bilayer. While the $\bk.\bp$ theory is formally applicable to such a case, it will result in a large and non-intuitive Hamiltonian, thus negating the principle advantages of the method. The weak interlayer bonding inherent to the emerging class of van der Walls bonded few layer systems\cite{geim13} makes such non-perturbation interlayer deformations - of which there are a rich variety e.g. twist faults and partial dislocations to name just two - likely to be the generic case\cite{ald13,butz14}. As such deformations exist on length scales that render conventional electronic structure approaches prohibitive there is an urgent need for an effective theory by which they may be treated.

In this paper we present a general method that goes beyond $\bk.\bp$ theory in that for both high and no symmetry situations, as well as for perturbative and non-perturbative deformations, it yields compact and physically transparent effective Hamiltonians with equal ease. Our approach is based on the surprising fact that, as we will show, there exists an exact map from the two centre tight-binding Hamiltonian to a compact continuum effective Hamiltonian. We deploy this method on three standout problems in the field of low dimensional materials: (i) effective Hamiltonians for the complex carbon allotropes\cite{bau87,nar98,en11,mal12,kim12,liu12,yue13,mi14,mi14a,yang15} (ii) a systematic treatment of deformations in single layer graphene, and (iii) and a general theory of interlayer deformations in bilayer systems.

The first two of these problems are cases of high symmetry 2d systems and their perturbative deformations, and an application of the method we present here to this general problem results in a formalism valid for all 2d materials. The structure of this theory is based around a rather simple \emph{connection formula} that relates the sublattice space of the crystal to the pseudospin space of the effective Hamiltonian. In conjunction with a set of \emph{universal functions} composed the basic variables of the problem, the position and momentum operators and the deformation tensor, this results in a very efficient scheme for treating the general case of 2d materials under weak (or no) structural perturbation. To demonstrate this we consider a number of all-carbon 2d allotropes: graphene, graphdiyne, $\gamma$-graphyne, and 6,6,12-graphyne. In each case we provide both an effective Hamiltonian for the high symmetry phase, as well as, in the case of graphene and graphdiyine, the corrections that arise from arbitrary deformations (with the proviso that the deformation is slow on the scale of the lattice constant).

To the case of deformations in graphene, which has attracted enormous attention in the literature\cite{mey07,voz08,gui08,sas08,kat10,gui10,gui10a,low10,ju10,can11,ju11,he12,ju12,warn12,kitt12,much12,kli12,ol13,lop13,peet13,peet13a,ju13,voz13,bar13,qu14,san14,lop14,car14,cross14,chang14,qi14,gong15,ol15,ol16}, we devote particular attention. We provide a systematic expansion in terms of the momentum operator and the deformation field, recovering all known results from the literature as well as several extensions. These are (i) an imaginary scalar potential that is the $\sigma_0$ partner of the very interesting imaginary geometric gauge first reported in Ref.~\onlinecite{ju12}, (ii) trigonal warping corrections due to the deformation (terms quadratic in momentum and linear in the deformation tensor), and (iii) scaler, gauge potentials, and Fermi velocity renormalization due to second order in the deformation tensor. We show that such higher order terms are, perhaps surprisingly, essential for the effective Hamiltonian approach to agree with corresponding tight-binding calculations, but that if they are included the agreement is almost perfect. We pay particular attention to the question of Hermiticity, showing that the effective Hamiltonian remain hermitian up to second order derivatives of the deformation tensor. For deformations stronger than this the emergent pseudospin degree of freedom is itself destroyed by the deformation, and the Dirac-Weyl framework no longer can provide an adequate description of the deformed material. For the more complex 2d allotropes the same formalism yields rather different physics, and in the case of graphdiyne (the only complex carbon allotropes thus far to have apparently been experimentally synthesized\cite{li10}) we find a low energy Dirac equation describes the high symmetry state, with deformations entering as a complex gap function field.

Our second primary example of the theory is provided by interlayer deformations in bilayer graphene. It is striking that despite the intense study of stacking defects in this material, for instance rotational faults\cite{lop07,hass08,lop13,bist11,mel12,shall13,shall16} and partial dislocations\cite{ald13,butz14,kiss15}, there exists \emph{no general theory of interlayer deformations comparable to that for deformations in single layer graphene}. In other words, a theory that relates an arbitrary stacking deformation to an effective Hamiltonian, in the way that the Dirac-Weyl Hamiltonian is linked to arbitrary deformations in single layer graphene via a deformation induced gauge field. The underlying reason for the absence of such a general theory is that this problem represents an example of a non-perturbative deformation for which $\bk.\bp$ theory fails. An application of the method presented here, however, immediately yields a general description from which, for the case of bilayer graphene, the well known Hamiltonians of the Bernal and twist bilayers emerge as special cases.

Deploying this general formalism we study, as a final example, partial dislocation networks in bilayer graphene recently reported in TEM images\cite{ald13,butz14} and already shown to have a significant impact on transport in these materials\cite{kiss15}. As an example of the power of the effective Hamiltonian approach we consider an experimentally derived system (taken from TEM images of bilayer graphene grown on the Si-face of SiC) consisting of the order of $10^8$ carbon atoms which, it goes without saying, could not be calculated by any other method. We find that partial dislocations are associated with localized current carrying states, with the current density propagating with both intralayer and interlayer components along the partials. Interestingly, different types of partial dislocations (there are just three partial Burgers vectors for bilayer graphene) develop these current carrying localized states at different energies, and this is independent of the network geometry. Finally, near the Dirac point we find a strong charge inhomogeneity in the form of charge pooling on the different segments of the mosaic of AB and AC stacked domains, a phenomena recently treated in Ref.~\onlinecite{kiss15} using a preliminary version of the theory presented in this manuscript.

% why start from TB

%%%%%%%%%%%%%%%%%%
% GENERAL THEORY %
%%%%%%%%%%%%%%%%%%

\section{Mapping the tight-binding method to a continuum description}
\label{GT}

The aim of this section is to demonstrate a general mapping between the two-centre tight-binding method and a continuum Hamiltonian, i.e., one involving only position $\hat{\br}$ and momentum $\hat{\bp}$ operators. (In this section it will prove notationally advantageous to explicitly denote operators, an approach that we do not use in the remainder of the paper.) 

We begin with a standard two centre tight binding Hamiltonian

\begin{equation}
 H_{TB} = \sum_{ij} t_{ij} c_j^\dagger c_i
\end{equation}
which is assumed to describe a system that is close to some high symmetry system $H_{TB}^{(0)}$ in that $H_{TB}^{(0)}$ could be structurally deformed to create $H_{TB}$, the system of interest. The sense in which this high symmetry system is ``close to'' $H_{TB}$ will presently be made clear. As a basis for the solution of $H_{TB}$ we take the Bloch states of the high symmetry $H_{TB}^{(0)}$ system, which are

\begin{equation}
 \ket{\psi_{\bk_I\alpha}} = \frac{1}{\sqrt{N}} \sum_{\bR_i} 
  e^{i \bk_I.(\bR_i + \boldsymbol{\nu}_\alpha)} \ket{\bR_i + \boldsymbol{\nu}_\alpha}
\end{equation}
where $\ket{\bR_i + \boldsymbol{\nu}_\alpha}$ are the localized tight-binding orbitals in which $\bR_i$ denotes a lattice site and $\bnu_\alpha$ a basis site. Other possible atomic labels, such as angular momentum or spin variables, are suppressed into the $\alpha$ label. We now consider an unknown Hamiltonian $H(\brh,\bph)$ that acts on a basis of free particle states

\begin{equation}
 \braket{\br}{\phi_{\bp_I\alpha}} = {\bf 1}_\alpha e^{i \bp_I\br}
\end{equation}
where $\bp_I = \bk_I - \bK_1$ is the crystal momentum measured from some expansion point $\bK_1$ in the Brillouin zone of $H_{TB}^{(0)}$. This will play a similar role to the expansion point in $\bk.\bp$ theory, i.e., this is the point in the Brillouin zone at which there exists some low energy spectrum of interest, for example in the case of graphene $\bK_1$ would be one of the high symmetry $K$ points. The vector ${\bf 1}_\alpha$ describes a general pseudospin degree of freedom and is defined as $\left[{\bf 1}_\alpha\right]_i = \delta_{i\alpha}$. Taking again graphene as an example these would be the pseudospin up, $(1, 0)^T$, and pseudospin down, $(0, 1)^T$, vectors. We now require $H(\brh,\bph)$ to be the operator equivalent of $H_{TB}$:

\begin{equation}
 \mel{\psi_{\bk_I\alpha}}{H_{TB}}{\psi_{\bk_J\beta}} = \mel{\phi_{\bp_I\alpha}}{H(\brh,\bph)}{\phi_{\bp_J\beta}}
 \label{Hequiv}
\end{equation}
%
% isomorphic
This equation contains only one unknown, the effective continuum Hamiltonian $H(\brh,\bph)$. Our strategy will be to derive an exact form for this Hamiltonian by manipulating this operator equivalence expression.
The left hand side of Eq.~(\ref{Hequiv}) is given by

\begin{eqnarray}
 \mel{\psi_{\bk_I\alpha}}{H_{TB}}{\psi_{\bk_J\beta}} & = & \frac{1}{N} \sum_{\bR_i\bR_j}
 e^{-i\bk_I.(\bR_i+\bnu_\alpha)}e^{i\bk_J.(\bR_j+\bnu_\beta)} \nonumber \\
 &\times&  \mel{\bR_i+\bnu_\alpha}{H_{TB}}{\bR_j+\bnu_\beta}
 \label{meltb}
\end{eqnarray}
where

\begin{eqnarray}
 \mel{\bR_i+\bnu_\alpha}{H_{TB}}{\bR_j+\bnu_\beta} = \nonumber \\ 
 \begin{cases}
  t_{\alpha\beta}(\bR_i + \bnu_\alpha,\bR_j+\bnu_\beta-\bR_i-\bnu_\alpha) \\
  t_{\beta\alpha}(\bR_j + \bnu_\beta,\bR_i+\bnu_\alpha-\bR_j-\bnu_\beta)
 \end{cases}
 \label{thop}
\end{eqnarray}
with $t_{\alpha\beta}(\br,\bdel)$ the electron hopping between position $\br$ on sublattice $\alpha$ and position $\br+\bdel$ on sublattice $\beta$. Note that these two spatial variables are very different in nature: $\br$ is a position in the lattice while $\bdel$ describes a hopping vector from point $\br$. We now introduce the Fourier transform of this hopping function

\begin{equation}
 t_{\alpha\beta}(\br,\bdel) = \frac{1}{(2\pi)^{2d}}
 \int d\bq'd\bq\, e^{-i\bq'.\br}e^{-i\bq.\bdel} t(\bq',\bq)
\end{equation}
(where $d$ the dimension of space) into Eq.~(\ref{meltb}) to find

\begin{eqnarray}
 \mel{\psi_{\bk_I\alpha}}{H_{TB}}{\psi_{\bk_J\beta}} & = & \frac{1}{(2\pi)^{2d}}\frac{1}{N} \sum_{\bR_i\bR_j} \int d\bq'd\bq\, t(\bq',\bq) \nonumber \\
 & \times & e^{-i(\bk_I+\bq'-\bq).(\bR_i+\bnu_\alpha)} \nonumber \\
 & \times & e^{i(\bk_J-\bq).(\bR_j+\bnu_\beta)}.
\end{eqnarray}
Use of both Poisson sum relation $\sum_\bR e^{i\bk.\bR} = \Omega_{BZ} \sum_{\bG} \delta(\bk+\bG)$ ($\Omega_{BZ}$ is the Brillouin zone volume of the high symmetry reference system) and the integral representation of the Dirac delta function $\delta(\bk) = 1/(2\pi)^d \int d\br e^{i\bk.\br}$ sends the double sum over direct space lattice vectors to a double sum over reciprocal lattice vectors:

\begin{eqnarray}
 \mel{\psi_{\bk_I\alpha}}{H_{TB}}{\psi_{\bk_J\beta}} & = & \frac{1}{V} \int d\br
 e^{i(\bk_J-\bk_I).\br}  \label{Gdouble} \\
 & \times &
 \frac{1}{V_{UC}} \sum_{\bG_i\bG_j} \int d\bq' e^{-i(\bG_i-\bG_j+\bq').\br} \nonumber \\
 & \times & t(\bq',\bk_J+\bG_j) e^{i\bG_i.\bnu_\alpha} e^{-i\bG_j.\bnu_\beta} \nonumber
\end{eqnarray}
We now make the only approximation of this derivation: that $t(\bq',\bq)$ is negligible for $|\bq'|$ comparable to the magnitude of the reciprocal lattice primitive vectors, implying that the double sum $\{\bG_i,\bG_j\}$ in Eq.~\eqref{Gdouble} can be reduced to a single sum $\{\bG_i\}$. An examination of the implications of this assumption we postpone to the end of this section. Given this assumption we find, by performing an inverse Fourier transform for the variable $\bq'$, a compact form for the tight-binding matrix element

\begin{eqnarray}
 \mel{\psi_{\bk_I\alpha}}{H_{TB}}{\psi_{\bk_J\beta}} & = & \frac{1}{V} \int d\br
 e^{i(\bk_J-\bk_I).\br}  \label{DUR} \\
 & \times &
 \frac{1}{V_{UC}} \sum_{i} \left[M_i\right]_{\alpha\beta} 
 t_{\alpha\beta}(\br,\bK_i+\bp_J) \nonumber
\end{eqnarray}
in this expression the ``$M$-matrices'' $M_i$ are given by

\begin{equation}
 \left[M_i\right]_{\alpha\beta} = e^{i(\bK_i-\bK_1).(\bnu_\alpha-\bnu_\beta)}
 \label{M}
\end{equation}
and the \emph{mixed space hopping function} $t_{\alpha\beta}(\br,\bq)$ is defined as

\begin{equation}
 t_{\alpha\beta}(\br,\bq) = \int d\bdel e^{i\bq.\bdel} t_{\alpha\beta}(\br,\bdel)
 \label{tq}
\end{equation}
Note also that the sum is taken over the \emph{translation group of the expansion point of the high symmetry reference system}: $\bK_i = \bK_1 + \bG_i$. This is very different in character from $\bk.\bp$ theory in which it is the point group that plays the central role.

Noting that

\begin{equation}
 \mel{\phi_{\bp_I\alpha}}{H(\brh,\bph)}{\phi_{\bp_J\beta}} =
 \frac{1}{V} \int d\br e^{i(\bp_J-\bp_I).\br} [H(\br,\bp)]_{\alpha\beta}
\end{equation}
we then can immediately ``read off'' the effective Hamiltonian as

\begin{equation}
 \left[H(\brh,\bph)\right]_{\alpha\beta} = \frac{1}{V_{UC}}\sum_{i}  \left[M_i\right]_{\alpha\beta} 
 t_{\alpha\beta}(\brh,\bK_i+\hat{\bp}/\hbar)
 \label{HM}
\end{equation}
where we have simply used the fact that $\bk_J-\bk_I=\bp_J-\bp_I$ (the expansion point $\bK_1$ cancels on the left hand side of this equality) and raised the $\bp_J$ variable to an operator $\bph/\hbar$. In this way when the Hamiltonian Eq.~\eqref{HM} acts to the right on the ket $\ket{\phi_{\bp_J\beta}}$ the operator equivalence equation $\mel{\psi_{\bk_I\alpha}}{H_{TB}}{\psi_{\bk_J\beta}} = \bra{\phi_{\bp_I\alpha}}\big(H(\brh,\bph)\ket{\phi_{\bp_J\beta}}\big)$ is satisfied (to see this we may simply reverse the steps leading to Eq.~\eqref{HM} to go from $\bra{\phi_{\bp_I\alpha}}\big(H(\brh,\bph)\ket{\phi_{\bp_J\beta}}\big)$ back to the tight-binding matrix element).
This equation is the central result of this section and proves that under the assumption $|\bq'| < |\bG_i|$ there exists a direct map between the two-centre tight-binding method and a continuum description. The operator $H(\brh,\bph)$ must, of course, satisfy associativity and Hermiticity. The former implies that $\big(\bra{\phi_{\bp_I\alpha}}H(\brh,\bph)\big)\ket{\phi_{\bp_J\beta}}$ must also equal Eq.~(\ref{DUR}) which is easily proved by repeating the derivation using the second of the two equivalent forms of the real space hopping function, Eq.~\eqref{thop}.

The central object in Eq.~\eqref{HM} is the mixed space hopping function $t_{\alpha\beta}(\br,\bq)$, and how to obtain this holds the key to applicability of method. For intrinsically perturbative cases, such as a lattice deformation, $t_{\alpha\beta}(\br,\bq)$ is obtained by Taylor expansion with respect to the small parameter of the deformation (the strain tensor) as we will show in the next section. However, for non-perturbative cases, such as twist faults and partial dislocations in bilayer systems, $t_{\alpha\beta}(\br,\bq)$ can, crucially, be obtained non-perturbatively (Section IV).

We close this section with an examination of the conditions under which the effective Hamiltonian $H(\brh,\bph)$ is hermitian. Evidently, the Hermiticity requirements of this object must be much stronger than those of the underlying tight-binding theory: while the tight binding Hamiltonian $H_{TB}$ is always hermitian if, for example, we attempted to find an effective Hamiltonian for amorphous carbon using the diamond structure as the high symmetry reference state, we would not expect the method to work, and this would be revealed by the effective Hamiltonian not being hermitian. For Hermiticity of $H(\brh,\bph)$ we require

\begin{equation}
 \left[H(\brh,\bph)\right]_{\alpha\beta} = \left[H(\brh,\bph)\right]_{\beta\alpha}^\ast
\end{equation}
which, since $\left[M_i\right]_{\alpha\beta} = \left[M_i\right]_{\beta\alpha}^\ast$, implies $t_{\alpha\beta}(\br,\bq) = t_{\beta\alpha}^\ast(\br,\bq)$. This in turn implies that for the real space hopping function we must have 

\begin{equation}
 t_{\alpha\beta}(\br,\bdel) = t_{\beta\alpha}(\br,-\bdel).
 \label{herm}
\end{equation}
Hopping between sites on sublattices $\alpha$ and $\beta$ in the high symmetry reference state defines a Bravais lattice that, as all Bravais lattices must, possesses inversion symmetry. Equation \eqref{herm} then states that this property must, within the range of the electron hopping for which $t_{\alpha\beta}(\br,\bdel)$ is non-zero,  hold \emph{at each point $\br$ of the lattice after deformation}. In other words, the sublattice structure of the high symmetry crystal must hold good as a local description of the crystal after deformation for the effective Hamiltonian $H(\brh,\bph)$ to be Hermitian. This is different to $\bk.\bp$ theory in which the reference state must, for a useful $\bk.\bp$ Hamiltonian to be obtained, be \emph{globally} close to the system of interest. This difference between local and global closeness is, as we will see, one of the key reasons behind the usefulness of the approach we espouse here.

%%%%%%%%%%%%%%%%%%%%%%
% DEFORMATION THEORY %
%%%%%%%%%%%%%%%%%%%%%%

\section{Theory of deformations in 2d materials}

In this section we will treat the case of deformations in 2d materials, in which the deformation is slowly varying on the scale of the lattice constant. We will consider systems that, within a minimal basis tight-binding scheme, can be treated as having a single orbital per site; this includes all the 2d carbon allotropes. Based on a perturbative approach we will, beginning from Eq.~\eqref{HM}, derive a theory of deformations (of which obviously no deformation is a special case) applicable to \emph{all} such 2d materials. We will then deploy this theory to determine the (i) effective Hamiltonians of a range 2d carbon allotropes and (ii) a theory of deformations in single layer graphene. Generalizations of this theory to 3d and more than one orbital per site are straightforward, but will not be given here.

%%% General theory %%%%

\subsection{General theory of deformations in 2d materials}
\label{dmeth}

\begin{table}
\begin{tabular}{c|c|c||c}
$m_1$ & $m_2$ & r &  $c^{(r)}_{m_1 m_2}$ \\ \hline \hline
2 & 0 & 1 & $2u_{xx} + (\partial_x \bu)^2$ \\
1 & 1 & 1 & $4u_{xy} + \partial_x \bu.\partial_y \bu$ \\
0 & 2 & 1 & $2u_{yy} + (\partial_y \bu)^2$ \\
3 & 0 & 1 & $\partial_x u_{xx} + \frac{1}{2}\partial_x(\partial_x\bu)^2$ \\
2 & 1 & 1 & $\partial_yu_{xx} + 2\partial_xu_{xy} + \frac{1}{2} \partial_x(\partial_x\bu.\partial_y\bu) + \frac{1}{2}\partial_y(\partial_x\bu)^2$ \\
1 & 2 & 1 & $\partial_xu_{yy} + 2\partial_yu_{xy}  + \frac{1}{2} \partial_y(\partial_x\bu.\partial_y\bu) + \frac{1}{2}\partial_x(\partial_y\bu)^2$ \\
0 & 3 & 1 & $\partial_y u_{yy} + \frac{1}{2}\partial_y(\partial_y\bu)^2$ \\
4 & 0 & 1 & $\frac{1}{3} \partial_x^2 u_{xx} + \frac{1}{4} (\partial_x^2 \bu)^2 + \frac{1}{3} \partial_x \bu . \partial_x^3 \bu$ \\
3 & 1 & 1 & $\partial_x^2 u_{xy} + \frac{1}{3} \partial_x^2 u_{yx} + \partial_y (\partial_x \bu . \partial_x^2 \bu) + \frac{1}{3} \partial_y \bu . \partial_x^3 \bu $ \\
2 & 2 & 1 & $\partial_y^2 u_{xx} + \partial_x^2 u_{yy} + \partial_x \partial_y (\partial_x \bu . \partial_y \bu) - \frac{1}{2} \partial_x^2 \bu . \partial_y^2 \bu$ \\
1 & 3 & 1 & $\partial_y^2 u_{xy} + \frac{1}{3} \partial_y^2 u_{xy} + \partial_x (\partial_y \bu . \partial_y^2 \bu) + \frac{1}{3} \partial_x \bu . \partial_y^3 \bu $ \\
0 & 4 & 1 & $\frac{1}{3} \partial_y^2 u_{yy} + \frac{1}{4} (\partial_y^2 \bu)^2 + \frac{1}{3} \partial_y \bu . \partial_y^3 \bu$ \\
4 & 0 & 2 & $\frac{1}{2}c_{20}^2$\\
3 & 1 & 2 & $c_{20}c_{11}$ \\
2 & 2 & 2 & $\frac{1}{2}c_{11}^2+c_{20}c_{02}$ \\
1 & 3 & 2 & $c_{02}c_{11}$ \\
0 & 4 & 2 & $\frac{1}{2}c_{02}^2$
\end{tabular}
\caption{Coefficients $c^{(r)}_{m_1 m_2}$ that arise in the Taylor expansion of the tight-binding hopping function due to a deformation field $\bu(\br)$ applied to a two dimensional material, see Eq.~\eqref{1} of the text.
}
\label{c}
\end{table}

We first Taylor expand the term $t_{\alpha\beta}(\br,\bK_i+{\bp}/\hbar)$ in the general form of the effective Hamiltonian, Eq.~\eqref{HM}, with respect to $\bp$, i.e., we consider momenta only close to the expansion point:

\begin{eqnarray}
 t_{\alpha \beta}(\br,\bK_i+\bp/\hbar)& =& \sum_{n_1 n_2} \frac{1}{n_1!n_2!} \partial_{q_x}^{n_1} \partial_{q_y}^{n_2} \left.t_{\alpha\beta}(\br,\bq)\right|_{\bq=\bK_i} \nonumber \\
&& \times\left(\frac{{p_x}}{\hbar}\right)^{n_1} \left(\frac{{p_x}}{\hbar}\right)^{n_2}
 \label{0}
\end{eqnarray}
A material deformation is encoded in a 3-vector distortion field $\bu(\br)$ and thus we include the possibility of both in plane $(u_x(\br),u_y(\br)$) and out of plane $u_z(\br)$ deformations. This deformation will, at each point $\br$ in the crystal, change the hopping vector from $\bdel$ to $\bdel+\bu(\br+\bdel)-\bu(\br)$. This in turn sends the real space hopping function from that of the high symmetry reference state, $t_{\alpha\beta}(\bdel^2)$, to a more complex form describing the inhomogeneous electron hopping in the deformed material: $t_{\alpha\beta}\left([\bdel+\bu(\br+\bdel)-\bu(\br)]^2\right)$. Note that we assume $t_{\alpha\beta}(\bdel^2)$ has scalar $\bdel^2$ not $\bdel$ vector argument; a common assumption in single orbital tight-binding calculations but different, for example, from the Slonczewski-Weiss-McClure (SWM) method developed for graphite and often used in adopted form in calculations of the Bernal stacked graphene bilayer. In Sections V and IV, where we discuss in-plane and stacking deformations respectively, we will examine impact on the effective Hamiltonians of differences in the underlying tight-binding method from which they are derived. For the case of the Bernal bilayer we will show that the $t_{\alpha\beta}(\bdel^2)$ and SWM tight-binding methods lead to exactly the same low energy effective Hamiltonian.

Given this general form of the tight-binding hopping function we must Fourier transform the $\bdel$ variable of $t_{\alpha,\beta}(\br,\bdel) = t_{\alpha\beta}\left([\bdel+\bu(\br+\bdel)-\bu(\br)]^2\right)$ to obtain $t_{\alpha\beta}(\br,\bq)$ - the central object of Eq.~\eqref{0}. As both $\bu(\br)$ and $t_{\alpha\beta}(\bdel^2)$ are unknown functions this Fourier transform, self evidently, cannot be obtained in closed form. To make progress we recall that the function $\bu(\br)$ is assumed slow on the scale of the lattice constant, and is therefore also slow on the scale of the magnitude of $\bdel$ (the most significant hopping vectors will be on the order of the lattice constant). We may then perform a double Taylor expansion: (i) of $\bu(\br+\bdel)-\bu(\br)$ with respect to $\bdel$ and (ii) of $t_{\alpha\beta}([\bdel+\bu(\br+\bdel)-\bu(\br)]^2)$ with respect to $\bu(\br+\bdel)-\bu(\br)$. This results in the following expression:

\begin{equation}
 t(\br,\bdel) = \sum_r \frac{\partial^r t_{\alpha\beta}(\delta^2)}{\partial(\delta^2)^r} \sum_{m_1 m_2} c_{m_1 m_2}^{(r)} \delta_x^{m_1}\delta_y^{m_2}
 \label{1}
\end{equation}
where for the zeroth order in $r$ the expansion coefficients $c^{(0)}_{m_1 m_2}$ are zero except for the case $c^{(0)}_{00} = 1$ (this is the case of no deformation), while for $r>0$ the expansion coefficients depend on the deformation field $\bu(\br)$, and are presented in Table \ref{c}. Evidently the labels $m_1$ and $m_2$ are associated with the first of the two Taylor expansions described above, and $r$ with the second.

The Fourier transform of Eq.~\eqref{1} with respect to $\bdel$ is

\begin{equation}
 t_{\alpha\beta}(\br,\bq) = \sum_{\substack{r \\ m_1 m_2}} (-i)^{m_1+m_2} c^{(r)}_{m_1 m_2} \partial_{q_x}^{m_1} \partial_{q_y}^{m_2} t_{\alpha\beta}^{(r)} (q^2)
 \label{GOB}
\end{equation}
where

\begin{equation}
 t^{(r)}_{\alpha\beta}(q^2) = \int d\bdel e^{i\bq.\bdel} \frac{\partial^r t_{\alpha\beta}(\delta^2)}{\partial(\delta^2)^r}
\end{equation}
and inserting this back into Eq.~\eqref{0} we find the expression

\begin{eqnarray}
 t_{\alpha\beta}(\br,\bK_i+\bp/\hbar) & = & \sum_{\substack{r \\ n_1 n_2 \\ m_1 m_2}} \frac{ (-i)^{m_1+m_2} c^{(r)}_{m_1m_2}}{n_1!n_2!}\nonumber\\
 &\times&
 \partial_{q_x}^{n_1+m_1} \partial_{q_y}^{n_2+m_2}
 \left.t^{(r)}_{\alpha\beta}(q^2)\right|_{\bq=\bK_i} \nonumber \\
 &\times&\left(\frac{{p_x}}{\hbar}\right)^{n_1} \left(\frac{{p_x}}{\hbar}\right)^{n_2}
 \label{2}
\end{eqnarray}
Thus we have obtained the ``electronic'' part of the effective Hamiltonian Eq.~\eqref{HM} for a general deformation of a 2d membrane in terms of (i) the Fourier transform of the hopping function of the high symmetry system $t^{(r)}_{\alpha\beta}(q^2)$, and (ii) the deformation via the coefficients $c^{(r)}_{ij}$ of Table \ref{c}. Substitution of this into Eq.~\eqref{HM} will now give us the result we seek: a general effective continuum Hamiltonian for deformations in 2d systems. This will be a compact but not particularly transparent expression and to render the result into a clear form we must separate the scalar $q^2$ and vector $q_i$ dependencies in Eq.~\eqref{2}. This may be achieved simply by repeated application of the chain rule to the derivatives in Eq.~\eqref{2}:

\begin{equation}
 \frac{\partial}{\partial q_i} = \frac{\partial(q^2)}{\partial q_i} \frac{\partial}{\partial(q^2)} =
 \left(\frac{2\pi}{a}\right) 2q_i \frac{\partial}{\partial(q^2)}
\end{equation}
where, with a slight abuse of notation, the vector component $q_i$ is chosen to be dimensionless (hence the prefactor of $2\pi/a$) while the scalar quantity $q^2$ is dimensionfull.
Working this out (a tedious though entirely trivial task) we find the Hamiltonian factors into two parts

\begin{equation}
 H = \sum_{\substack{i r \\ o_1 o_2 p}} \bchi^r_{o_1 o_2 p}(\bK_i) \Phi^{(r)}_{o_1 o_2 p}(\bK_i)
\label{DH}
\end{equation}
the first factor

\begin{eqnarray}
 \left[\bchi^r_{o_1 o_2 p}(\bq)\right]_{\alpha\beta} & = & \frac{1}{V_{UC}} \left(\frac{2\pi}{a}\right)^{2p-o_1-o_2} \frac{\partial^p t^{(r)}_{\alpha\beta}(q^2)}{\partial(q^2)^p}\nonumber\\
 &\times&e^{i(\bq-\bK_1).(\bnu_\alpha-\bnu_\beta)}
 \label{CHI}
\end{eqnarray}
is matrix valued and contains all material specific information. It is labeled by 4 numbers that arise from the various Taylor expansions involved in the theory: $r$ the order of the expansion of the tight-binding hopping function (``electronic expansion''); $o_1=m_1+m_2$ the order of the Taylor expansion in the hopping vector $\delta$ (Eq.~\eqref{1}, ``geometric expansion''); $o_2 = n_1+n_2$ the order of the momentum Taylor expansion in Eq.~\eqref{0}; and $p$ the order of the derivative $\partial_{q^2}$. This expression, despite the multiple indices due to its general nature, is nevertheless evidently rather easy to calculate and requires only the Fourier transform of the hopping function of the high symmetry lattice, $t_{\alpha\beta}(q^2)$, the expansion point choice $\bK_1$, and geometric information of the high symmetry system through the translation group of $\bK_1$ and the basis vectors of the high symmetry system $\bnu_\alpha$.

\begin{table*}[thbp]
\begin{tabular}{c|c|c|l}
$o_1$ & $o_2$ & $p$ & $\Phi^{(r)}_{o_1 o_2 p}$ \\ \hline
0 & 0 & 0 & 1 \\
0 & 1 & 1 & $\bp.\bq$ \\
0 & 2 & 1 & $p^2$ \\
0 & 2 & 2 & $2(\bp.\bq)^2$ \\
0 & 3 & 2 & $2p^2\bp.\bq$ \\
0 & 3 & 3 & $\frac{4}{3}(\bp.\bq)^3$ \\
0 & 4 & 2 & $\frac{1}{2}p^4$ \\
0 & 4 & 3 & $2p^2(\bp.\bq)^2$ \\
0 & 4 & 4 & $\frac{2}{3}(\bp.\bq)^4$ \\
0 & 5 & 3 & $p^4\bp.\bq$ \\
0 & 5 & 4 & $\frac{4}{3}p^2(\bp.\bq)^3$ \\
0 & 5 & 5 & $\frac{4}{15}(\bp.\bq)^5$ \\
2 & 0 & 1 & $-2(c^{(r)}_{20} + c^{(r)}_{02})$ \\
2 & 0 & 2 & $-4(c^{(r)}_{20}q_x^2 + c^{(r)}_{11}q_x q_y + c^{(r)}_{02} q_y^2)$ \\
2 & 1 & 2 & $-4\left( [3c^{(r)}_{20} + c^{(r)}_{02}]p_xq_x + c^{(r)}_{11}[p_yq_x + p_xq_y] + [c^{(r)}_{20} + 3c^{(r)}_{02}]p_yq_y  \right)$ \\
2 & 1 & 3 & $-8\bp.\bq(c^{(r)}_{20}q_x^2+c^{(r)}_{11}q_xq_y + c^{(r)}_{02}q_y^2)$ \\
2 & 2 & 2 & $-2\left( [3c^{(r)}_{20} + c^{(r)}_{02}]p_x^2 + c^{(r)}_{11}p_xp_y + [c^{(r)}_{20} + 3c^{(r)}_{02}]p_y^2  \right)$ \\
2 & 2 & 3 & $-4\Big( 6c^{(r)}_{20}p_x^2q_x^2 + [3c^{(r)}_{11}p_x^2 + 6[c^{(r)}_{20}+c^{(r)}_{02}] p_xp_y + 3c^{(r)}_{11} p_y^2]q_xq_y + 6c^{(r)}_{02}p_y^2q_y^2-[c^{(r)}_{02}p_x^2 + 2c^{(r)}_{11}p_xp_y + c^{(r)}_{20} p_y^2]\bq^2 \Big)$ \\
2 & 2 & 4 & $-8(\bp.\bq)^2(c^{(r)}_{20}q_x^2+c^{(r)}_{11}q_xq_y + c^{(r)}_{02}q_y^2)$\\
 3 & 0 & 2 & $i4[(c^{(r)}_{12}+3c^{(r)}_{31})q_x + (c^{(r)}_{21}+3c^{(r)}_{03})q_y]$ \\
 3 & 0 & 3 & $i8(c^{(r)}_{30} q_x^3 + c^{(r)}_{21}q_x^2q_y + c^{(r)}_{12}q_x q_y^2 + c^{(r)}_{03} q_y^3)$ \\
 4 & 0 & 2 & $4(3c^{(r)}_{04}+c^{(r)}_{22} + 3c^{(r)}_{40})$ \\
 4 & 0 & 3 & $8q_x^2(c^{(r)}_{22}+6c^{(r)}_{40}) + 24q_xq_y(c^{(r)}_{13}+c^{(r)}_{31}) + 8q_y^2(6c^{(r)}_{04} + c^{(r)}_{22})$ \\
 4 & 0 & 4 & $16(q_x^4c^{(r)}_{40}+q_x^3q_yc^{(r)}_{31} + q_x^2q_y^2c^{(r)}_{22} + q_xq_y^3c^{(r)}_{13} + q_y^4c^{(r)}_{04})$ \\
 4 & 1 & 3 & $24q_x\left[(c^{(r)}_{04}+c^{(r)}_{22}+5c^{(r)}_{40})p_x + (c^{(r)}_{13}+c^{(r)}_{31})p_y\right] + 24q_y\left[(c^{(r)}_{13}+c^{(r)}_{31})p_x+(c^{(r)}_{40}+c^{(r)}_{22}+5c^{(r)}_{04})p_y\right]$ \\
\multirow{2}{*}{4} & \multirow{2}{*}{1} & \multirow{2}{*}{4} & $16q_x^3[(c^{(r)}_{22}+10c^{(r)}_{40})p_x+c^{(r)}_{31}p_y]+48q_x^2q_y[(c^{(r)}_{13}+2c^{(r)}_{31})p_x+(c^{(r)}_{22}+2c^{(r)}_{40})p_y]$ \\
& & & $+48q_xq_y^2[(c^{(r)}_{22}+2c^{(r)}_{04})p_x+(2c^{(r)}_{13}+c^{(r)}_{31})p_y]+16q_y^3[c^{(r)}_{13}p_x+(c^{(r)}_{22}+10c^{(r)}_{04})p_y]$ \\
 4 & 1 & 5 & $32\bp.\bq(q_x^4c^{(r)}_{40} + q_x^3q_yc^{(r)}_{31} + q_x^2q_y^2c^{(r)}_{22} + q_xq_y^3c^{(r)}_{13} + q_y^4c^{(r)}_{04}$)
\end{tabular}
\caption{\emph{Universal functions for two dimensional systems}: The universal functions $\Phi^{(r)}_{o_1 o_2 p}$ for slow perturbative deformations in 2d materials. The index $o_1$ is the order of the Taylor expansion in the hopping vector $\delta$ (see Eq.~\eqref{1}, ``geometric expansion''); the index $o_2$ is the order of the momentum Taylor expansion (see Eq.~\eqref{0}, ``momentum expansion''); and $p$ the order of the derivative $\partial_{q^2}$ applied to the high symmetry hopping function. The index $r$ is the order of the expansion of the tight-binding hopping function (``electronic expansion'') and enters only through the $c^{(r)}_{ij}$ which are given in terms of components of the deformation tensor and derivatives of the deformation field, $\bu(\br)$, as indicated in Table \ref{c}, see also Eq.~\eqref{1}. These functions are rotational invariants of the basic variables of the problem: the momentum operator $\bp$, the deformation field (via the $c^{(r)}_{ij}$), and a reciprocal space vector $\bq$. They are universal under the (quite general) assumptions of: (i) a two dimensional material, (iii) deformations slow on the scale of the lattice constant, (iii) a single orbital per site, and (iv) a tight-binding hopping function that has a scalar dependence on the hopping vector, i.e., $t(\bdel^2)$. These assumptions allow for a very accurate description of the electronic structure of graphene, few layer graphenes, and the many complex all carbon 2d allotropes. The generalization to more than one orbital per site, that allows for the treatment of all 2d materials, is straightforward (see text for details).
}
\label{uni}
\end{table*}

\begin{table*}[t!]
\begin{small}
\begin{tabular}{c|c|c|c} 
  Material & $H_0$ 
           & $H_x$
           & $H_y$ \\ \hline \hline
  Graphene & $\alpha_0 \sigma_0$ & $v_F \sigma_x$ & $v_F \sigma_y$ \\
  Graphdiyne & $\Delta\begin{pmatrix} -\sigma_0 & 0 \\ 0 & \sigma_0 \end{pmatrix}$
             & $v\begin{pmatrix} 0  & \sigma_x \\ \sigma_x & 0 \end{pmatrix}$ 
             & $v\begin{pmatrix} 0  & \sigma_y \\ \sigma_y & 0 \end{pmatrix}$ \\
  6,6,12-graphyne & \multirow{2}{*}{$\alpha_0\sigma_0$} 
                  & \multirow{2}{*}{$v^F_x \sigma_x + v^F_0 \sigma_0$} 
                  & \multirow{2}{*}{$v^F_y \sigma_y$} \\
  Cone I          & & & \\
  6,6,12-graphyne & \multirow{4}{*}{$\begin{pmatrix} \epsilon_1 & 0 & 0 & 0 \\ 0 & 0 & 0 & 0 \\ 0 & 0 & 0 & 0 \\ 0 & 0 & 0 & \epsilon_2 \end{pmatrix}$} 
                  & \multirow{4}{*}{$\begin{pmatrix} 0 & v_x^{(1)} & v_x^{(2)} & 0 \\ v_x^{(1)} & 0 & 0 & v_x^{(3)} \\ v_x^{(2)} & 0 & 0 & v_x^{(4)} \\ 0 & v_x^{(3)} & v_x^{(4)} & 0 \end{pmatrix}$} 
                  & \multirow{4}{*}{$\begin{pmatrix} 0 & 0 & 0 & v_y^{(1)} \\ 0 & v_y^{(2)} & v_y^{(3)} & 0 \\ 0 & v_y^{(3)} & v_y^{(4)} & 0 \\ v_y^{(1)} & 0 & 0 & 0 \end{pmatrix}$} \\
  cone II         & & & \\
  &&& \\
  &&& \\
  &&& \\
  $\gamma$-graphyne & \multirow{4}{*}{$\begin{pmatrix} \epsilon_1 & 0 & 0 & 0 \\ 0 & -\Delta & 0 & 0 \\ 0 & 0 & \Delta & 0 \\ 0 & 0 & 0 & \epsilon_2 \end{pmatrix}$} 
                    & \multirow{4}{*}{$\begin{pmatrix} 0 & 0 & 0 & v_x^{(1)} \\ 0 & 0 & v_x^{(2)} & 0 \\ 0 & {v_x^{(2)}}^\ast & 0 & 0 \\ {v_x^{(1)}}^\ast & 0 & 0 & 0 \end{pmatrix}$} 
                    & \multirow{4}{*}{$\begin{pmatrix} 0 & 0 & v_y^{(1)} & 0 \\ 0 & 0 & 0 & v_y^{(2)} \\ {v_y^{(1)}}^\ast & 0 & 0 & 0 \\ 0 & {v_y^{(2)}}^\ast & 0 & 0 \end{pmatrix}$} \\
  &&& \\
  &&& \\
  &&&
\end{tabular}
\end{small}
\vspace{0.3cm}
\caption{\emph{Pseudospin algebra for carbon allotropes}: Low energy continuum Hamiltonians $H = H_0 + H_x p_x + H_y p_y$ for a selection of two dimensional all carbon allotropes. These Hamiltonians are obtained from the lattice to pseudospin connection formula Eq.~\eqref{XXXX} and Eq.~\eqref{CHI}.}
\label{pseu}
\end{table*}

The second factor $\Phi^{(r)}_{o_1 o_2 p}(\bq)$ consists of polynomials in $\bq$, $\bp$, $\{c^{(r)}_{m_1 m_2}\}$, and are tabulated in Table \ref{uni}. Strikingly, \emph{these polynomials are universal}: they depend on none of the system specific objects of the $\bchi^r_{o_1 o_2 p}(\bq)$ matrix and, once calculated, may be used for any 2d system possessing (in a tight-binding basis) a single orbital per site. They are, in fact, the complete set of rotational invariants that may be constructed from the basic variables of the theory $\bp$, $\bq$, and $\{c^{(r)}_{m_1 m_2}\}$, most easily seen for the polynomials containing only the vectors $\bp$ and $\bq$. This property results from the underlying rotation symmetry of the mixed space hopping function: $t_{\alpha\beta}(\br,\bq) = t_{\alpha\beta}(R\br,R\bq)$ that is preserved term by term in the Taylor expansions involved in the derivation above. Evaluated over the translation group of the expansion point in Eq.~\eqref{DH}, $\{\bK_i\}$, the rotational symmetry of these polynomials is then reduced to that of the point group symmetry of the translation group $\{\bK_i\}$.
We stress that these polynomials are universal only under the following conditions: (i) a two dimensional material; (ii) a single orbital per site; and (iii) a hopping function that has a scalar dependence on the hopping vector.

The combination of $\bchi^r_{o_1 o_2 p}(\bK_i)$ and $\Phi^{(r)}_{o_1 o_2 p}(\bK_i)$ in Eq.~\eqref{DH} leads to the requirement to evaluate the following general form:

\begin{equation}
 h^{(m,n)}_{r o_1 o_2 p} = \sum_i \bchi^r_{o_1 o_2 p}(\bK_i) K_{ix}^m K_{iy}^n.
\label{XXXX}
\end{equation}
For the purposes of later reference we will refer to such an expression as the ``connection formula'': the right hand side consists of lattice information through $\bchi^r_{o_1 o_2 p}(\bK_i)$ and the $\{\bK_i\}$, the left hand side is an object in the pseudospin space of the effective Hamiltonian. It thus encodes the link between the lattice degree of freedom of the material and the pseudospin degree of freedom of effective Hamiltonian describing the material. 

The degree to which Eqs.~\eqref{DH}-\eqref{XXXX} are analytically tractable depends on the complexity of the system under consideration, but in particular on whether the sublattice and pseudospin spaces are isomorphic. For graphene, in which the number of sublattice degrees of freedom (2) is equal to the dimension of the pseudospin space required to describe the Dirac cone, these equations (particularly if the sum $i$ over the translation group is truncated) can be manipulated very easily.
However, for more complex materials - represented here by the more complex 2d allotropes of carbon - the number of sublattice degrees of freedom exceeds the number of pseudospin degrees of freedom necessary for a description of the low energy band manifold. In the case of graphdiyne, $\gamma$-graphyne, and 6,6,12-graphyne we have 18, 12, and 18 basis atoms respectively, while the dimension of pseudospin space required to describe the low energy manifolds in these systems is (at most) 4. A second step (standard also in $\bk.\bp$ theory) is therefore required and this is to transform from the pseudospin space of the full Hamiltonian to a space of eigenfunctions at the expansion point. This is achieved simply by a unitary transform $U H U^\dagger$ with $U$ the matrix that diagonalizes the Hamiltonian at the expansion point $\bK_1$. The block of the Hamiltonian describing the low energy manifold may then  easily be extracted from the transformed $H$. Increasing the dimension of the extracted $H$ amounts to increasing the number of bands included in the effective Hamiltonian. Note that in such a case the resulting effective Hamiltonian is still analytical in the basic variables of the problem, $\bp$ and the deformation tensor for example, but all coefficients in the Hamiltonian are obtained numerically.

%%%% Lattices without deformation %%%%

\subsection{Lattices without deformation} \label{OINK}

%-------------------------- FIG --------------------------
\begin{figure}[t!]
  \centering
  \includegraphics[width=0.93\linewidth]{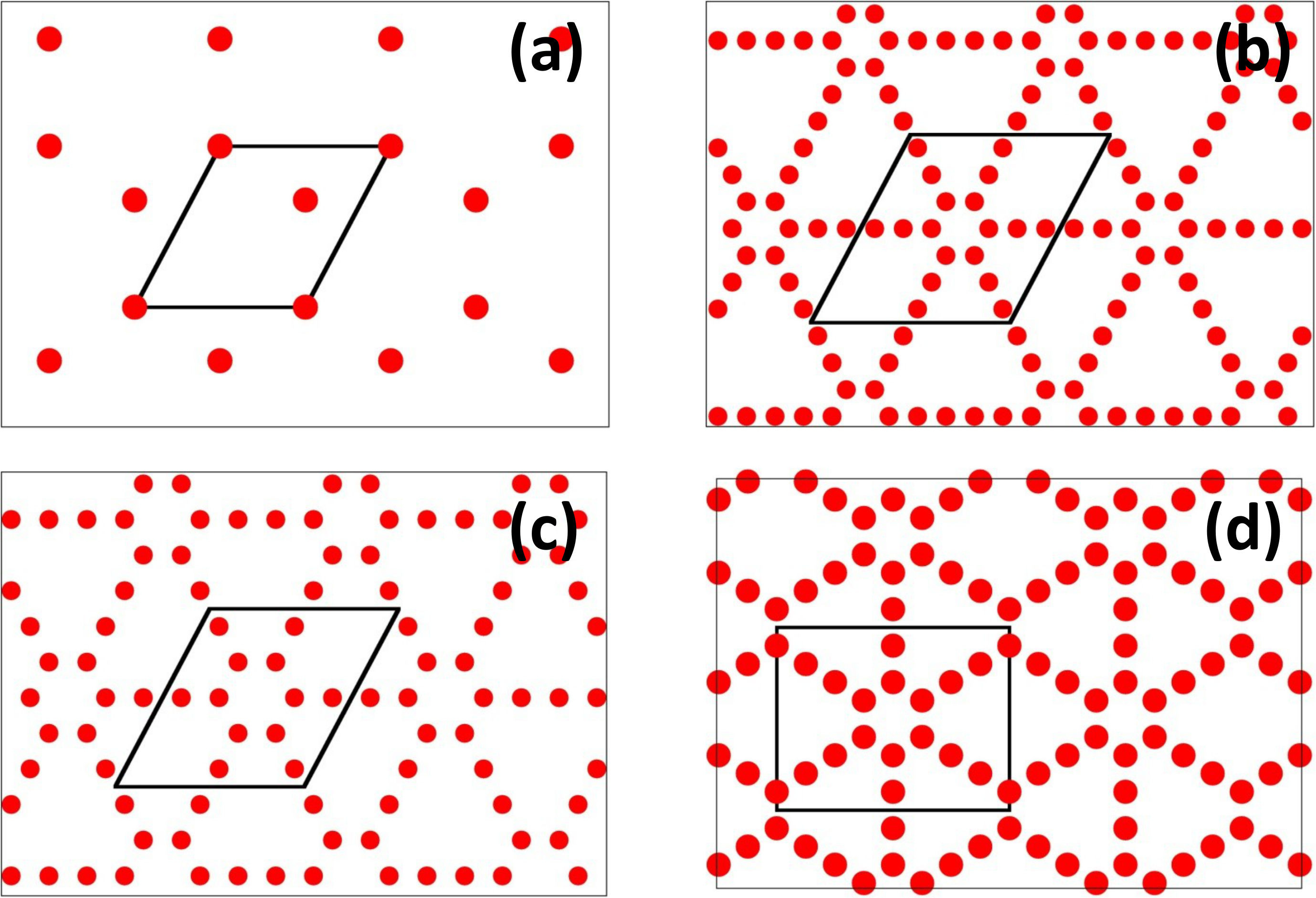}
  \caption{Real space lattices of the materials for which effective Hamiltonians are derived in Section \ref{OINK}: (a) graphene, (b) graphdiyine, (c) $\gamma$-graphyne, and (d) 6,6,12-graphyne.
  }
  \label{Fig1}
\end{figure}
%-------------------------- FIG --------------------------

We now consider the ideal lattices of a number of 2d carbon allotropes: graphene, graphdiyne, 6,6,12-graphyne, and $\gamma-$graphyne; real space lattices of the materials are presented in Fig.~\ref{Fig1}. Of these only the first two have been produced experimentally\cite{nov04,li10}, although it should be mentioned that graphdiyne has not yet been characterized with STM and is certainly very far from the ease of production enjoyed by graphene. The remaining two members of this list are to be found not in experiment, but merely in a theoretical zoo by now well populated with potential all-carbon 2d materials\cite{bau87,nar98,li10,kang11,en11,mal12,kim12,liu12,cui13,ivan13,chen13,huang13,xi14,pad14,wang14,mi14,mi14a,yang14,yang15}, see especially the reviews Refs.~\onlinecite{en11,yang15}. These four materials are chosen as together they contain many of the features seen generically in all-carbon 2d systems: Dirac cones; gapped low energy manifolds; Dirac points at high as well as low symmetry points in the Brillouin zone; and $C_4$ as well as $C_6$ point group symmetry of the lattice.

\emph{Underlying tight-binding method}: We must first specify our tight binding scheme and for the general form of the hopping function we choose a Gaussian form

\begin{equation}
 t(\bdel) = A \exp(-B\bdel^2).
 \label{hop}
\end{equation}
In the case of the more complex 2d allotropes that contain acetylene bonds we require 3 such hopping functions to describe: (i) electron hopping between atoms of the acetylene bond; (ii) between an atom that has an acetylene bond and another atom with only $sp^2$ bonds; (iii) between atoms with $sp^2$ bonds only\cite{mi14}. For graphene we therefore have a theory with 2 unknown constants, and for graphdiyne, $\gamma$-graphyne, and 6,6,12-graphyne a theory with unknown 6 constants. 

It is useful at this point to once again draw a contrast with $\bk.\bp$ theory. In $\bk.\bp$ theory it is the optical matrix elements that must be fitted to \emph{ab-initio} (or experimental) band data, and one relies on symmetry considerations to reduce - if possible - the number of such unknown matrix elements. On the other hand in the method we describe here however many constants arise in our final theory, and there may be many if the material is complex or the Taylor expansion in momentum or deformation taken to high order, the unknown constants to be fitted occur at the more fundamental level of the tight-binding hopping function, not on the level of individual matrix elements. The number of unknowns in the theory is thus sharply curtailed \emph{independent of particular symmetries that the problem may or may not have}, a fact that one expects to be highly advantageous in the treatment of either low symmetry systems of systems for which high orders in the momentum or deformation tensor are required. For example, for the pristine graphene lattice we have, when taking the momentum expansion to third order, 6 numerical constants; in the full treatment of deformations in graphene 12 numerical constants appear. However, these are all derived via the connection formula from the 2 basic parameters of the underlying tight-binding method and thus the number of ``fitting parameters'' is the same whether one obtains the simplest effective Hamiltonian for graphene, $H=v_F \bsig.\bp$, or one derives the much more complex correction terms in momentum and deformation. In what follows we will not clutter the presentation with explicit values of these numerical parameters; the \emph{form} of the effective Hamiltonians is the crucial issue and while the numerical coefficients depend on the particular flavour of the underlying tight-binding method, these Hamiltonian forms do not.

\begin{table}[htbp]
\begin{tabular}{ccc}
$M_1$ & $M_2$ & $M_3$ \\ \hline\hline
 $\begin{pmatrix} 1 & 1  \\ 1 & 1 \end{pmatrix}$ & 
 $\begin{pmatrix} 1 & e^{-i2\pi/3} \\ e^{i2\pi/3} & 1 \end{pmatrix}$ & 
 $\begin{pmatrix} 1 & e^{i2\pi/3}  \\ e^{-i2\pi/3} & 1 \end{pmatrix}$ \\
\hline
\end{tabular}
\caption{$M_i$ matrices for the first star of the translation group of the expansion point in single layer graphene.}
\label{MSLG}
\end{table}

\begin{table}[htbp]
 \begin{tabular}{c|c|c|c}
  $m$ & $n$ & First star only & Full summation \\ \hline\hline
   0 & 0 & $C_0 \sigma_0$ & $C_0 \sigma_0$  \\ \hline
   1 & 0 & $C_1 \sigma_x$ & $C_1 \sigma_x$  \\
   0 & 1 & $C_1 \sigma_x$ & $C_1 \sigma_y$ \\ \hline
   2 & 0 & $2C_2 \sigma_0 + C_2 \sigma_x$ & $C_2\sigma_0+C_3 \sigma_x$ \\
   1 & 1 & $-C_2 \sigma_y$ & $C_3 \sigma_y$ \\
   0 & 2 & $2C_2\sigma_0 - C_2\sigma_x$ & $C_2\sigma_0-C_3 \sigma_x$ \\ \hline
   3 & 0 & $\frac{2}{3}C_3\sigma_0 + C_3\sigma_x$ & $C_4\sigma_0 + 3C_5\sigma_x$ \\
   2 & 1 & $\frac{1}{3}C_3\sigma_y$ & $C_5\sigma_y$ \\
   1 & 2 & $-\frac{2}{3}C_3\sigma_0 + \frac{1}{3}C_3\sigma_x$ & $-C_4\sigma_0+C_5\sigma_x$ \\
   0 & 3 & $C_3\sigma_y$ & $3C_5\sigma_y$
 \end{tabular}
 \caption{Sublattice to pseudospin connection relation for graphene for both a first star approximation (Eq.~\eqref{YOSIMPLE}) and full convergence over all stars of the translation group of the expansion point (Eq.~\eqref{CV}). The coefficients in these expressions are denoted by a generic notation simply to indicate which coefficients are the same and which different; the precise values that the coefficients $C_i$ depend on the particular values of that $r$, $o_1$, $o_2$, and $p$ take, see Section \ref{OINK}.}
 \label{g}
\end{table}

\emph{Single layer graphene}: For graphene the general formalism described in the previous section undergoes considerable simplification as the electron hopping function $t(\bdel^2)$ is identical for both sublattices. As a result we may write the general Hamiltonian as

\begin{equation}
 H= \sum_{r\, o_1\, o_2\, p\, i} C_{o_1o_2p}^{(r)}(K_i) M_i \Phi^{(r)}_{o_1o_2p}(\bK_i)
\end{equation}
where the constants $C_{o_1o_2p}^{(r)}(K_i)$ are given by

%-------------------------- FIG --------------------------
\begin{figure}[t!]
  \centering
  \includegraphics[width=0.93\linewidth]{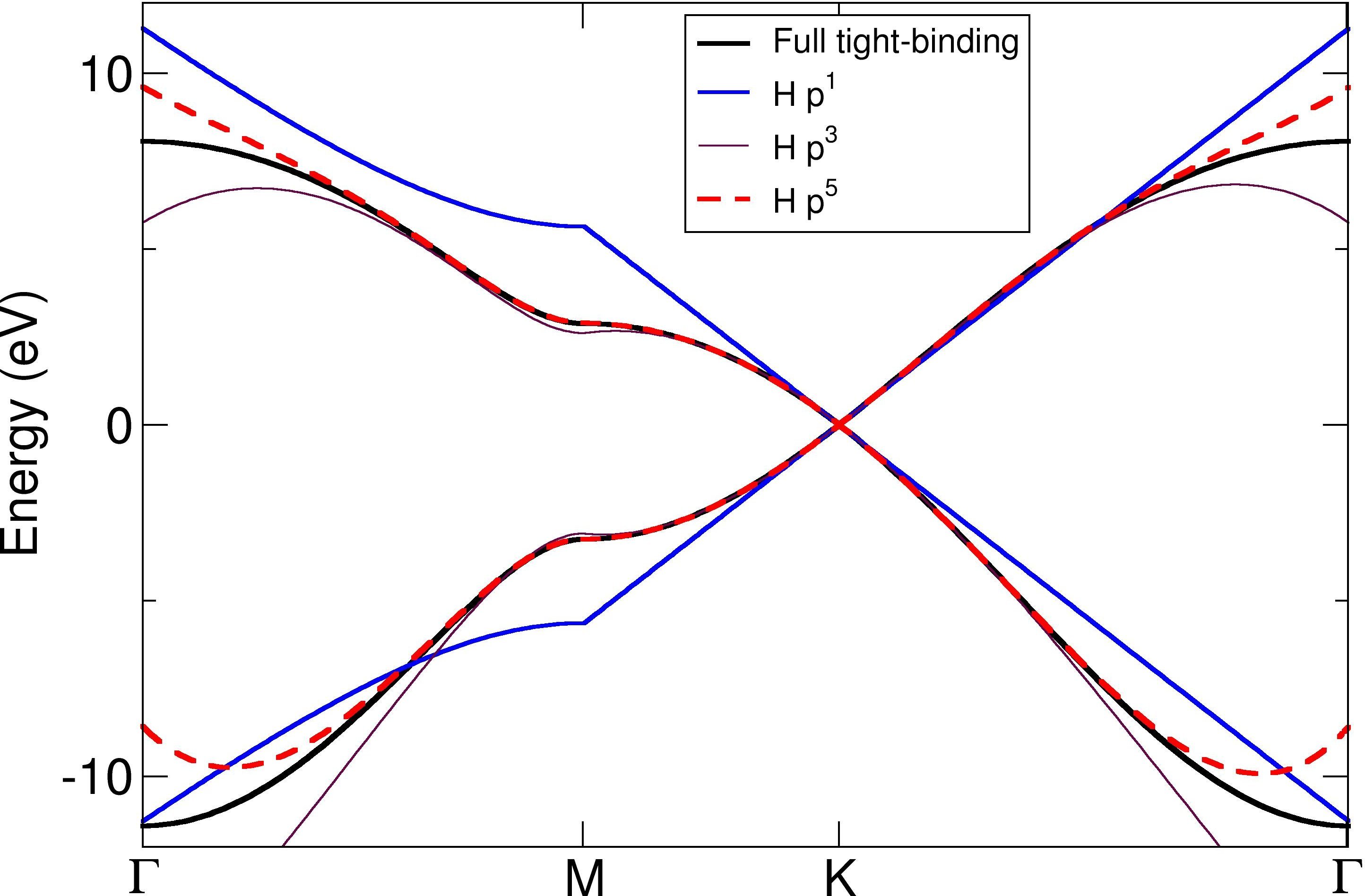}
  \caption{\emph{Graphene}: Band structure from the full-tight binding method, dark (black) full line, and three effective Hamiltonians: (i) the linear in momentum Dirac-Weyl equation, indicated by the dark (blue) full line, (ii) an effective Hamiltonian third order in momentum indicated by the thin full (violet) line and given by Eq.~\eqref{BIG3}, (iii) an fifth order in momentum effective Hamiltonian indicated by the dashed (red) line.
  }
  \label{Fig1a}
\end{figure}
%-------------------------- FIG --------------------------

%-------------------------- FIG --------------------------
\begin{figure}[t!]
  \centering
  \includegraphics[width=0.93\linewidth]{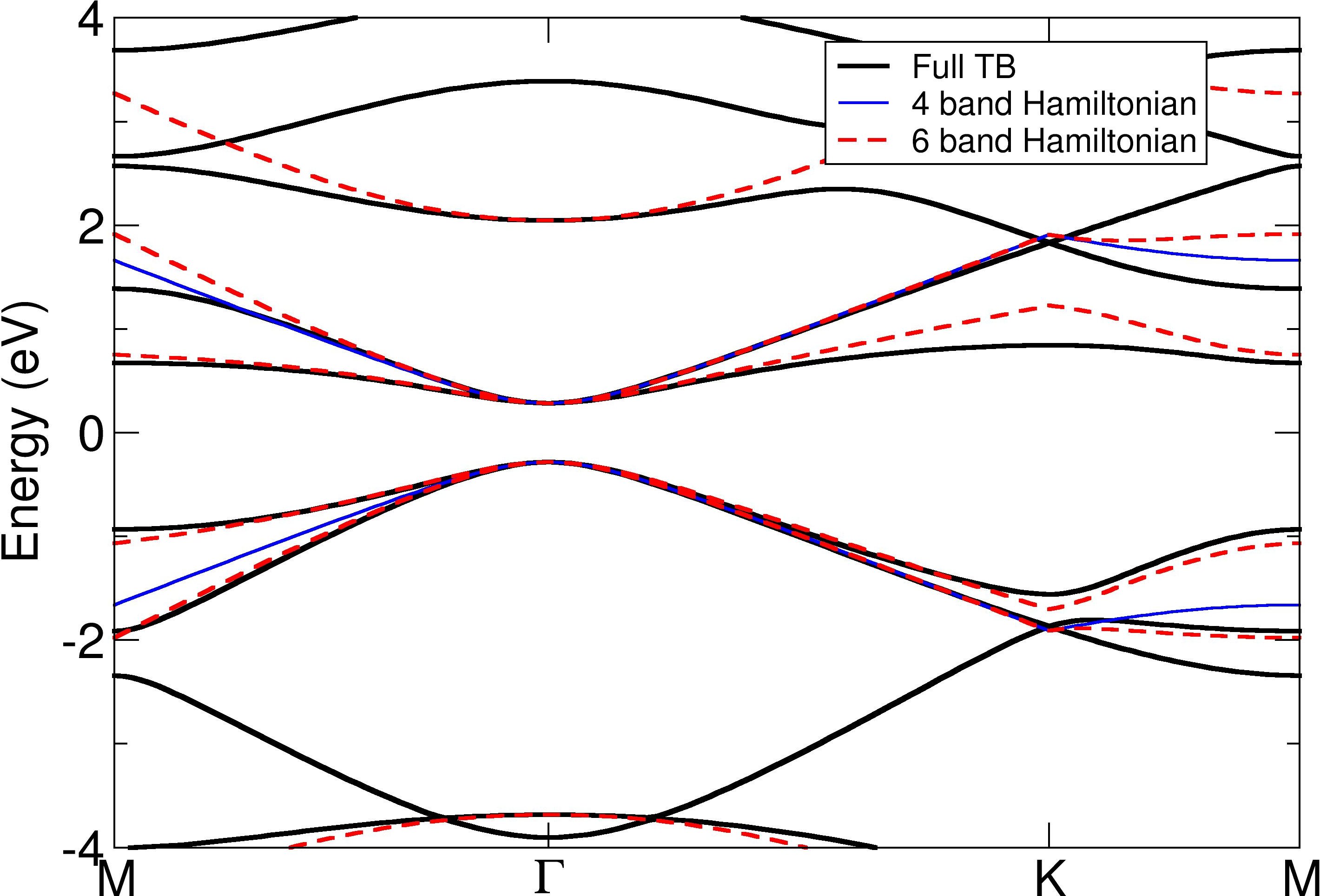}
  \caption{\emph{Graphdiyne}: Band structure from the full-tight binding method, dark (black) full line, and two effective Hamiltonians: (i) an effective Dirac equation (i.e., 4-band model), Eq.~\eqref{GG}, indicated by the dark (blue) full line, and (ii) a six band model given by Eq.~\eqref{BIG6} and indicated by the dashed (red) line. The real space lattice of this structure may be found in Fig.~1(b).
  }
  \label{Fig1b}
\end{figure}
%-------------------------- FIG --------------------------

\begin{equation}
 C_{o_1o_2p}^{(r)}(K_i) = \frac{1}{V_{UC}} \left(\frac{2\pi}{a}\right)^{2p-o_1-o_2} \frac{\partial^p t^{(r)}(q^2)}{\partial(q_i^2)^p}
\end{equation}
The connection formula thus simplifies to

\begin{equation}
  h^{(m,n)}_{r o_1 o_2 p} = \sum_i C_{o_1o_2p}^{(r)}(K_i) M_i K_{ix}^m K_{iy}^n
  \label{CV}
\end{equation}
and if we consider only the ``first star'' of the translation group of the special $K$ point then the coefficient $C_{o_1o_2p}^{(r)}(K_i)$ can be taken out of the sum which then runs from $i=1,3$

\begin{equation}
  h^{(m,n)}_{r o_1 o_2 p} = C_{o_1o_2p}^{(r)}(K) \sum_i M_i K_{ix}^m K_{iy}^n.
  \label{YOSIMPLE}
\end{equation}
In this expression $K = 4\pi/(3a)$ and the reciprocal vectors $\bG_i$ that form the first star of the translation group are $\bG_1={\bf 0}$, $\bG_2=(-\sqrt{3},1/3)$, and $\bG_3=(-\sqrt{3},-1/3)$.
The M matrices for this case are given in Table \ref{MSLG} and for such a circumstance the connection formula may be evaluated analytically quite straightforwardly. This is useful for model calculations, but here we will instead fully converge over the translation group as this is an essential requirement for testing our theory against tight-binding calculations. The number of stars of the translation group that must be included we find to be 3. In both cases the connection formula sends the sub-lattice space to a Pauli matrix algebra (see Table \ref{g}), with the only difference between the first star approximation and full convergence residing in the pre-factors to these algebraic forms, which are more numerous for the fully converged case. For the single star approach the coefficients of the Paul matrices (indicated by the general notation $C_i$ in Table \ref{g}) are just the coefficients $C_{o_1o_2p}^{(r)}(K)$ see Eq.~\eqref{YOSIMPLE}, while for the fully converged case they are the result of the sum over the $C_{o_1o_2p}^{(r)}(K_i)$ in Eq.~\eqref{CV}.

We first consider the effective Hamiltonian for graphene to linear order in momentum which, of course, results in the well known Dirac-Weyl equation $H = v_F \bsig.\bp$ with $v_F=41.6$~eV\AA~ (corresponding to a Fermi velocity of 10$^6$~ms$^{-1}$). This result is derived from the $\Psi^{(1)}_{011}$ universal polynomial in Table \ref{uni} in combination with the linear connection formula results exhibited in Table \ref{g}. In Fig.~\ref{Fig1a} we display the band structure of both the effective Hamiltonian and the results of a full tight-binding calculation using the hopping function Eq.~\eqref{hop}. As must be the case, the agreement is perfect at the expansion point, and seen to be very good for -1eV $< E <$ +1eV around the Dirac point. To improve the agreement further from the Dirac point higher orders in momentum are required. To that end we consider a third order Hamiltonian which is arrived at via the polynomials $\Psi^{(1)}_{011}$ to $\Psi^{(1)}_{033}$ in Table \ref{uni}, in conjunction once again with the connection results of Table \ref{g}. This yields, after a few lines of algebra, the still compact form

%\begin{eqnarray}
% &&H = v_F\bsig.\bp + m_1\sigma_0 p^2 \\
% && +\begin{pmatrix} m_2p^2 & \kappa(p_x+ip_y)^2 \\ \kappa(p_x-ip_y)^2 & m_2p^2 \end{pmatrix} \nonumber \\
% && +\begin{pmatrix} \kappa_1(p_x^3\!-\!3p_xp_y^2) & \kappa_2(p_x^3\!-\!ip_x^2p_y\!+\!p_xp_y^2\!-\!ip_y^3) \\
%                       \kappa_2(p_x^3\!+\!ip_x^2p_y\!+\!p_xp_y^2\!+\!ip_y^3) & \kappa_1(p_x^3\!-\!3p_xp_y^2) 
%       \end{pmatrix} \nonumber
%       \label{BIG3}
%\end{eqnarray}
%
\begin{widetext}
\begin{equation}
 H = v_F\bsig.\bp + m_1\sigma_0 p^2
  +\begin{pmatrix} m_2p^2 & \kappa(p_x+ip_y)^2 \\ \kappa(p_x-ip_y)^2 & m_2p^2 \end{pmatrix}
  +\begin{pmatrix} \kappa_1(p_x^3\!-\!3p_xp_y^2) & \kappa_2(p_x^3\!-\!ip_x^2p_y\!+\!p_xp_y^2\!-\!ip_y^3) \\
                       \kappa_2(p_x^3\!+\!ip_x^2p_y\!+\!p_xp_y^2\!+\!ip_y^3) & \kappa_1(p_x^3\!-\!3p_xp_y^2) 
       \end{pmatrix},
       \label{BIG3}
\end{equation}
\end{widetext}
which now involves 6 numerical parameters. A second order in momentum expansion is already enough to describe the high energy trigonal warping in graphene and, as may be seen from Fig.~\ref{Fig1b}, the third order Eq.~\eqref{BIG3} provides a very good description of the low energy manifold in a large energy range of $\approx10$~eV about the Dirac point. This provides a numerical confirmation of the fact that increasing the order of $\bp$ in the effective Hamiltonian must extend further from the expansion point the agreement with tight-binding calculation. Further convergence of the effective Hamiltonian in $p$ is, however, rather slow, as may be seen from the $O(p^5)$ band structure shown in Fig.~\ref{Fig1b}.

\emph{Graphdiyne}: Turning to the case of graphdiyne, the real space lattice of which may be found in Fig.~\ref{Fig1}(b), we encounter a low energy spectrum dramatically different from that of graphene: the low energy manifold is both gapped as well as situated at the $\Gamma$ point in the Brillouin zone, see Fig.~\ref{Fig1b}. Evaluating the connection formulas, Eq.~\eqref{CHI} and Eq.~\eqref{XXXX}, and retaining 6 bands in the Hamiltonian we find

\begin{widetext}
\begin{equation}
H = 
\begin{pmatrix}
\epsilon_1 & -v_1^1 p_x + v_2^1 p_y & -v_2^1 p_x -v_1^1 p_y & 0 & 0 & 0 \\
v_1^1 p_x - v_2^1 p_y & -\Delta & 0 & -v_1^2p_x+v_2^2p_y & -v_2^2p_x+v_1^2p_y & 0 \\
v_2^1p_x+v_1^1p_y & 0 & -\Delta & v_2^2p_x+v_1^2p_y & -v_1^2p_x+v_2^2p_y & 0 \\
0 & v_1^2p_x-v_2^2p_y & -v_2^2p_x-v_1^2p_y & \Delta & 0 &  -v_2^3p_x-v_1^3p_y \\
0 & v_2^2p_x-v_1^2p_y & v_1^2p_x-v_2^2p_y & 0 & \Delta & -v_1^3p_x+v_2^3p_y \\
0 & 0 & 0 & v_2^3p_x+v_1^3p_y & v_1^3p_x-v_2^3p_y & \epsilon_2
\end{pmatrix}
\label{BIG6}
\end{equation}
\end{widetext}
The description of the gapped low energy manifold provided by this Hamiltonian is rather good, as may be seen from Fig.~\ref{Fig1b}, however there is obviously ``one band more than needed'' to describe the low energy manifold. To see if the quasiparticles close to the $\Gamma$ point are governed by a more intuitive Hamiltonian we consider a 4-band Hamiltonian, which is just the central $4\times4$ block of the Hamiltonian presented in Eq.~\ref{BIG6}. Applying a spin space transformation to this block we find

\begin{equation}
 H = \begin{pmatrix} -\Delta \mathbbm{1} & v \bsig.\bp \\ v \bsig.\bp & \Delta \mathbbm{1} \end{pmatrix},
 \label{GG}
\end{equation}
and thus close to the $\Gamma$ point the quasiparticles are governed by an effective Dirac equation.
The appropriate unitary transformation to arrive at this equation is given by $U = (\sigma_0\otimes R_z(\theta))\,(\sigma_0\otimes R_y(\pi/2))\,\text{Diag}(1,i,i,1)$, with $R_i$ the $SU(2)$ rotation operators and $\theta$ the angle such that the $p_x$ and $p_y$ operators are associated with the $\sigma_x$ and $\sigma_y$ matrices respectively. A comparison between the band structure of the 4- and 6-band effective Hamiltonians indicates that the splitting of the $\Gamma$ point degeneracy is a result of interaction with the neighboring band manifolds, and not due to higher order momentum terms (both Hamiltonians are linear in momentum).

\begin{figure}[t!]
  \centering
  \includegraphics[width=0.93\linewidth]{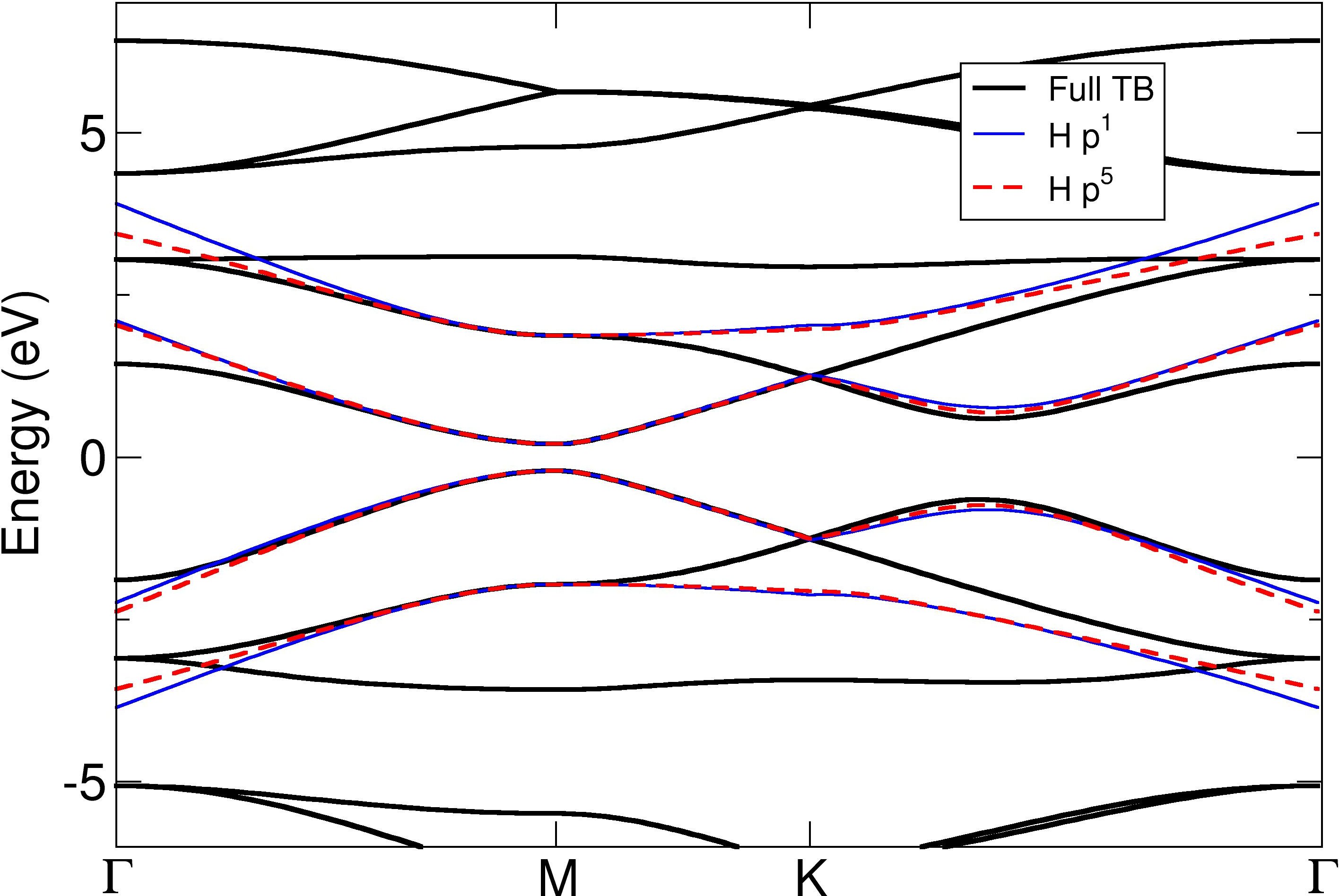}
  \caption{\emph{$\gamma$-graphyne}: Band structure from the full-tight binding method, dark (black) full line, and two effective Hamiltonians of first order and fifth order in momentum indicated by, respectively, the dark (blue) full line and the dashed (red) line. The effective Hamiltonian that generates this low energy band structure, for the first order in $p$ case, may be found in Table \ref{pseu}. The real space lattice of this structure may be found in Fig.~1(c).
  }
  \label{Fig1c}
\end{figure}

\emph{$\gamma$-graphyne}: The electronic spectrum of $\gamma$-graphyne differs from that of graphdiyne in two significant ways: (i) the gapped low energy manifold is situated at the $M$ point of the hexagonal Brillouin zone rather than the $\Gamma$-point and (ii) there is no band degeneracy\cite{mal12,mi14}. These features may be seen in the tight-binding band structure shown in Fig.~\ref{Fig1c}. Unsurprisingly therefore, the form of the effective Hamiltonian differs completely from that of the Dirac equation found for graphdiyne, see Table \ref{pseu}. The agreement between full tight-binding and the effective Hamiltonian is very good close to the expansion point and, for the low energy band manifold quite good throughout the Brillouin zone (see Fig.~\ref{Fig1c}).

An important point to note is that in a global coordinate system the effective Hamiltonians at each $M$ point differ significantly and it is only in a local $M$ point coordinate system, in which the Cartesian $k_y$ axis is aligned along the direction $\Gamma$-$M$, that the form displayed in Table \ref{pseu} is found identically (up to phases) at each $M$ point. This is quite different from the case of graphene in which, within the same global coordinate system, a $\bsig.\bp$ Hamiltonian (up to phases) is found at each high symmetry $K$ point. The underlying reason for this is a strong anisotropy in the effective mass tensor, a fact that is clear from the presence of the $p_y$ operator, but not the $p_x$ operator, in the central $2\times2$ block of the Hamiltonian. This situation is very similar to that encountered in the $\bk.\bp$ description of the VI-VI semiconductors SnTe and PbTe where a Dirac equation is found at each $M$ point only if a local $M$ point coordinate system is used\cite{pan83,pan83a,pan87}.

%-------------------------- FIG --------------------------
%-------------------------- FIG --------------------------

%-------------------------- FIG --------------------------
\begin{figure}[t!]
  \centering
  \includegraphics[width=0.93\linewidth]{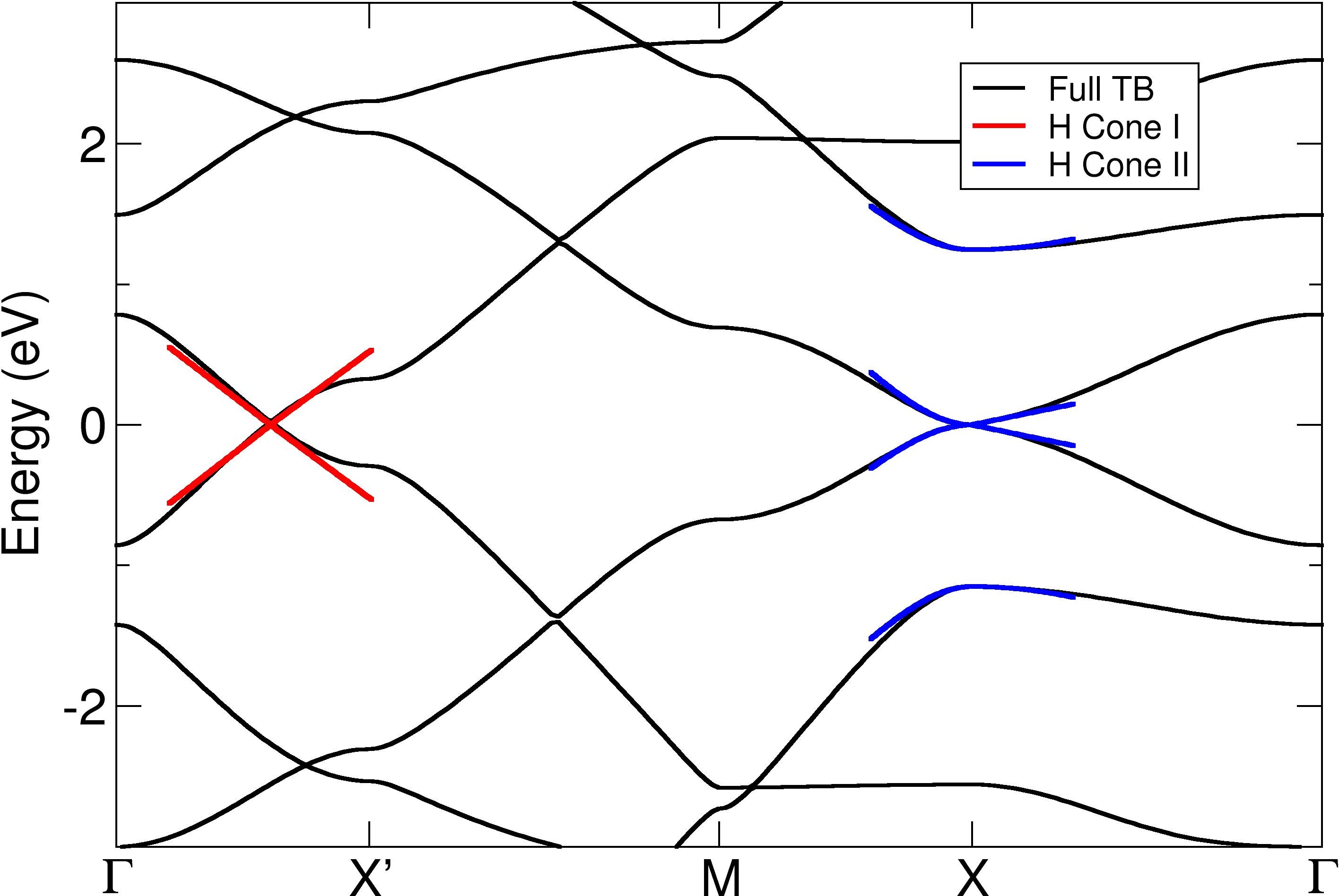}
  \caption{\emph{6,6,12-graphyne}: Band structure from the full-tight binding method, dark (black) full line, and effective Hamiltonians for cone I and cone II, indicated by the light (red) and dark (blue) full lines respectively. The effective Hamiltonians that generate these low energy band structures can be found in Table \ref{pseu}. The real space lattice of this structure may be found in Fig.~1(d).
  }
  \label{Fig1d}
\end{figure}
%-------------------------- FIG --------------------------

\emph{6,6,12-graphyne}: We finally consider 6,6,12-graphyne, which differs from all of the previous 2d allotropes in that the system has a rectangular lattice with two quite distinct low energy spectra in the rectangular Brillouin zone: (i) on the $\Gamma$-$X$' high symmetry line and (ii) at $X$ point. These are known in the literature as cone I and cone II\cite{mal12}. 

% What is ``cone shear''?
For cone I of 6,6,12-graphyne we find (after a spin-space transformation) a Dirac-Weyl Hamiltonian with anisotropic velocities and a pseudospin diagonal ``cone shear term'': $H = v_0 \sigma_0 p_x + v_x \sigma_x p_x + v_y \sigma_y p_y$ with velocities in the $k_x$-direction of +27.5~eV\AA~and -27~eV\AA, and in the $k_y$-direction of $\pm26$~eV\AA. The Dirac point is found at 0.61$\bX'$ along the $\Gamma$-$X$' high symmetry line. To draw once again a contrast with $\bk.\bp$ it should be noted that the method presented here makes no reliance on the existence of a high order point-symmetry group, thus allowing  for the derivation of effective Dirac-Weyl equation both for the low symmetry expansion point here, as well as the high symmetry expansion point found in graphene.

In the case of cone II the low energy effective Hamiltonian is completely different. We find that it is not possible to describe this cone with a 2-vector pseudospin space and we must include neighbouring bands into the calculation. The reason for this can be seen in the form of the effective Hamiltonian, which can be read off from the third line of Table \ref{pseu}: 
there is no $p_y$ dependence in the lowest energy central $2\times2$ block of the effective Hamiltonian. This reflects a curious feature of the topology of cone-II in that it is linear close to the Dirac point in $k_x$ direction (we find a band velocity of 7eV\AA), but quadratic in the $k_y$ direction. 
In both cases the agreement between the low energy portions of the full tight-binding band structure, and spectrum generated by the effective Hamiltonian approach is, once again, see Fig.~\ref{Fig1d}, found to be excellent.

%%%% Deformations in graphene %%%%

\subsection{Deformations in graphene}

Deformations in graphene have been subject to a huge number of theoretical  studies\cite{mey07,voz08,gui08,sas08,kat10,gui10,gui10a,low10,ju10,can11,ju11,he12,ju12,warn12,kitt12,much12,kli12,ol13,lop13,peet13,peet13a,ju13,voz13,bar13,qu14,san14,lop14,car14,cross14,chang14,qi14,gong15,ol15,ol16} and we will devote a separate section to this material, before treating deformations in other 2d carbon allotropes in the subsequent section.
A deformation in graphene, slow on the scale of the lattice constant, generates a number of additional terms to the effective Dirac-Weyl Hamiltonian of pristine graphene, most famously a fictitious gauge field, $v_F \bsig.\bp \to v_F \bsig.(\bp+\bA/v_F)$, that generates experimentally measured zero field Landau levels\cite{crom10,chan12}. The prefix ``fictitious'' is necessary as, obviously, time reversal symmetry is not broken by a distortion induced $\bA$, which takes opposite signs at the conjugate high symmetry $K$ points thus preserving $T$ symmetry.

In addition, a deformation also sends the Fermi velocity $v_F$ to a Fermi velocity tensor $v_F \to v_F^{ij}$~\cite{ju12,peet13,peet13a} and generates a higher order gauge field known as a ``geometric'' gauge that is, remarkably, pure imaginary without breaking the Hermiticity of the Hamiltonian\cite{ju12}.

A number of disparate methods have been used to derive these results, including the $\bk.\bp$ method, derivations based on the tight-binding method, as well as - perhaps most elegantly - a ``space-time'' approach in which the deformation is treated by sending the flat space-time Dirac-Weyl equation to a curved space manifold\cite{voz08,ju11,ju12,voz13}. Not all works agree on the terms that a deformation induces however; for example Ref.~\onlinecite{ju12} and Ref.~\onlinecite{peet13} find slightly different form for the geometric gauge, while Ref.~\onlinecite{ol15} reports a somewhat different form for the Fermi velocity tensor that that found in the two aforementioned works. 
There are, furthermore and quite naturally, fundamental differences between the space-time approach and those based on an underlying tight-binding method. Most importantly, while the pseudospin degree of freedom is ``hard wired'' into the former approach it \emph{emerges} in the latter approach. Thus while in the space-time approach any deformation must preserve the existence of the pseudospin degree of freedom as a useful structure (in the sense of that for any metric - i.e., any deformation - the Hermiticity of the curved-space Hamiltonian is guaranteed), this is not the case in tight-binding derived effective Hamiltonians: sufficiently strong deformations will always destroy pseudospin as an emergent structure.

In the following we will present a fully unified treatment of deformations in graphene, based on the systematic expansion in momentum and deformation that the universal polynomials of Table \ref{uni} afford. A natural requirement for any effective Hamiltonian theory derived from an underlying tight-binding method is that it should numerically reproduce the results of the tight-binding theory. In what follows we will pay careful attention to this requirement and converge the expansion in momentum and deformation tensor until it is met. Having satisfied this condition, one can then be confident that the results of the effective Hamiltonian theory are indeed correct, in the minimal sense of correctly representing the underlying tight-binding method from which they are derived. A more profound statement, that they are fundamental to the physics of the material under consideration, touches on the question of how dependent the emergent effective Hamiltonian structures are on the particular form of the tight-binding method. We first note that the only assumption we make regarding the tight-binding method is that the hopping function has a scalar dependence on the hopping vector: $t(\bdel^2)$. In a local bond frame the fundamental overlap integrals of the tight-binding method have exactly such a scalar dependence, and this this constitutes an approximation only in the context of a single orbital tight-binding method as the ``down-folding'' of the neglected orbital dependence into a single orbital scheme could lead to some angular dependence in the single orbital hopping function. The question can therefore be re-framed as: do deformations in graphene invoke a substantial role of the in-plane $\sigma$-bond system and lead to new effective Hamiltonian forms? This can be investigated directly by any general method (such as that espoused here), and we will do so in the subsequent section. We now take the various universal polynomials of Table \ref{uni} and, case by case, examine the contributions to the deformation modified Dirac-Weyl operator that they generate.

%-------------------------- FIG --------------------------
\begin{figure}[tbp]
  \centering
  \includegraphics[width=0.98\linewidth]{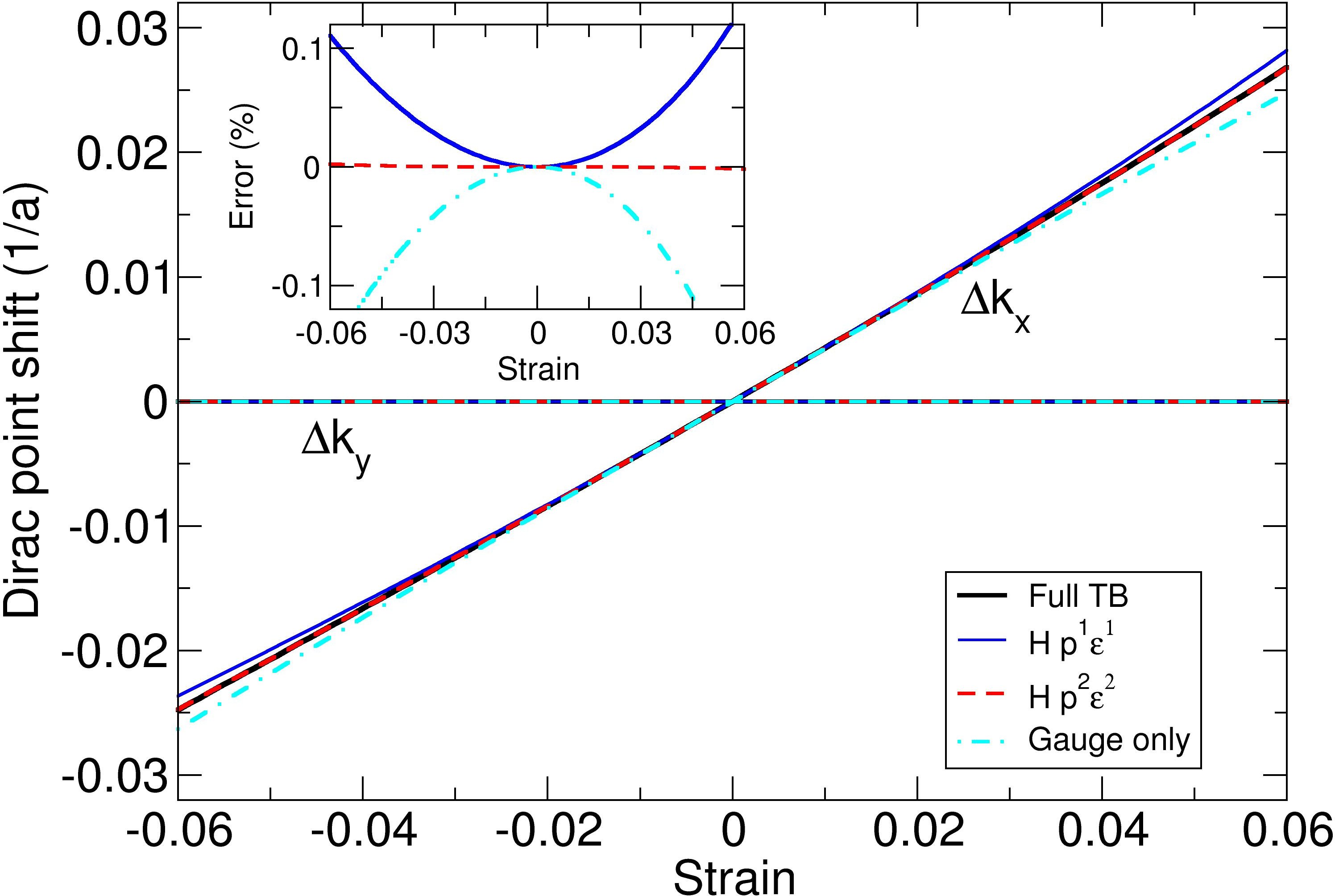}
  \caption{Shift of the Dirac cone off the high symmetry $K$ point due to strain in the $x$ direction. The dark full (black) line represents the full tight binding result with the light full line the effective Hamiltonian approach to first order in momentum and deformation tensor (labeled by $p^1\epsilon^1$), and the red dashed line the results of the an effective Hamiltonian approach to second order in both the momentum and deformation tensor (labeled $p^2\epsilon^2$). The case in which \emph{only} the gauge term $\bA = \alpha_2 (u_{yy}-u_{xx},2u_{xy})$ is included in the effective Hamiltonian is indicated by the dot-dashed cyan line. In inset displays the \% error from the tight-binding result in each case. The inclusion of only the gauge term already captures very well the Dirac point shift for a constant strain.
  }
  \label{gstrain}
\end{figure}
%-------------------------- FIG --------------------------

From the universal polynomials $\Phi^{(1)}_{201}$ and $\Phi^{(1)}_{202}$, i.e. zeroth order in momentum and first order in the deformation tensor, we find, after a few trivial lines of algebra, the Hamiltonian correction

\begin{equation}
 H^{(2,0)} = \sigma_0 V + \bsig.\bA
 \label{X}
\end{equation}
where

\begin{eqnarray}
 V & = & \alpha_1 (u_{xx} + u_{yy}) \label{A} \\
 \bA & = & \alpha_2 (u_{yy}-u_{xx},2u_{xy}) \label{B}
\end{eqnarray}
In Eq.~\ref{X} the superscript $H^{(o_1,o_2)}$ indicates respectively the order of the deformation (in terms of the Taylor expansion parameter $\delta$ not the deformation tensor itself) and the order of the momentum operator. In this expression, and in all subsequent, we will from the deformation coefficients $c^{(1)}_{ij}$ of Table \ref{c} include only the lowest order terms, and thus in Eqs.~\eqref{A} and \eqref{B} there appears only the deformation tensor elements $u_{ij} = (\partial_j u_i+\partial_i u_j)/2$, and not the higher order terms involving both derivatives and powers of the deformation field $\bu(\br)$ that may be seen in Table \ref{c}. The gauge field $\bA$ will, for a constant strain, lead to a shift of the Dirac cone of the high symmetry point of the (distorted) Brillouin zone, and in Fig.~\ref{gstrain} we present this shift calculated both from Eqs.~\eqref{A} and \eqref{B} (the dash-dotted line labeled ``Gauge only'') along with the result of a full tight-binding calculation. As may be seen, the agreement between the two is rather good, especially for small strain.

At first order in momentum and first order in the deformation tensor, i.e. from the polynomials $\Phi^{(1)}_{212}$ and $\Phi^{(1)}_{213}$ given in Table \ref{uni}, we find (in agreement with Refs.~[\onlinecite{peet13}] and [\onlinecite{ju12}]) a deformation induced Fermi velocity tensor $v^{ij}_F$

%-------------------------- FIG --------------------------
\begin{figure}[t!]
  \centering
  \includegraphics[width=0.98\linewidth]{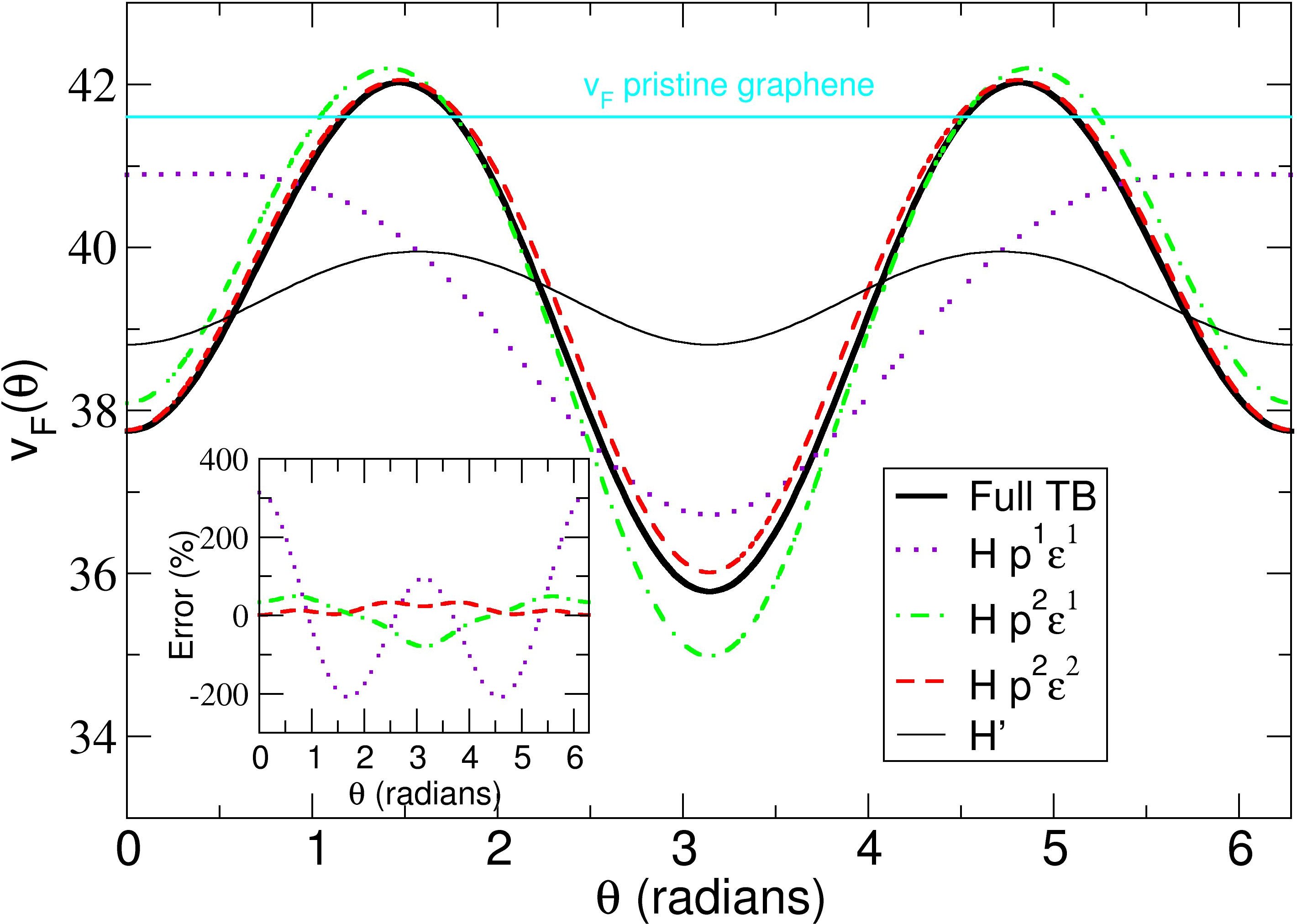}
  \caption{\emph{Fermi velocity renormalization in strained graphene: comparison of tight-binding and effective Hamiltonian results}. Velocity renormalization that results from a constant strain in the $x$ direction of $\Delta x/x = 0.06$. The Fermi velocity of the positron cone of the deformed system, $v(\theta)$, is plotted as a function of polar angle $\theta$ (with the origin of the coordinate system at the Dirac point), while the Fermi velocity of the pristine lattice is indicated by the horizontal line. The full (black) lines are the results of a full tight binding calculation, with the remaining lines the results of increasingly accurate effective Hamiltonian calculations in which the expansion in momentum and deformation tensor is taken to higher order (as indicated by the captions $p^n\epsilon^m$). In detail: the dotted (violet) line the effective Hamiltonian to first order in momentum and deformation tensor (labeled $p^1\epsilon^1$ and which includes Eqs.~\eqref{A}-\eqref{ferm}), the dash-dotted (green) line the effective Hamiltonian to second order in momentum and first order the deformation tensor (labeled $p^2\epsilon^1$ and which includes Eqs.~\eqref{A}-\eqref{H22}), and the dashed (red) line the effective Hamiltonian to second order in momentum and deformation tensor (labeled $p^2\epsilon^2$ and which includes Eqs.~\eqref{A}-\eqref{OMG}). The dark (black) full thin line labeled $H'$ represents the first order in momentum and deformation tensor result in which the pseudospin diagonal terms are excluded; this corresponds to the standard scalar and gauge fields, and $v_F$ renormalization term, that are found in the literature, see e.g. Refs.~\onlinecite{ju12,peet13}. The inset panel indicates the \% error from full tight-binding of each of the results in the main panel.
  }
  \label{gv}
\end{figure}
%-------------------------- FIG --------------------------

\begin{eqnarray}
 H^{(2,1)} & = & \alpha_3 \begin{pmatrix} \sigma_x &\!\!\! \sigma_y \end{pmatrix} 
     \begin{pmatrix} 3u_{xx} + u_{yy} & 2u_{xy} \\ 2u_{xy} & u_{xx} + 3u_{yy} \end{pmatrix} 
     \begin{pmatrix} p_x \\ p_y \end{pmatrix} \nonumber \\
     & + & \alpha_4 \sigma_0 \left[p_x(u_{xx} - u_{yy}) - p_y 2u_{xy}\right],
     \label{ferm}
\end{eqnarray}
and in addition a second term consisting of \emph{pseudospin diagonal momentum operators} not found in the aforementioned references. (Note that for the space-time approach the commutator structure of the spin connection implies that it is, in principle, not possible to generate $\sigma_0$ type terms.) For a constant strain the Fermi velocity tensor results in anisotropic Fermi velocities (the Fermi surface will distort from a circle to an ellipse), while the pseudospin diagonal momentum term results in a shearing of the Dirac cone.
In Fig.~\ref{gv} we present the band velocity at the Dirac point, $v_F(\theta)$, as a function of polar angle for the case of constant strain in the $x$ direction ($\Delta x/x = 0.06$). As may be seen the agreement between tight-binding (full line) and the result of the effective Hamiltonian (dark shaded dotted line labeled ``$H \,\,p^1 \epsilon^1$'') is not particularly good.
% Coordinate transform!
Before pursuing this point we pause to note that in order to compare the effective Hamiltonian theory, Eq.~\eqref{ferm}, and results from the underlying tight-binding theory we must in Eq.~\eqref{ferm} (and all others involving $p$) make the substitution $\bp \to (\epsilon^{-1})^T\bp$. This arises from the fact that we have in the tight-binding theory for the case of strain or shear applied a linear coordinate transformation $\br'=\epsilon\br$ to real space, and have therefore applied a transformation $\bk'=(\epsilon^{-1})^T \bk$ to reciprocal space. The coordinate system of the effective Hamiltonian theory is that of the undistorted system, and to compare the two approaches the same coordinate system must be used, thus enforcing the change of variables. For the more general case of a spatially dependent deformation the $\bq={\bf 0}$ component of the deformation $\epsilon(\br)$ must be used.

To see if the poor agreement with tight-binding may be improved we now go to quadratic order in momentum while retaining first order in the deformation tensor, the resulting terms will therefore describe deformation corrections to the curvature (i.e., trigonal warping) terms in graphene. For such terms the approprioate universal polynomials are $\Phi^{(1)}_{222}$, $\Phi^{(1)}_{223}$, and $\Phi^{(1)}_{224}$ from Table \ref{uni}. The first two of these generate the requiredd second order in momentum term which is given by

\begin{eqnarray}
 \label{H22}
 && H^{(2,2)} = \\
 && \alpha_5 \sigma_0[(3u_{xx}+u_{yy})p_x^2 + 4u_{xy}p_xp_y + (u_{xx}+3u_{yy})p_y^2]\nonumber \\
 && +\alpha_6\sigma_x[2u_{xx}p_x^2-2u_{yy}p_y^2]\nonumber \\
 && +\alpha_6\sigma_y[u_{xy}p_x^2+(u_{xx}-u_{yy})p_xp_y+u_{xy}p_y^2] \nonumber
\end{eqnarray}
while the third generates a more complex form that we do not show explicitly here (we will return to this point at the end of this discussion). As may be seen from Fig.~\ref{gv} the agreement with tight-binding theory for the angle dependent renormalized Fermi velocity is much improved by the inclusion of these terms, as is indicated by the green dot-dashed line. (Note that while these terms are second order in momentum, and thus offering no direct contribution to a velocity evaluated at $\bp={\bf 0}$, the Dirac point is shifted to a finite momentum and thus these terms do contribute.)

In an attempt to converge towards the tight-binding result we now consider terms that are second order in the deformation tensor, and look for terms similar to those found at first order in the deformation tensor, i.e. scalar and gauge fields and Fermi velocity renormalization. For the gauge and scalar fields the relevant polynomials are now $\Phi^{(2)}_{402}$, $\Phi^{(2)}_{403}$, and $\Phi^{(2)}_{404}$ (see Table \ref{uni}) and we find

\begin{equation}
 H^{(4,0)} = \sigma_0 V^{(2)} + \bsig.\bA^{(2)}
\end{equation}
with

\begin{equation}
 V^{(2)} = \alpha_7(3u_{xx}^2+2u_{xy}^2+u_{xx}u_{yy}+3u_{yy}^2)
\end{equation}
and

\begin{equation}
 \bA^{(2)} = \alpha_8(u_{yy}^2-u_{xx}^2,[u_{xx}+u_{yy}]u_{xy}).
\end{equation}
(As before the highest order polynomial $\Phi^{(2)}_{404}$ yields a more complex structure, it has 6 distinct numerical coefficients, and we do not show it explicitly.)
Similarly, we find Fermi velocity renormalization and pseudospin diagonal terms at second order in deformation tensor for which the relevant polynomials are $\Phi^{(2)}_{413}$, $\Phi^{(2)}_{415}$, and $\Phi^{(2)}_{416}$:

\begin{widetext}
\begin{eqnarray}
\label{OMG}
 H^{(4,1)} & = & \alpha_9 \begin{pmatrix} \sigma_x & \sigma_y \end{pmatrix}
 \begin{pmatrix}
  5u_{xx}^2+2u_{xy}^2+u_{xx}u_{yy} + u_{yy}^2 & 2(u_{xx}+u_{yy})u_{xy} \\
  2(u_{xx}+u_{yy})u_{xy} & 5u_{yy}^2+2u_{xy}^2+u_{xx}u_{yy}+u_{xx}^2 
 \end{pmatrix}
\begin{pmatrix} p_x \\ p_y \end{pmatrix} \\
& + & \alpha_{10} \sigma_0\left[(5u_{xx}^2-2u_{xy}^2-u_{xx}u_{yy}-3u_{yy}^2)p_x -2u_{xy}(3u_{yy}+u_{xx})p_y\right] \nonumber
\end{eqnarray}
\end{widetext}
As may be seen from Fig.~\ref{gv}, the inclusion of both second order in momentum and deformation tensor (indicated by the red dashed line) finally brings the results very close to those of the tight-binding theory. Note that, as in the other cases we have once again not shown the term arising from the highest order polynomial, in this case $\Phi^{(2)}_{416}$, and we now comment on these terms that we have hidden from the reader. As may be seen from Table \ref{uni}, the polynomials $\Phi^{(2)}_{224}$, $\Phi^{(2)}_{404}$, and $\Phi^{(2)}_{415}$ involve, respectively, 4$^\text{th}$, 4$^\text{th}$, and 5$^\text{th}$ order terms in the sublattice to pseudospin connection formula. At orders 0-3, shown in Table \ref{g}, there are at most two unknown coefficients in the Pauli matrix forms, while at 4$^\text{th}$ order (and any higher order) a plethora of distinct coefficients occur and the elegant correction forms to the Dirac-Weyl Hamiltonian found at lower orders can therefore no longer be found. The expressions, while cumbersome, can easily be worked out by the interested reader and we have therefore omitted them.

\emph{The geometric gauge field}: A curious point, immediately clear from Eq.~\eqref{GOB}, is that if the deformation order $o_1 = m_1+m_2$ is odd (note we are not referring here to the deformation tensor, but to the parameter of the Taylor expansion $\delta$) then the mixed space hopping function, and any fields derived from it, are pure imaginary. Such terms, it would seem, should destroy the Hermiticity of the effective Hamiltonian and indicate (as discussed in Section \ref{GT}) that the deformation is then so large that the pseudospin description itself breaks down. Interestingly, for graphene this is not the case.

To see this we now consider the contribution from zeroth order in momentum and third order in deformation, i.e., the polynomials $\Phi^{(r)}_{302}$ and $\Phi^{(r)}_{303}$ in Table \ref{uni}. Using the third order sublattice to pseudospin connection formula results, see Table \ref{g}, we find

\begin{equation}
 H^{(3,0)} = i \sigma_0 \phi + i \bsig.\bGam
\end{equation}
with

\begin{eqnarray}
 \phi & = & \alpha_{11}\big(\partial_x u_{xx} - 2\partial_y u_{xy} + 3\partial_x u_{yy}\big) \\
 \bGam & = & \alpha_{12} \big(3\partial_x u_{xx} + 2\partial_y u_{xy} + \partial_x u_{yy}, \nonumber \\
       &  &    3\partial_y u_{yy} + 2\partial_x u_{xy} + \partial_y u_{xx}\big)
\end{eqnarray}
the second of these terms is very similar to that found using the space-time approach in Ref.~\onlinecite{ju12}, and termed by those authors a ``geometric'' gauge field, and is identical with that derived using a tight-binding based method in Ref.~\onlinecite{peet13}. In addition to the imaginary geometric gauge, however, we also find an \emph{imaginary scalar potential} term $i\phi$ not found in Ref.~\onlinecite{ju12} or \onlinecite{peet13}.

The fact that Hermiticity of the effective Hamiltonian is not destroyed by the imaginary geometric gauge, a remarkable result, was first demonstrated in Ref.~\onlinecite{ju12} using a relation between the Fermi velocity tensor and the geometric gauge $\bGam$. However, the results here are more general, as we find both a ``geometric'' scalar potential $i\phi$ as well as a pseudospin diagonal momentum term, and thus we must revisit the question of Hermiticity.

The general form of the second order (in deformation) contribution to the mixed space hopping function is, see Eq.~\eqref{GOB},

\begin{eqnarray}
t^{(2)}(\br,\bq) & = & \sum_{m_1+m_2=2} i^{-m_1-m_2} c^{(1)}_{m_1m_2} \partial^{m_1}_{q_x} \partial^{m_2}_{q_y} t^{(1)}(q^2) \nonumber \\
           & = & \sum_{m_1+m_2=2} c^{(1)}_{m_1m_2} T_{m_1m_2}
\end{eqnarray}
where we have defined a function $T_{m_1m_2}$ in the second line. 

All terms in the effective Hamiltonian that are linear in momentum $\bp$ are generated from the first order term of the Taylor expansion of Eq.~\ref{HM}:

\begin{equation}
v_F^{ij}\sigma_i p_j = \frac{1}{V_{UC}} \sum_i M_i \left.\bnab_\bq t(\br,\bq)\right|_{\bq=\bK_i}.\frac{\bp}{\hbar}
\end{equation}
Note that on the left hand side of this expression the $i$ sum runs over 0-2, which is necessary as the first order in $\bp$ terms generate not only a Fermi velocity tensor $v_F^{ij}$ but also pseudospin diagonal momentum terms that arise from $\sigma_0$.

Now inserting the gradient $\bnab_\bq$ of $t^{(2)}(\br,\bq)$ 

\begin{equation}
\bnab_\bq t^{(2)}(\br,\bq) = i \sum_{m_1+m_2=2} c^{(1)}_{m_1m_2} \left(T_{m_1+1,m_2},T_{m_1,m_2+1}\right)
\end{equation}
into this expression and taking the spatial derivative we find

\begin{eqnarray}
\partial_j v_F^{ij} \sigma_i & = & \frac{1}{V_{UC}}\frac{i}{\hbar} \sum_i M_i \!\!\!\sum_{m_1+m_2=2}
 \Big[\partial_x c^{(1)}_{m_1m_2} T_{m_1+1,m_2} \nonumber \\
 &&+ \partial_y c^{(1)}_{m_1m_2} T_{m_1,m_2+1}\Big]
 \label{XX}
\end{eqnarray}
To make further progress we note from Table ~\ref{c} a simple relation that exists between the derivatives of the expansion coefficients at second order in $\delta$, $m_1+m_2=2$, and at third order in $\delta$, $m_1+m_2=3$:

\begin{eqnarray}
\partial_x c^{(1)}_{2,0} & = & 2 c^{(1)}_{3,0} \label{ca} \\
\partial_y c^{(1)}_{0,2} & = & 2 c^{(1)}_{0,3} \\
\partial_x c^{(1)}_{1,1} + \partial_y c^{(1)}_{2,0} & = & 2 c^{(1)}_{2,1} \\
\partial_x c^{(1)}_{0,2} + \partial_y c^{(1)}_{1,1} & = & 2 c^{(1)}_{1,2} \label{cd}
\end{eqnarray}
substitution of these results into the right hand side of Eq.~\eqref{XX} gives

\begin{eqnarray}
-i\hbar \partial_j v_F^{ij} \sigma_i & = & 2 \frac{1}{V_{UC}} \sum_i M_i \sum_{n+m=3} c^{(1)}_{m_1m_2} T_{m_1m_2} \nonumber \\
& = & 2\sigma_i \Phi_i^{(3)}
\end{eqnarray}
and dropping $\sigma_i$ from this equation finally gives the relation between the Fermi velocity tensor at second order and the field terms at third order:

\begin{equation}
 -i\hbar \partial_j v_F^{ij} = 2 \Phi_i^{(3)}
\label{4}
 \end{equation}
In this expression $\Phi_0^{(3)}$ denotes the generalized third order field term that incorporates all previously derived third order terms, with the relation to the previously derived fields given by $\Phi_0^{(3)} = i\phi$, $\Phi_1^{(3)} = i\bGam_x$, and $\Phi_2^{(3)} = i\bGam_y$.

From this expression the Hermiticity of $H = v_F \bsig.\bp + H^{(2,0)} + H^{(2,1)} + H^{(3,0)}$ is easily proved using integration by parts. Without the imaginary ``geometric'' terms then $H$ is only Hermitian if the Fermi velocity tensor and pseudospin diagonal momentum terms of $H^{(2,1)}$ are also dropped. Note that Hermiticity is assured separately for (i) the imaginary scalar potential $i\phi$ in combination with the pseudospin diagonal momentum terms of $H^{(2,1)}$ and (ii) the imaginary gauge potential $i\bGam$ in combination with the Fermi velocity tensor $v_F^{ij}$ term of $H^{(2,1)}$. Thus in the geometric approach of Ref.~\onlinecite{ju12}, in which \emph{both} the pseudospin diagonal momentum terms as well as the imaginary scalar potential are absent, Hermiticity is guaranteed by a incomplete version of Eq.~\eqref{4}.

We find that a corresponding relation to Eqs. \eqref{ca}-\eqref{cd} does not exist at any higher order, and thus deformation corrections to the Dirac-Weyl Hamiltonian are not Hermitian above fourth order (recall terms from even order in the deformation are always real and thus always hermitian). The Dirac-Weyl framework can therefore treat effective fields from slow deformations that are \emph{up to second derivatives, or second powers, of the deformation tensor}. If higher order derivatives of the deformation tensor are substantial, it indicates the breakdown of the emergence of pseudospin structure from the honeycomb lattice. It should be stressed that it is relatively surprisingly that the pseudospin description of graphene exists to such higher order, and that for 2d materials in general order 2 in $\delta$ (i.e., first derivatives and linear powers of the deformation tensor) represents the breakdown point. In this context it should be noted that a Fermi velocity renormalization at order 3 in $\delta$ exists (we have not shown it), and this \emph{does} break Hermiticity: the strict cutoff for graphene is therefore order 3, one order higher than a general 2d material.

% Time reversal symmetry

\emph{Time reversal symmetry}: We briefly comment on the question of $T$ symmetry of the effective Hamiltonian. By expanding at conjugate high symmetry point the sublattice to pseudospin connection generates a somewhat different Pauli matrix algebra to that presented in Table \ref{g}. The results for the conjugate $K^\ast$ point are obtained from Table \ref{g} by the following transformations which depend on whether $n+m$ is even or odd:

\begin{equation}
\hspace{0.8cm}
\begin{matrix}
 \begin{cases}
 \begin{matrix} \sigma_0 \to \sigma_0 \\ \sigma_x \to \sigma_x \\ \sigma_y \to -\sigma_ y \end{matrix} & \text{even} 
 \end{cases} 
  &
  \,\,\,\,\,\,\,\,
 \begin{cases}
 \begin{matrix} \sigma_0 \to -\sigma_0 \\ \sigma_x \to -\sigma_x \\ \sigma_y \to \sigma_ y \end{matrix} & \text{odd}.
 \end{cases}
\end{matrix}
\end{equation}
Using these relations we find

\begin{eqnarray}
 H_K & = & v_F \bsig.(\bp+\bA/v_F+i\bGam/v_F) \nonumber \\
 & + & \sigma_0 (V+i\phi) \\
 H_{K^\ast} & = & -v_F \bsig^\ast.(\bp-\bA/v_F+i\bGam/v_F) \nonumber\\
 &+&\sigma_0 (V-i\phi),
\end{eqnarray}
thus while the lattice gauge $\bA$ changes sign, the geometric gauge $i\bGam$ does not; consistent with the fact that the $T$ operator, as well as changing the sign of momenta (and therefore magnetic field), also involves the complex conjugation operator $K$. An opposite behaviour is found for the scalar field: the geometric scalar field changes sign under $K\to K^\ast$ while the real scalar field $V$ does not - also consistent with $T$ symmetry.

% Out of plane deformations

\emph{Out of plane deformations and $\sigma$-bonds}: Thus far only in-plane deformations have been explicitly considered: the expressions derived above include only $u_{xx}$, $u_{yy}$, and $u_{xy}$. Out of plane deformations of graphene are, however, trivially incorporated into the expressions by inclusion of the higher order terms in the $c^{(1)}_{ij}$ presented in Table \ref{c}. These higher order terms involve the full deformation field $\bu(\br)$ which naturally includes a possible out of plane component $u_z(\br)$. For example, in the case of the second order scalar potential $V$ and gauge potential $\bA$ we find

\begin{eqnarray}
 V & = & \alpha_1 (u_{xx} + u_{yy} + \frac{1}{2}\bnab.\bu)^2 \\
 \bA & = & \alpha_2 \big(u_{yy}-u_{xx} + \frac{1}{2}\left((\partial_y\bu)^2-(\partial_x\bu)^2\right), \nonumber \\
       &  & 2u_{xy} + \frac{1}{2} \partial_x\bu.\partial_y\bu\big)
 \end{eqnarray}
with similar corrections easily obtained for all other expressions derived in this section.
However, once out of plane deformations are included into the formalism we must treat both $\pi$- and $\sigma$-electron hopping. This may be seen from the $t_{p_zp_z}$ element (responsible for the low energy band manifold) of the full tight-binding hopping matrix 

\begin{equation}
 t_{p_zp_z} = t_\pi(\delta^2) + (t_\sigma(\delta^2)-t_\pi(\delta^2)) n^2
 \label{5}
\end{equation}
where the directional cosine $n = \delta_z/\delta$. Evidently, once out of plane deformations occur then $n \ne 0$. The general theory of Section \ref{dmeth} may be deployed on the second term in Eq.~\eqref{5} (which will of course generate new polynomials $\Phi^{(r)}_{o_1 o_2 p}$ as we are no longer considering the case of one orbital per site) and we find at second order in deformation the result

\begin{equation}
 H^{(2,0)} = \sigma_0 V_z + \bsig.\bA_z
\end{equation}
where

\begin{eqnarray}
 V_z & = & \gamma_1 (\bnab u_z)^2 \\
 \bA_z & = & \gamma_2 ((\partial_y u_z)^2-(\partial_x u_z)^2,2\partial_x u_z \partial_y u_z)
\end{eqnarray}
each of these depend on two more coefficients $\gamma_i$ that now involve the Fourier transform of $(t_\sigma(\delta^2)-t_\pi(\delta^2))/\delta^2$. We thus see that introduction of $\sigma$-orbitals does not lead to corrections to the effective Hamiltonian that are lower order in $u_{ij}$ than occur due to $\pi$-orbitals. Out of plane deformation of the graphene lattice therefore \emph{always} leads to terms that are higher order in $u_{ij}$ than those generated by in-plane deformation.

% possibility of gap opening from slow deformation

\emph{Mass generating deformation}: Generation of a mass term, i.e. a $\sigma_z$-type potential (corresponding to a mass in the two dimensional Dirac-Weyl equation) might be expected to require a deformation that has an ``optical component'' in addition to an ``acoustic component'', in the sense that both the fields $\bu_{\pm}(\br) = \bu_2(\br) \pm \bu_1(\br)$ are both non-zero ($\bu_{1,2}$ are the deformation fields acting on the two basis atoms of the graphene lattice). As we now show, in the absence of an optical component to the deformation the connection formula cannot yield a $\sigma_z$ potential. To see this we can simply derive a general formula for the sublattice to pseudospin connection, which for the first star takes the form

\begin{widetext}
\begin{eqnarray}
\sum_{i=1}^3 M_i K_{ix}^m K_{iy}^n = \xi^m
 \begin{cases}
  3^{-m}\left[2^m + 2(-1)^m\right]\sigma_0 - 3^{-m}\left[2^m - (-1)^m\right]\sigma_x & n=0 \\
  2(-1)^m 3^{-n/2-m}\sigma_0 -(-1)^m 3^{-n/2-m} \sigma_x & n>0 \text{ and } n \text{ even} \\
  (-1)^m 3^{-(n-1)/2-m} \sigma_y & n>0 \text{ and } n \text{ odd}
 \end{cases}
\end{eqnarray}
\end{widetext}
(where $\xi=\pm1$ indicates the $K$ or conjugate $K$ expansion point has been used). Similar formulas may be found for any star of the translation group of the expansion point, thus demonstrating that slow acoustic deformations cannot generate a gap opening $\sigma_z$ potential in the Dirac-Weyl framework. On the other hand, slow acoustic deformations for which the second derivative of the deformation tensor cannot be neglected breaks, as we have shown above, the Hermiticity of the effective Hamiltonian and so could, in principle, present behaviour corresponding to mass generation. The treatment of such deformations, however, falls outside the ambit of a pseudospin description of graphene.

%%% Arbitrary deformations in graphdiyne %%%

\subsection{Arbitrary deformations in graphdiyne}

While the impact of strain on the electronic structure of the complex 2d all-carbon allotropes has been the subject of a number of \emph{ab-initio} and tight-binding investigations there does not exist, to the best of our knowledge, a theory describing the impact of an arbitrary deformation field $\bu(\br)$ on the low energy spectrum, i.e., a theory corresponding to that of the gauge field induced by deformations in graphene. However, the general theory of Section \ref{dmeth} is as easily applicable to these more complex 2d allotropes as it is to graphene (one evidently deploys the same universal polynomials with only the connect formula producing a different pseudospin structure), and in this section we will apply it to the semiconductor graphdiyne. We choose this material as an example as (i) its low energy spectrum differs strongly from that of graphene (gapped as opposed to a topologically protected cone) and (ii) it has apparently been synthesized experimentally. Although we present our results here merely as an example of the wide applicability of the theory of Section \ref{dmeth}, a general theory of deformations in 2d allotropes beyond graphene is of considerable interest: investigations of the interesting transport properties of these materials relies on an understanding of electron-phonon coupling, and the electronic perturbation induced by a general $\br$-dependent deformation is exactly the theory required to elucidate this in a systematic way.

\begin{figure}[tbp]
  \centering
  \includegraphics[width=0.98\linewidth]{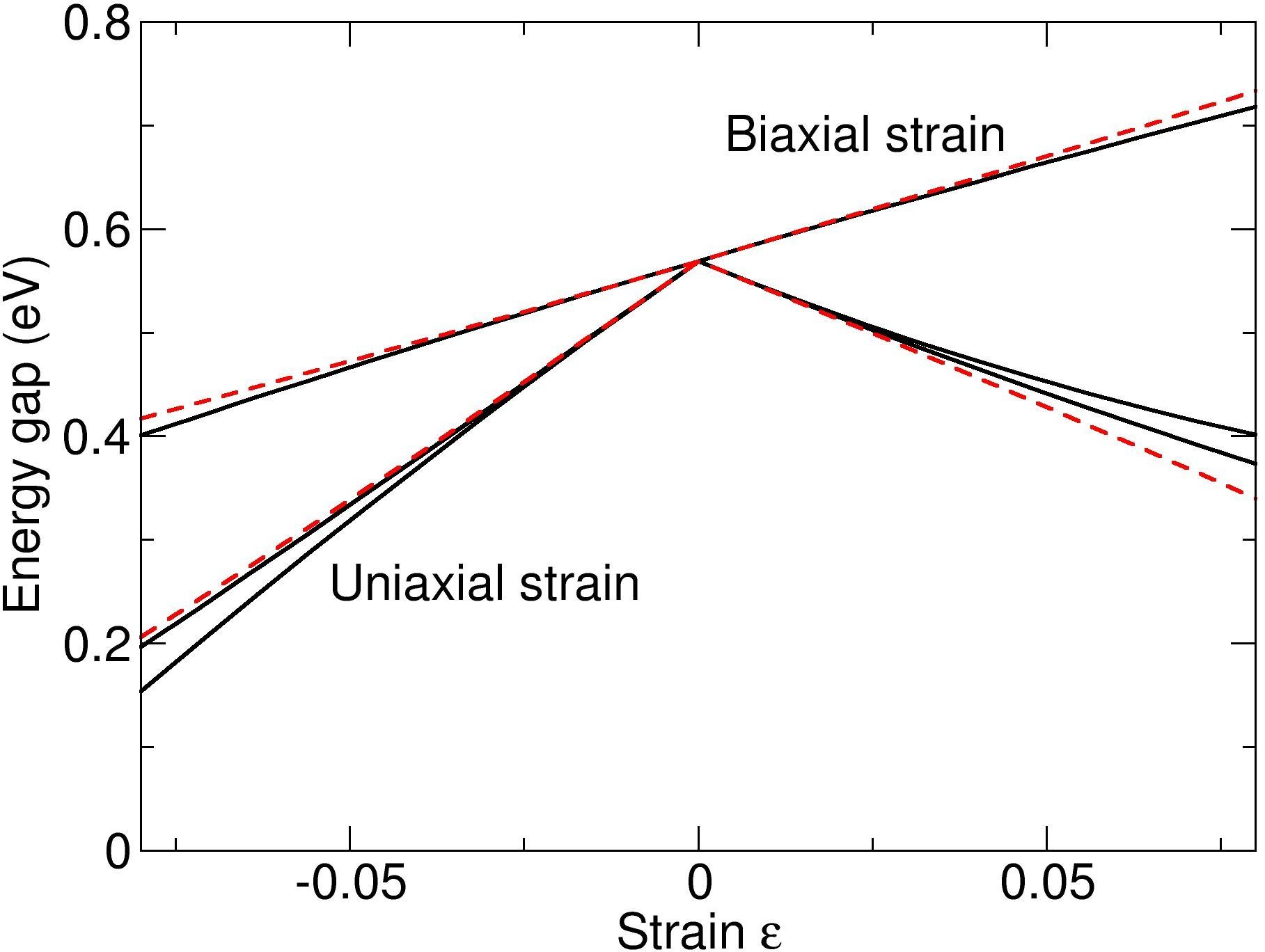}
  \caption{Change in the $\Gamma$ point band gap of graphdiyne due to biaxial as well as uniaxial strain. While biaxial strain shows a monotonic behaviour with strain $\epsilon = \Delta x_i/x_i$, uniaxial strain, unusually, decreases the gap for \emph{both} positive and negative strain. Full (black) lines are the result of a tight-binding calculation with dashed (red) lines the result of the effective Hamiltonian approach using the strain induced gap field, Eq.~\eqref{DDD}, in conjunction with a low energy Dirac equation, Eq.~\eqref{GG}. The small splitting found in the tight-binding for uniaxial strain corresponds to strain in the $x$ (armchair) and $y$ (zigzag) directions.
  }
  \label{syne}
\end{figure}

First principles and tight-binding calculations report that while positive biaxial strain increases the band gap (as one would expect), positive uniaxial strain (applied in either the armchair or zigzag directions) leads, in unusual contrast, to a reduction in the size of the band gap\cite{yue13}. On the other hand, both negative biaxial and uniaxial strain result in a reduction in the band gap - the expected behaviour for compressing a material. Employing the connection formula, Eq.~\eqref{CHI}, and the lowest order polynomial $\Phi^{(1)}_{011}$ from Table \ref{uni} we find that the deformation field $\bu(\br)$ enters the effective Dirac Hamiltonian of this material not as a gauge field (which this polynomial produced in the case of graphene) but as a \emph{gap function}:

%\begin{eqnarray}
% H^{(2,0)} &=& \begin{pmatrix} \alpha_1 & \alpha_2 & 0 & 0 \\ \alpha_2^\ast & \alpha_3 & 0 & 0 \\ 0 & 0 & \alpha_4 & \alpha_5 \\ 0 & 0 & \alpha_5^\ast & \alpha_6 \end{pmatrix} u_{xx} +
% \label{DDD}\\
%  &&\begin{pmatrix} \alpha_3 & \alpha_2^\ast & 0 & 0 \\ \alpha_2 & \alpha_1 & 0 & 0 \\ 0 & 0 & \alpha_6 & \alpha_5^\ast \\ 0 & 0 & \alpha_5 & \alpha_4 \end{pmatrix} u_{yy} + \nonumber\\
%    &&\begin{pmatrix} \text{Im}\alpha_2 & \beta_1 & 0 & 0 \\ \beta_1^\ast & -\text{Im}\alpha_2 & 0 & 0 \\ 0 & 0 & \text{Im}\alpha_5 & \beta_2 \\ 0 & 0 & \beta_2^\ast & -\text{Im}\alpha_5 \end{pmatrix} 2u_{xy} \nonumber
%\end{eqnarray}
%
\begin{widetext}
\begin{equation}
 \label{DDD}
 H^{(2,0)} =
 \begin{pmatrix} \alpha_1 & \alpha_2 & 0 & 0 \\ \alpha_2^\ast & \alpha_3 & 0 & 0 \\ 0 & 0 & \alpha_4 & \alpha_5 \\ 0 & 0 & \alpha_5^\ast & \alpha_6 \end{pmatrix} u_{xx}(\br) +
 \begin{pmatrix} \alpha_3 & \alpha_2^\ast & 0 & 0 \\ \alpha_2 & \alpha_1 & 0 & 0 \\ 0 & 0 & \alpha_6 & \alpha_5^\ast \\ 0 & 0 & \alpha_5 & \alpha_4 \end{pmatrix} u_{yy}(\br) +
 \begin{pmatrix} \text{Im}\alpha_2 & \beta_1 & 0 & 0 \\ \beta_1^\ast & -\text{Im}\alpha_2 & 0 & 0 \\ 0 & 0 & \text{Im}\alpha_5 & \beta_2 \\ 0 & 0 & \beta_2^\ast & -\text{Im}\alpha_5 \end{pmatrix} 2u_{xy}(\br)
\end{equation}
\end{widetext}
where the $\alpha_i$ and $\beta_i$ are numerical constants derived from the underlying tight-binding method via the connection formula.
As a simple application of this result, we now consider the case of uniaxial and biaxial strain for which, as may be seen from Fig.~\ref{syne}, we find exactly the \emph{ab-initio} result\cite{yue13} of positive biaxial strain increasing the band gap, with positive uniaxial strain reducing it. The agreement between full tight binding and the low energy approach is seen to be very good, and comparable to that found in the case of graphene (see Fig.~\ref{gstrain}).

%%%%%%%%%%%%%%%%%%%%%%%%%%%%%%%%%%%%%%%%%%%%%
% STACKING DEFORMATIONS IN BILAYER GRAPHENE %
%%%%%%%%%%%%%%%%%%%%%%%%%%%%%%%%%%%%%%%%%%%%%

\section{Stacking deformations in bilayer graphene}

While deformations experienced by a single layer material are necessarily small on the scale of the lattice constant, systems of weakly (van der Walls) coupled layers are subject to stacking deformations that are, in scale, greatly in excess of the lattice constant. Often, these qualitatively change both the lattice geometry as well as electronic structure of the material. The most studied such example is the mutual rotation of two layers of bilayer graphene that, in the small angle limit, leads to the emergence of a moir\'e lattice of periodicity of $D = a/(2\sin\theta/2)$ and a novel low energy spectrum dramatically different from both single layer graphene and any ``simple'' stacking of the two layers\cite{lop07,hass08,lop13,shall10,bist11,mel12,shall16} (such as the graphitic AB stacking). Recently, partial dislocation networks have been imaged\cite{butz14} in bilayer graphene, which again represents a non-perturbative structural rearrangement of the lattice, as the bilayer segments into a mosaic of AB and AC stacked tiles. Such dislocation networks have been shown to have a profound impact on the physical properties of the material, notably in transport and magnetotransport\cite{kiss15}.

However, it is precisely for the rich non-perturbative physics of the stacking deformation that the $\bk.\bp$ method is guaranteed to fail to produce a compact and physically intuitive effective Hamiltonian and, for this reason, there is no well developed theory of deformations in bilayer graphene corresponding to that of single layer graphene. In particular, there does not exist a general ``interlayer gauge field'' corresponding to the deformation induced $\bA$ and $V$ fields that, for single layer graphene, allow one to treat any deformation within the same compact formalism. Using the formalism of Section \ref{GT}, we will provide such a general theory of interlayer deformations before deploying it to examine all of the possible stacking deformations of bilayer graphene: translations, rotations, and partial dislocations. As we will show, both the effective Hamiltonians of simply stacked bilayer graphene, as well as the twist bilayer, emerge as special cases of a general ``interlayer field''.

%%%% General theory %%%%

\subsection{General theory for weakly coupled layers}
\label{ST}

The general form of a bilayer Hamiltonian may be written as

\begin{equation}
 H(\br,\bp) = \begin{pmatrix} H_1 & S(\br,\bp) \\ S^\dagger(\br,\bp) & H_2 \end{pmatrix}
\label{YOLO}
\end{equation}
where $H_i$ are layer diagonal blocks (that may be of any dimension) and correspond to Hamiltonians of the type considered in the previous section while

\begin{equation}
 \left[S(\br,\bp)\right]_{\alpha\beta} = \frac{1}{V_{UC}} \sum_i \left[M_i\right]_{\alpha\beta} t^\perp_{\alpha\beta}(\br,\bK_i+\bp/\hbar)
\label{6}
\end{equation}
represents an effective field coupling the two layers. Equation \eqref{6} is nothing more than the general form of the effective Hamiltonian, Eq.~\eqref{HM}, but with the sublattice degrees of freedom restricted such that $\alpha$ resides on layer one and $\beta$ on layer two. To render this into a useful form we must (as in all other examples in the previous section) evaluate the mixed space hopping function $t^\perp_{\alpha\beta}(\br,\bq)$. An arbitrary hopping vector between the two layers, from $\br$ in the first layer to $\br+\bdel$ in the second, will, if a local shift $\bu_1(\br)$ is applied at every point $\br$ in the first layer, be transformed $\bdel \to \bdel - \bu_1(\br)$. The associated tight binding hopping function in consequence transforms from a function describing hopping in the high symmetry system, $t^\perp_{\alpha\beta}(\bdel^2)$, to a more complex function describing electron hopping in the system after interlayer deformation  $t^\perp_{\alpha\beta}\left([\bdel-\bu_1(\br)]^2\right)$. We therefore require the following Fourier transform

\begin{equation}
 t^\perp_{\alpha\beta}(\br,\bq) = \int d\bdel\, e^{i\bq.\bdel} t^\perp_{\alpha\beta}\left([\bdel-\bu_1(\br)]^2\right)
\label{FT2}
\end{equation}
In the case of \emph{intra}layer deformations the corresponding object to be Fourier transformed, Eq.~\eqref{1}, could be evaluated only by treating the deformation as a perturbation. For \emph{inter}layer deformations, that are by the nature non-perturbative effects, this approach will fail. Fortunately, with the simple change of variables $\bdel' = \bdel - \bu_1(\br)$ the integral Eq.~\eqref{FT2} can be taken \emph{exactly}:

\begin{equation}
 t^\perp_{\alpha\beta}(\br,\bq) = e^{i\bq.\bu_1(\br)} t^\perp_{\alpha\beta}(\bq^2)
\end{equation}
with $t^\perp_{\alpha\beta}(\bq^2)$ the Fourier transform of the hopping function of the high symmetry system before deformation.

Thus for the zeroth order in momentum term of Eq.~\eqref{6} we find

\begin{equation}
 \left[S_0(\br)\right]_{\alpha\beta} = \frac{1}{V_{UC}} \sum_i \left[M_i\right]_{\alpha\beta}\,\, t^\perp_{\alpha\beta}(K_i) e^{i\bK_i.\bu_1(\br)}
\label{S}
\end{equation}
This is the result we seek: an expression that connects an arbitrary interlayer deformation $\bu_1(\br)$ to the effective field coupling the Hamiltonians of each layer. Terms higher order in momentum are as easily obtained and for the first order in momentum we find

%-------------------------- FIG --------------------------
%-------------------------- FIG --------------------------

\begin{eqnarray}
 \left[S_1(\br)\right]_{\alpha\beta}& =& \frac{1}{V_{UC}} \sum_i \left[M_i\right]_{\alpha\beta} \nonumber \\
& \times & \left(2 t^{\perp'}_{\alpha\beta}(K_i) \bK_i + i t^\perp_{\alpha\beta}(K_i) \bu_1(\br)\right). \nonumber \\
 && \left(\frac{\bp}{\hbar}\right) \,\, e^{i\bK_i.\bu_1(\br)}.
\label{S1}
\end{eqnarray}
Note that, beyond the assumption of a bilayer geometry, we have thus far employed no assumptions concerning the nature of the two layers that are coupled, and Eqs.~\eqref{S} and \eqref{S1} are thus quite general. To specify to a particular material we must fix both the Fourier transform of the hopping function of the high symmetry system $t_{\alpha\beta}^\perp(\bq^2)$, as well as the $M_i$ matrices that encode purely geometric information (these matrices are given by Eq.~\eqref{M}).
 
%%%% Dirac pockets of bilayer graphene %%%%

\subsection{Choice of hopping function and Dirac pockets of bilayer graphene}

\begin{table}
\begin{tabular}{l|ccc}	
Stacking & $M_1$ & $M_2$ & $M_3$ \\ \hline\hline
 AB &
 $\begin{pmatrix} 1 & 1  \\ 1 & 1 \end{pmatrix}$ &
 $\begin{pmatrix} 1 & e^{i2\pi/3}  \\ e^{i2\pi/3} & e^{-i2\pi/3} \end{pmatrix}$ &
 $\begin{pmatrix} 1 & e^{-i2\pi/3}  \\ e^{-i2\pi/3} & e^{i2\pi/3} \end{pmatrix}$ \\
\hline
 AC &
 $\begin{pmatrix} 1 & 1  \\ 1 & 1 \end{pmatrix}$ &
 $\begin{pmatrix} e^{i2\pi/3} & e^{-i2\pi/3}  \\ e^{-i2\pi/3} & 1 \end{pmatrix}$ &
 $\begin{pmatrix} e^{-i2\pi/3} & e^{i2\pi/3}  \\ e^{i2\pi/3} & 1 \end{pmatrix}$
\end{tabular}
\caption{$M_i$ matrices for the first star of Bernal stacked AB and AC stacked bilayer graphene.}
\label{M2}
\end{table}

For the high symmetry hopping function we choose (as we did for the case of single layer graphene) the form

\begin{equation}
t_\perp(\bdel) = A\exp(-B\bdel^2),
\label{9}
\end{equation}
which assumes that all tight binding matrix elements depend only on the length of the hopping vector (this form of the hopping function is often used for tight-binding calculations of the graphene twist bilayer\cite{shall13,shall16}. This assumption distinguishes Eq.~\eqref{9} from, for example, the Slonczewski-Weiss-McClure (SWM) method used for graphite and often adapted for use in the Bernal stacked graphene bilayer. What difference, if any, will this choice make to the resulting low energy effective Hamiltonians? To investigate this we will consider the low energy spectrum of Bernal stacked graphene bilayer which consists of a Dirac point located at each high symmetry $K$-point, with each of these in turn trigonally decorated by three satellite Dirac points, at a separation of $\approx0.07$\AA$^{-1}$ away from the $K$-point. The SWM tight-binding method leads to a low energy Hamiltonian that perfectly captures this complex low energy manifold, and the purpose of this section is to demonstrate that Eq.~\eqref{9} leads to a Hamiltonian \emph{identical in form} to that derived from the SWM method, thus confirming the intuitive notion that the particular form of the tight-binding method should not qualitatively change the resulting effective Hamiltonian structure.

As we consider here a case of no deformation, $\bu_1 = {\bf 0}$, we do not need the formalism of the previous section and may directly use the connection formula, Eq.~\eqref{CHI}, along with the universal polynomials of Table \ref{uni}. Evaluating the connection formula for an interlayer geometry will evidently result in a different pseudospin structure from that found in the case of single layer graphene. In fact, this interlayer pseudospin structure turns out to be just that of the single layer graphene pseudospin structure with the substitution $\sigma_0 \rightarrow \tau_0$, $\sigma_x \rightarrow \tau_+$, and $\sigma_y \rightarrow \tau_-$ made to the results of Table \ref{pseu}. These $\tau$-matrices, which define the interlayer pseudospin algebra, are given by

\begin{equation}
\begin{array}{ccc}
 \tau_0 = \begin{pmatrix} 1 & 0 \\ 0 & 0 \end{pmatrix}, &
 \tau_+ = \begin{pmatrix} 0 & 1 \\ 1 & 1 \end{pmatrix}, &
 \tau_- = \begin{pmatrix} 0 & i \\ i & -i \end{pmatrix}
\end{array}
\end{equation}
Evaluating the first 4 of the universal polynomials of Table \ref{uni}, i.e. we consider up to $\bp^2$ in the momentum expansion, we find the bilayer Hamiltonian is given by

\begin{equation}
 H_{AB} = \begin{pmatrix} \bsig.\bp & S_{AB}(\bp) \\ S_{AB}^\dagger(\bp) & \bsig^\ast.\bp \end{pmatrix}
 \label{HAB}
\end{equation}
where the layer off-diagonal blocks are given by

\begin{eqnarray}
 S_{AB} & = & t_\perp \begin{pmatrix} 1 & 0 \\ 0 & 0 \end{pmatrix} +
 v \begin{pmatrix} 0 & p_x + i p_y \\ p_x + i p_y & p_x - i p_y \end{pmatrix} +\\
 && \kappa_1 \begin{pmatrix} p^2 & 0 \\ 0 & 0 \end{pmatrix} +
 \kappa_2 \begin{pmatrix} 2p^2 & (p_x - i p_y)^2 \\ (p_x - i p_y)^2 & (p_x + i p_y)^2 \end{pmatrix}. \nonumber
\end{eqnarray}
$H_{AB}$ describes not only the low energy spectrum, but also the high energy bonding and anti-bonding band manifolds, and a down-folding procedure is required to eliminate these high energy manifolds from the Hamiltonian. A standard down-folding procedure (which follows very closely that detailed in Ref.~[\onlinecite{mcc13}]) allows us to obtain from Eq.~\eqref{HAB} a $2\times2$ Hamiltonian describing only the low energy manifold:

\begin{equation}
 H_{eff} = \begin{pmatrix} 0 & h \\ h^\ast & 0 \end{pmatrix}
\end{equation}
where

\begin{equation}
 h = \frac{v_F^2}{t_\perp}p_+^2 + v p_- + \kappa_2 p_+^2
\label{10}
\end{equation}
(with $p_\pm = p_x \pm i p_y$).
This operator $h$ is exactly that found by the SWM tight-binding method\cite{mcc13} and thus the choice of underlying tight-binding method, in this case, impacts only the coefficients of the effective Hamiltonian and not the basic form.

%%%% Mutual translation of the layers of bilayer graphene %%%%

\subsection{Mutual translation of the layers of bilayer graphene}
\label{SHIFT}

%-------------------------- FIG --------------------------
\begin{figure}[tbp]
  \centering
  \includegraphics[width=0.98\linewidth]{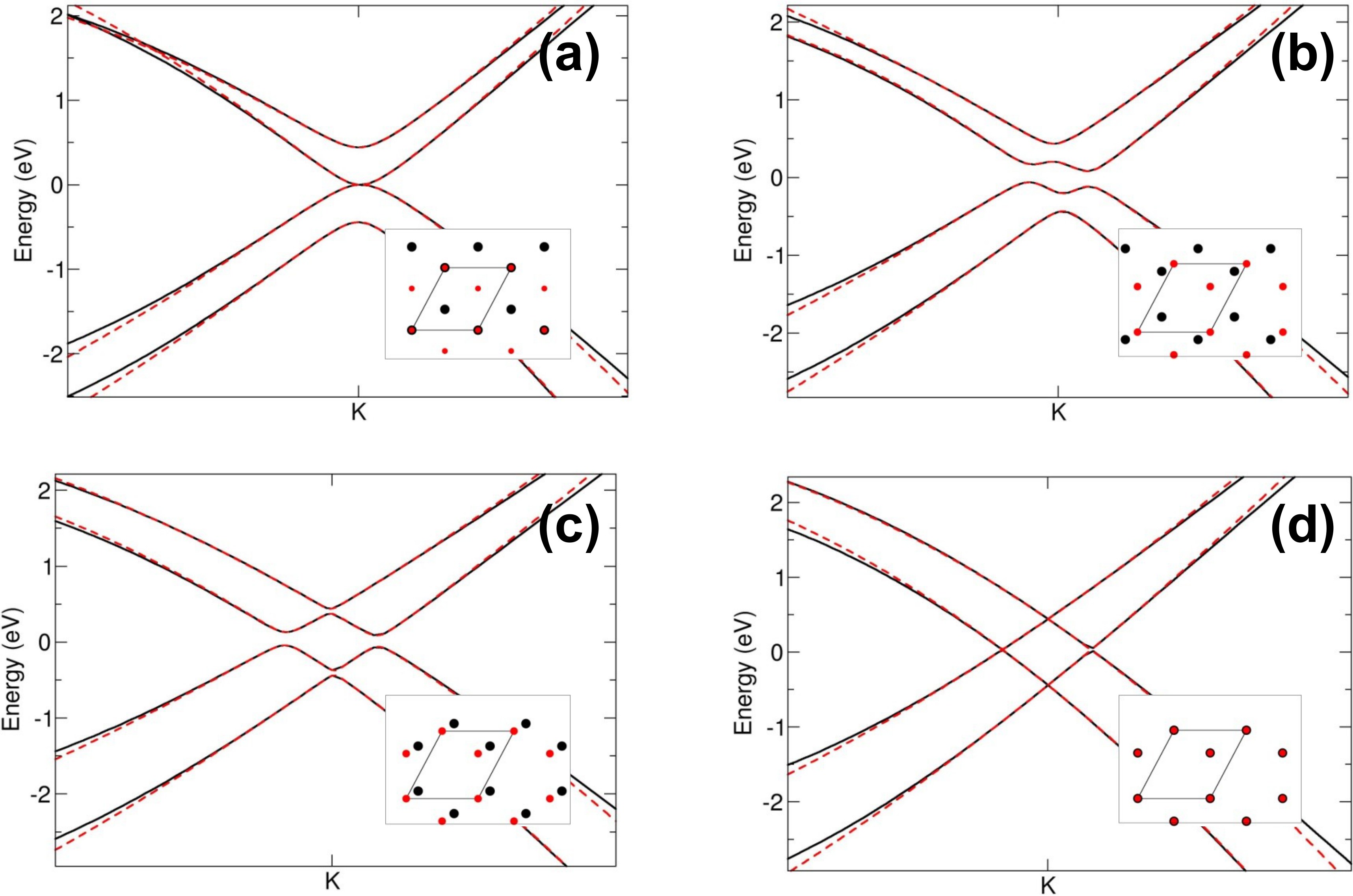}
  \caption{Low energy electronic structure of bilayer graphene for 4 different mutual translations of the two layers. (a) $\bu_1 = {\bf0}$ (AB stacking), (b) $\bu_1 = 1/10(\ba_1+\ba_2)$, (c) $\bu_1 = 1/5(\ba_1+\ba_2)$, (d) $\bu_1 = 1/3(\ba_1+\ba_2)$ (AA stacking). The full (black) lines are the result of a tight binding calculation, while the dashed (red) lines display results obtained from the interlayer gauge, Eq.~\eqref{HU}. The inset in each picture displays the lattice structure of the bilayer.
  }
  \label{shift}
\end{figure}
%-------------------------- FIG --------------------------

Having established the efficacy of our choice of tight-binding method we return to the question of interlayer stacking deformations, and first consider uniform translations of the two layers. For the case of a graphene bilayer the sublattice independence of the hopping form Eq.~\eqref{9} means that the interlayer fields, Eqs.~\eqref{S} and \eqref{S1}, take the simpler forms

\begin{equation}
 S_0(\br) = \frac{t_\perp(K)}{V_{UC}} \sum_i M_i \,\, e^{i\bK_i.\bu_1(\br)}
\label{Sg}
\end{equation}
and

\begin{eqnarray}
 S_1(\br) & =& \frac{1}{V_{UC}} \sum_i M_i \left(2 t_{\perp}'(K) \bK_i + i t_\perp(K) \bu_1(\br)\right). \nonumber \\
 && \left(\frac{\bp}{\hbar}\right) \,\, e^{i\bK_i.\bu_1(\br)}.
\label{S1g}
\end{eqnarray}
where the $M_i$ matrices are given in Table \ref{M2} and $K=4\pi/(3a)$ with $a$ the lattice constant of graphene. We have restricted the sum over the translation group $\bK_i$ of the expansion point to only the first star; this, as we will show, is enough to treat accurately the interlayer deformation physics of the graphene bilayer.

For a constant shift the $\bu_1(\br)$ field in Eqs.~\eqref{Sg} and \eqref{S1g} becomes simply a constant vector $\bu_1$ and the effective Hamiltonian is given by

\begin{equation}
 H_{\bu_1} = \begin{pmatrix} \bsig.\bp & S(\bu_1,\bp) \\ S(\bu_1,\bp)^\dagger(\bp) & \bsig^\ast.\bp \end{pmatrix}
 \label{HU}
\end{equation}
with $S(\bu_1,\bp) = S_0(\bu_1) + S_1(\bu_1,\bp)$, and may be directly diagonalized. In Fig.~\ref{shift} we show the low energy spectrum of the bilayer for four shift vectors on a path that translates the bilayer from AB to AA stacking: $\bu_1 = {\bf 0}$ (the AB stacked bilayer), $\bu_1 = 1/10(\ba_1+\ba_2)$, $\bu_1 = 1/5(\ba_1+\ba_2)$, and $\bu_1 = 1/3(\ba_1+\ba_2)$ (AA stacked). Full tight-binding results are shown by the full (black) lines with the results of the effective Hamiltonian presented with broken (red) lines. As may be seen, an excellent agreement exists between the two methods, and it is clear that the effective Hamiltonian Eq.~\eqref{HU} captures the low energy electronic structure for all mutual translations of the bilayer.

There is one quite remarkable feature of the interlayer field $S(\br)$, Eq.~\eqref{Sg}, that mutual translation of the layers of bilayer graphene reveals in simple form, and which we now comment on. Translation of the bilayer by a lattice vector will, obviously, leave the real space lattice unchanged however it \emph{does not leave the Hamiltonian invariant}. Instead the layer off-diagonal blocks of the Hamiltonian acquire a phase $e^{i n\pi/3}$, with $n=-1,0,1$ depending on the lattice vector (the precise stacking phases that occur we collate in Table \ref{stackshift}). The spectrum is, as it must be, completely unaffected by this phase. 

This behaviour represents a simple example of a more general feature of the interlayer stacking field $S(\br)$. The exponential of Eq.~\eqref{Sg} contains the $\bK_i$ vectors, the translation group of the high symmetry $K$ point, and as $\bK_i.\ba_j \ne 2\pi n$, but rather is equal to $2\pi n/3$, then an unusual 3-fold relation between the deformation field $\bu_1(\br)$ and the interlayer field $S(\br)$ is implied. While this feature of $S(\br)$ may seem highly counter intuitive (in the sense that one expects the effective fields of a continuum approach to inherit the translation group of the underlying lattice) it has, as we will show in the next section, been noticed before in the context of the graphene twist bilayer\cite{bist11}.

\begin{table}[htbp]
 \begin{tabular}{c|c|c|c|c} \hline
 initial to final stacking & s & $s\bd_1$ & $s\bd_2$ & $s\bd_3$ \\ \hline
 AB $\to$ AC & -1 & $e^{-i2\pi/3}$ & $e^{+i2\pi/3}$ & 1 \\
 AB $\to$ AA & +1 & $e^{+i2\pi/3}$ & $e^{-i2\pi/3}$ & 1 \\
 AC $\to$ AA & +1 & $e^{+i2\pi/3}$ & $e^{-i2\pi/3}$ & 1 \\
 AC $\to$ AB & -1 & $e^{-i2\pi/3}$ & $e^{+i2\pi/3}$ & 1 \\ \hline
 \end{tabular}
\caption{Stacking phases that occur upon mutual translation of the layers of a bilayer by a nearest neighbour vector of graphene $s\bd_i$, with the change in stacking type indicated in the first column. The nearest neighbour vectors, shown in Fig.~\ref{expi}(b) are $\bd_1 = a(1/2,1/(2\sqrt{3}))$, $\bd_2 = a(-1/2,1/(2\sqrt{3}))$, and $\bd_3 = -a(0,1/\sqrt{3})$.  
}
\label{stackshift}
\end{table}

%%%% Linear deformations: example of twist bilayer graphene %%%%

\subsection{Linear deformations: mutual rotation of the layers of bilayer graphene}

The twist bilayer represents a system with a simple structural variable, the rotation angle $\theta$, that nevertheless encompasses a very broad range of electronic structure phenomena\cite{lop07,hass08,lop13,shall10,bist11,mel12,shall16}. There are essentially two quite distinct regimes: at large $\theta$ the layers are electronically decoupled, while at small $\theta$ the layers strongly couple resulting in a novel low energy electronic structure that features: (i) charge localization on AA stacked regions of the emergent moir\'e lattice\cite{shall13} and (ii) an extraordinarily rich series of changes in the Fermi surface topology as a function of energy\cite{shall16}. The \emph{diverging} size of the moir\'e lattice unit cell as $\theta\to0$ implies that all atomistic based approaches will ultimately fail in this limit, as well as indicating a natural role for a continuum approach in which the lattice is replaced by some $\br$-dependent moir\'e field. Just such a Hamiltonian was derived in Ref.~\onlinecite{bist11} that, however, was subsequently shown to agree with tight-binding calculations only after rescaling\cite{shall16}. This has been attributed to the use of an incorrect momentum scale on which the single layer states couple, and indeed a revised Hamiltonian, which has an identical form but with a different momentum scale, has been shown to yield almost perfect agreement with tight-binding calculations\cite{shall16}. In this section we will derive \emph{both} of these Hamiltonians and their associated momentum scales as special cases of the theory of section \ref{ST}. As we will see, contrary to the suspicion evoked in recent papers that the Hamiltonian of Ref.~\onlinecite{bist11} is in some way in error, it turns out that both of these effective Hamiltonians are equivalent provided they are deployed in conjunction with the correct basis. 

We will first consider the Hamiltonian of Bistritzer \emph{et al.} which we will show to be simply a special case of the general interlayer field $S(\br)$, Eq.~\eqref{Sg}.
However, rather than restrict to pure rotations we will consider the more general case of a linear transformation such that a point $\br$ in layer one is linearly transformed to $\epsilon_1\br$. The deformation field is then $\bu_1(\br) = \epsilon_1\br - \br$. Taking a matrix element $\mel{\phi_{\bp_I\alpha}}{S(\br)}{\phi_{\bp_J\beta}}$ of the interlayer field $S(\br)$ yields

\begin{eqnarray}
 \mel{\phi_{\bp_I\alpha}}{S(\br)}{\phi_{\bp_J\beta}} & = & \frac{1}{V}\int d\br\, e^{i(\bp_J-\bp_I).\br}  \\
 & \times & \frac{t_\perp(K)}{V_{UC}} \sum_i \left[M_i\right]_{\alpha\beta} e^{i\bK_i.(\epsilon_1\br-\br)} \nonumber
\end{eqnarray}
which upon the change of variables $\br' = \epsilon_1\br$, and switching the action of $\epsilon_1$ from the real to the reciprocal space part of the scalar product $\bK_i.(\epsilon_1\br-\br)$, transforms to

\begin{eqnarray}
 \mel{\phi_{\bp'_I\alpha}}{S(\br')}{\phi_{\bp'_J\beta}} & = & \frac{1}{V}\int d\br'\, e^{i(\bp'_J - \bp'_I).\br'} \label{XY} \\
 & \times & \frac{t_\perp(K)}{V_{UC}}\frac{1}{|\epsilon_1|} \nonumber \\
 &\times& \sum_i \left[M_i\right]_{\alpha\beta} e^{i(\bK_i - \left(\epsilon_1^{-1}\right)^T\bK_i).\br'} \nonumber
\end{eqnarray}
where we have introduced the shorthand notation $\bp'_I = (\epsilon_1^{-1})^T \bp_I$ and 
$\bp'_J = (\epsilon_1^{-1})^T \bp_J$. From this result we may read off the interlayer field as

\begin{equation}
S_{\epsilon}(\br') = \frac{t_\perp(K)}{V_{UC}}\frac{1}{|\epsilon_1|} 
\sum_i  M_i e^{i(\bK_i - \left(\epsilon_1^{-1}\right)^T\bK_i).\br'}
\end{equation}

Specializing to the case of a pure rotation, and noting that the determinant of the rotation operator is unity, $|R|=1$, we then find for the moir\'e field of the twist bilayer

\begin{equation}
 S_{twist}(\br') = \frac{t_\perp(K)}{V_{UC}} \sum_i M_i e^{i(\bK_i - R\bK_i).\br'}
\label{bis}
\end{equation}
This is exactly the result first derived by Bistritzer and MacDonald\cite{bist11} which, as we have stated, apparently yields results that do not agree with TB calculations unless scaled\cite{shall16}. It is also striking, and was noted by the original authors, that the coupling momentum of the exponential, $|\bK_i - R\bK_i|=\frac{8\pi}{3a}\sin\frac{\theta}{2}$, generates a real space moir\'e field that \emph{does not have the periodicity of the real space moir\'e lattice} - the period of $S_{twist}(\br)$ is $\sqrt{3}$ times greater than that of the moir\'e. This behaviour is simply a manifestation of the deeper fact of the 3-fold structure to the relation between an arbitrary deformation field $\bu_1(\br)$ and the general interlayer field $S(\br)$.

To see how these two distinct Hamiltonians may have come about we now derive a Hamiltonian for the twist bilayer directly from Eq.~\ref{DUR}. For the layer off-diagonal block of the Hamiltonian this will yield

\begin{eqnarray}
\label{Y}
\mel{\phi_{\bk_I\alpha}}{S_\epsilon(\br,\bp/\hbar)}{\phi_{\bk_J\beta}} & = & 
 \frac{1}{V}\int d\br\, e^{i(\bk_J-\bk_I).\br} \\
 & \times & \frac{1}{V_{UC}} \sum_i \left[M_i\right]_{\alpha\beta} e^{i(\bk_J+\bG_i).\bu_1(\br)} \nonumber \\
 & \times & t_\perp(\bp_J+\bG_i) \nonumber
\end{eqnarray}
We now employ the zeroth order in momentum approximation only for the hopping function in Eq.~\eqref{Y}, setting $t_\perp(\bk_J+\bG_i) \approx t_\perp(\bK_i)$, and treat the phase terms exactly. Making the rearrangement $(\bk_J-\bk_I).\br + (\bk_J+\bG_i).(\epsilon_1\br-\br) = \bG_i.(\epsilon_1\br-\br) + \bk_J.\epsilon_1\br - \bk_I.\br$ for the phases in Eq.~\eqref{Y} then leads, after the same change of variables $\br' = \epsilon_1\br$ and trick with the exponential, to the result

\begin{eqnarray}
\mel{\phi_{\bk'_I\alpha}}{S_\epsilon(\br')}{\phi_{\bk_J\beta}} & = & 
\frac{1}{V}\int d\br'\, e^{i(\bk_J-\bk'_I).\br'} \\
& \times & \frac{t_\perp(K)}{V_{UC}}\frac{1}{|\epsilon_1|} \nonumber \\
& \times & \sum_i \left[M_i\right]_{\alpha\beta} e^{i(\bG_i - (\epsilon_1^{-1})^T\bG_i).\br'} \nonumber
\end{eqnarray}
where $\bk'_I = (\epsilon_1^{-1})^T \bk_I$. For the case of a pure rotation $\epsilon = R$ we then have

\begin{eqnarray}
\mel{\phi_{R\bk_I\alpha}}{S_{twist}(\br')}{\phi_{\bk_J\beta}} & = & \frac{1}{V}\int d\br'\, e^{i(\bk_J-R\bk_I).\br'} \\
&\times&\frac{t_\perp(K)}{V_{UC}} \sum_i \left[M_i\right]_{\alpha\beta} e^{i(\bG_i - R\bG_i).\br'} \nonumber
\label{shall0}
\end{eqnarray}
and so we would insist that the moir\'e field is given not by Eq.~\eqref{bis} but instead by

\begin{equation}
S_{twist}(\br') = \frac{t_\perp(K)}{V_{UC}} \sum_i M_i e^{i(\bG_i - R\bG_i).\br'}
\label{shall}
\end{equation}
This is exactly the form of the moir\'e field derived by Weckbecker \emph{et al.} and the momentum in the exponential $|\bG_i - R\bG_i|=\frac{8\pi}{\sqrt{3}a}\sin\frac{\theta}{2}$, now involving reciprocal lattice vectors $\bG_i$ not the $\bK_i$, yields a moir\'e field with a periodicity exactly that of the moir\'e lattice\cite{shall16}.

The reason that these two effective Hamiltonians will, as they must, yield the same electronic structure is simply that they are evaluated in each case with a different basis: Eq.~\eqref{bis} must be evaluated in a basis of rotated single layer graphene states, whereas Eq.~\eqref{shall} must be evaluated using a basis of unrotated single layer states from the unrotated layer, and rotated single layer states from the rotated layer (this is clearly seen by inspection of the relevant matrix elements that $S_{twist}(\br)$ is exacted from in each case). Both approaches are, therefore, correct, and the unusual mismatch between the translational symmetry of the continuum and lattice Hamiltonians found in the approach of Ref.~\onlinecite{bist11} may be removed, in this case, simply by changing to a ``more natural'' basis in which the single layer states follow the rotational geometry of the bilayer (i.e., basis states from the unrotated layer are unrotated single layer states, and from the rotated layer are rotated single layer states).

%%%% Partial dislocations in bilayer graphene %%%%

\subsection{Complex stacking disorder: partial dislocations in bilayer graphene}

%-------------------------- FIG --------------------------
\begin{figure}
  \centering
  \includegraphics[width=0.98\linewidth]{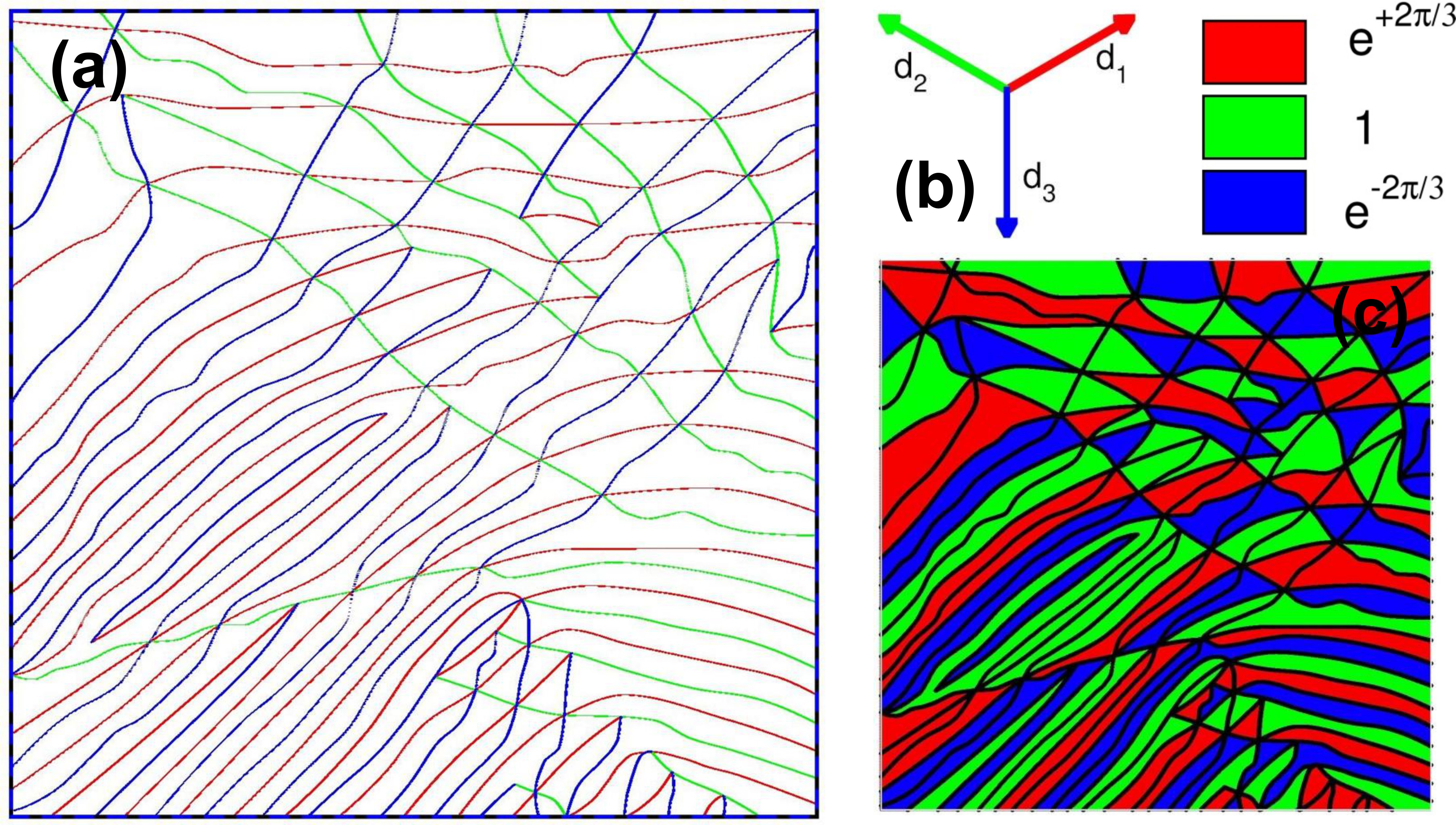}
  \caption{(a); Partial dislocation network found in bilayer graphene grown on the Si-face of SiC, the corresponding TEM image from which the network data is extracted may be found in Ref.~\onlinecite{kiss15}. The colour of each partial indicates the partial Burgers vector of the partial, with the Burgers vectors indicated in panel (b); (c) The phase structure of the partial dislocation network, with the colour of each segment indicating the stacking phase $e^{i 2\pi n/3}$ that arises from the 3-fold nature of the general interlayer stacking potential, see Section \ref{SHIFT}.
  }
  \label{expi}
\end{figure}
%-------------------------- FIG --------------------------

The Bernal (AB) stacked graphene bilayer is generally assumed - with the exception of possible point defects - to be structurally perfect. Recently, this has been shown not to be the case, and TEM images of the bilayer (grown by sublimation of Si from the Si-face of SiC) have been shown to exhibit a dense network of partial dislocations\cite{ald13,butz14,kiss15}. These arise because of a hidden structural degeneracy in the bilayer: there are two equivalent stacking choices, conventionally referred to as AB and AC stacking. In an infinite crystal these are, of course, physically equivalent, however they may also coexist as domains in a single crystal, at which point they become physically distinct. The requirement of an continuous graphene membrane in each layer then leads to the condition that such domains be connected by one of three possible partial Burgers vectors $\bd_1 = a(1/2,1/(2\sqrt{3}))$, $\bd_2 = a(-1/2,1/(2\sqrt{3}))$, and $\bd_3=a(0,-1/\sqrt{3})$, shown in Fig.~\ref{expi}(b). Traversing from one domain to the other then involves a local shift of one layer by one of these partial Burgers vectors. Such a mosaic of AB and AC domains  represents a quite different material to that of the structurally perfect bilayer, and indeed transport measurements find a number of very distinct properties for the mosaic material as compared to the perfect bilayer\cite{kiss15}.

In Ref.~[\onlinecite{kiss15}] a partial dislocation network taken from experiment was calculated using a preliminary version of the method described in this paper, and in this section we will consider in further detail the theoretical treatment of such networks. The experimental network we will investigate is illustrated in Fig.~\ref{expi}(a) with the various partial dislocations coloured according to the their Burgers vector, compare with Fig.~\ref{expi}(b). For the original TEM images we refer the reader to Ref.~[\onlinecite{kiss15}]. The area of the TEM image in experiment was 1$\mu m^2$, equivalent to $\approx\!10^8$ carbon atoms. Calculating such a system within an atomistic approach is, obviously, completely out of the question.

%-------------------------- FIG --------------------------
\begin{figure}
  \centering
  \includegraphics[width=0.85\linewidth]{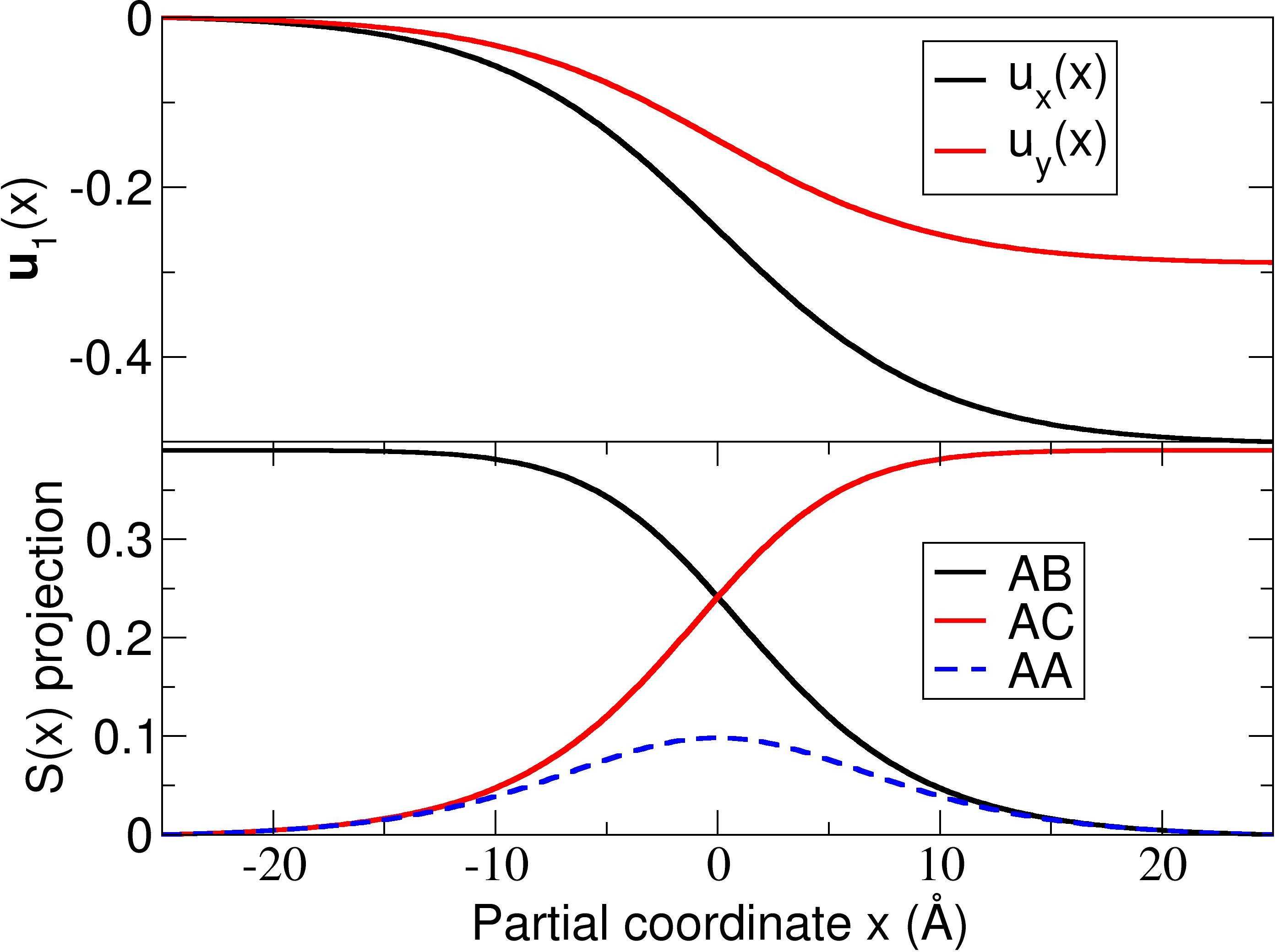}
  \caption{\emph{Upper panel}: Deformation field $\bu_1(x)$ of a partial dislocation that smoothly connects regions of AB and AC stacking; $x$ is a coordinate perpendicular to the partial. On the left hand side $\bu_1 = {\bf 0}$ (AB stacking) and on the right hand side $\bu_1 = -\bd_1 = a(-1/2,-1/(2\sqrt{3}))$ (AC stacking), the connection between the two is effected by a partial Burgers vector $-\bd_1 = a(-1/2,-1/(2\sqrt{3}))$. \emph{Lower panel}: The corresponding effective interlayer field $S(x)$ required to describe the electronic structure of this partial dislocation. This matrix valued field is shown as a projection onto the three high symmetry stacking types of the graphene bilayer: AB, AC, and AA stacking, see Eqs.~\eqref{OH}-\eqref{yesss} for the projection matrices.
  }
  \label{uphi}
\end{figure}
%-------------------------- FIG --------------------------

%-------------------------- FIG --------------------------
\begin{figure*}
  \centering
  \includegraphics[width=0.85\linewidth]{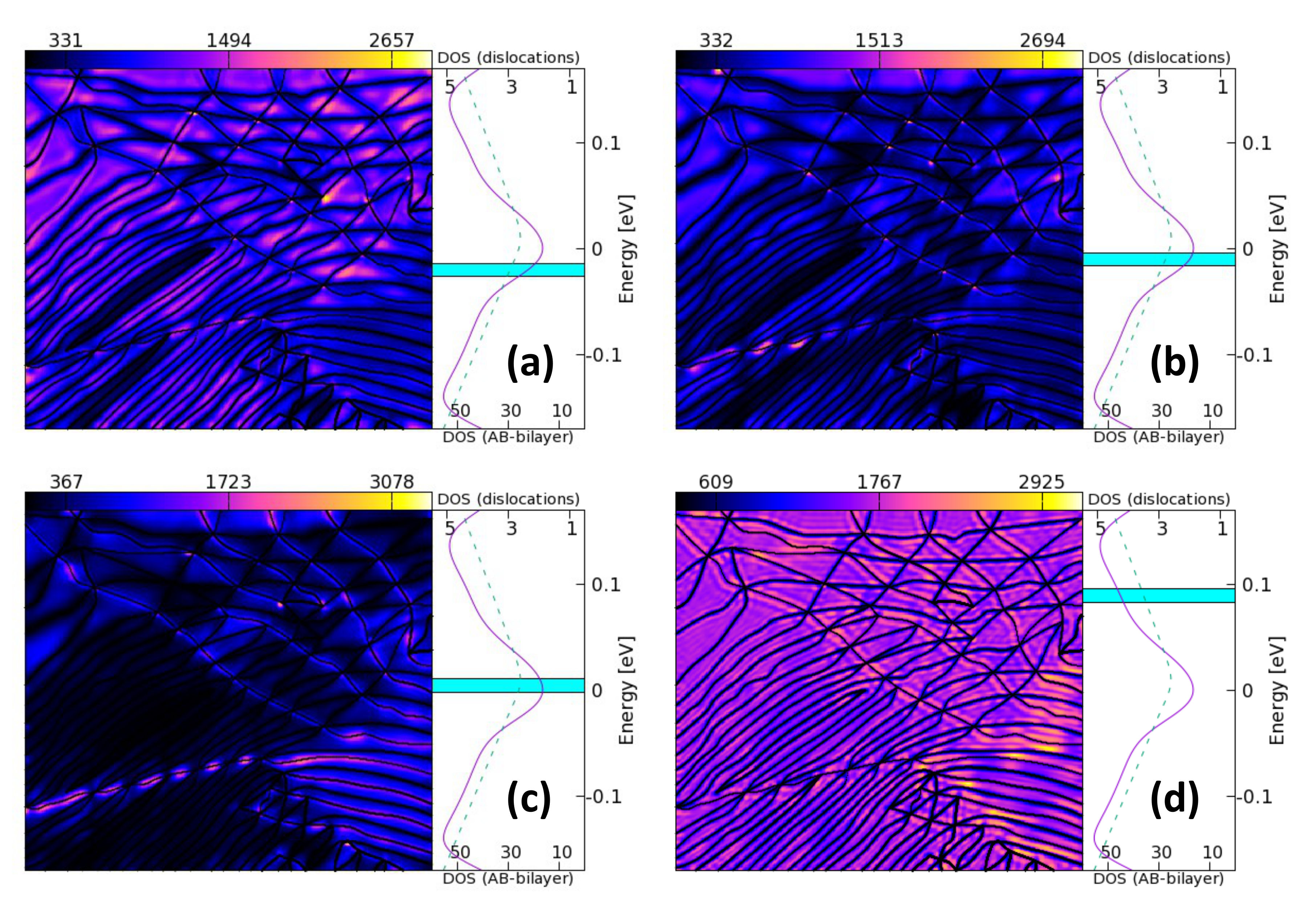}
  \caption{Electron density obtained by integrating over all states within a 13meV energy window situated at 4 different energies in the density of states (DOS) of the partial dislocation network: -20~meV (a), -5~meV (b), +5~meV (c), and +90~meV (d). Close to the Dirac point, panels (b) and (c), charge pooling on the segments of the mosaic network can clearly be seen. In panel (a) charge localization associated with partials of type 3 may be observed which, above the Dirac point energy, switches to a charge accumulation associated with partials of type 2, see especially panel (d). The right hand panel in each case indicates the energy window within which the wavefunctions are accumulated in to give the electron density, with the full line the network DOS and the dashed line the Bernal bilayer DOS presented for comparison.
  }
  \label{expd}
\end{figure*}
%-------------------------- FIG --------------------------

This problem can, however, straightforwardly be treated with the general interlayer field $S(\br)$, Eq.~\eqref{Sg}, in which the deformation field $\bu_1(\br)$ now simply has to encode the mutual translation of the layers that occurs on crossing a partial (within the domains of the mosaic structure the function $\bu_1(\br)$ will obviously be constant). This transition occurs, according to experiment, over a width of $\approx 5$nm and the detailed atomic structure of this transition region has been carefully investigated via semi-empirical tight-binding calculations\cite{butz14}. From the data of Ref.~\onlinecite{butz14} we are able to extract a model form of $\bu_1(\br)$, and this is shown in Fig.~\ref{uphi} for the $AB \to AC$ transition mediated by a partial Burgers vector $-\bd_1$. Also shown is the interlayer field $S(\br)$, projected onto the 3 distinct high symmetry stacking types that exist through this transition, $S_{AB}$, $S_{AC}$, and $S_{AA}$:

\begin{eqnarray}
 S_{AB} & = & \begin{pmatrix} 1 & 0 \\ 0 & 0 \end{pmatrix}, \label{OH}\\
 S_{AC} & = & e^{-i2\pi/3}\begin{pmatrix} 0 & 0 \\ 0 & 1 \end{pmatrix}, \label{YES} \\
 S_{AA} & = & e^{-i\pi/3}\begin{pmatrix} 0 & 1 \\ 1 & 0 \end{pmatrix} \label{yesss}.
\end{eqnarray}
(Deployed as a constant interlayer block these matrices will generate the standard AB/AC and AA stacked band structures.)
The matrix function $S(\br)$ can be seen to transition between AC and AB stacking with the maximum of the AA component in the middle of the partial. Note that as the partial dislocations "wander" through the lattice, the angle between the partial tangent and the Burgers vector will, in general, take on all values between pure screw ($90^\circ$) and pure edge ($0^\circ$).

The peculiar stacking phases described in section \ref{SHIFT}, and the existence of 3 AB and AC types that differ by a phase $e^{2n\pi/3}$, with $n=-1,0,1$, cannot now be removed through a change of basis as was the case in the example of the twist bilayer. If all partial dislocations extended through the sample, this would not cause any complication, however the annihilation of partial dislocations at point defects in the lattice leads to a certain complexity in the phase structure of the network. This can most clearly be seen by imagining an AC stacked island bounded by partials that annihilate at two point dislocations in the lattice, a geometry that can in fact be seen in Fig.~\ref{expi}(a). Crossing two partial dislocations will (for partials type 1 and 2) lead to the accumulation of a phase $e^{2n\pi/3}$, with $n=\pm1$ depending on the particular partial type. This leads to a contradiction as, if one encircles the enclosed AC island through the perfect AB material, the stacking phase of the Hamiltonian obviously cannot change while, on the other hand, if one traverses this AC island, and thus intersects two partials, a phase $e^{2n\pi/3}$ must be accumulated: there is no consistent way to treat this situation. In Ref.~\onlinecite{kiss15} it was assumed that the point defects that create and annihilate partial dislocations also create and annihilate a phase contribution to the stacking phase of the Hamiltonian, and with this assumption the phase structure can be mapped out over the partial network, as shown in Fig.~\ref{expi}(c).

\emph{Numerical details}: Even within the effective Hamiltonian approach a $1\mu m^2$ area represents a substantial computational burden. We utilize a basis of single layer graphene states that, as in the case of the twist bilayer for which this basis has also been deployed\cite{shall10,shall13,shall16}, has the advantage that to capture the low energy spectrum requires single layer graphene states of energy approximately double the energy window one is interested in calculating. Even so, a basis of 20,000 states must be employed leading to a Hamiltonian matrix of dimension 80,000 that requires massive parallel calculation to efficiently (and iteratively) diagonalize.

Of principle interest is the form of the wavefunctions of the mosaic network, which are expected to be very different from the uniform density wavefunction of the structurally perfect bilayer. In Fig.~\ref{expd}(a-d) we show the density integrated over a 13~meV window (of the order of the Fermi smearing at 150 K) with this window placed at four different energies: -20~meV, -5~meV, +5~meV, +90~meV. Even within this small energy window of the order of $10^3$ individual eigenstates contribute to the probability density. Quite clearly, the mosaic structure of the bilayer has a dramatic impact on the wavefunctions. For the density integrated in the window situated at -20meV one notices that there is charge accumulation associated with the type 3 partials (compare with Fig.~\ref{expi}(a)), but not on the partials of type 1 and 2. Above the Dirac point this switches to a charge accumulation associated with type 2 partials, see panel (d). Close to the Dirac point, see panels (b) and (c), one also notices a substantial charge pooling as some segments of the mosaic network have significantly higher density than others, a point first noticed in Ref.~\onlinecite{kiss15}. For the energy window of +90meV, shown in panel (d), the pronounced charge pooling seen near the Dirac point is absent, although one still notes substantial density inhomogeneity.

\begin{figure}
  \centering
  \includegraphics[width=0.85\linewidth]{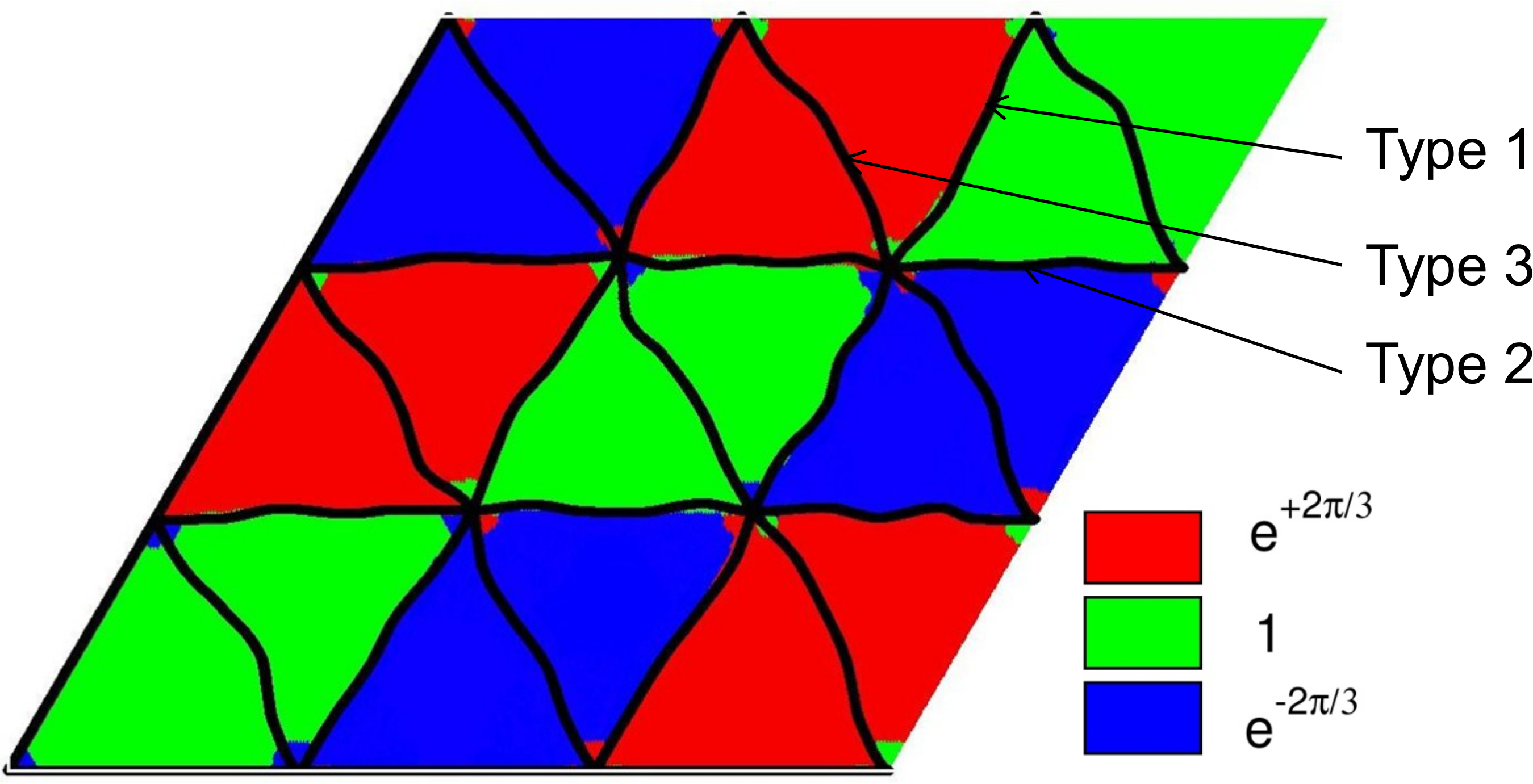}
  \caption{The stacking phase structure of a hexagonal partial network, with the type of partial dislocation indicated by the labels. Note that these dislocations extend through the sample (the unit cell is periodically repeated) and, in contrast to the experimental network (see Fig.~\ref{expi}) which features creation and annihilation of partials and has a very complex phase structure, the phase field here is a simple periodic tiling. The partial Burgers vector associated with each partial dislocation type is illustrated in panel (b) of Fig.~\ref{expi}.
  }
  \label{hexi}
\end{figure}

How much of this structure, and in particular the existence of "hot" partials on which density is accumulated, is related to the specific partial network shown in Fig.~\ref{expd}? To investigate this we consider an designed hexagonal network of partial dislocations shown in Fig.~\ref{hexi}. We introduce some random disorder into the partials such that they are not perfectly straight, however all partials are now non-terminating (the area shown is periodically repeated) and the problems with mapping the phase structure of the experimental network do not exist. In Fig.~\ref{hexi} we show the phase structure for this network.

%-------------------------- FIG --------------------------
%-------------------------- FIG --------------------------

%-------------------------- FIG --------------------------
\begin{figure}
  \centering
  \includegraphics[width=0.98\linewidth]{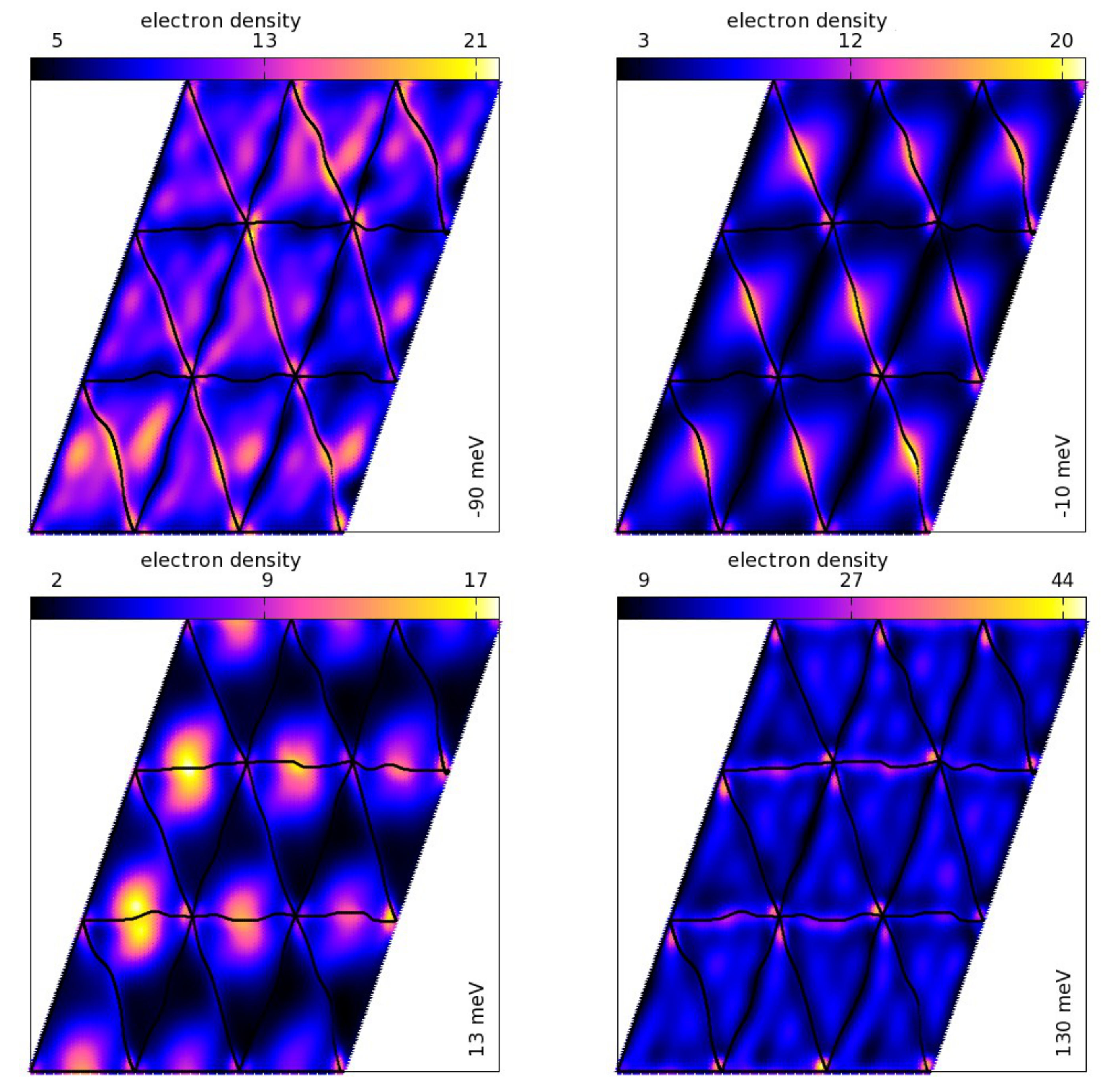}
  \caption{Electron density obtained by integrating all states over a 13~meV energy window situated at 4 different energies: -90~meV (a), -10~meV (b), +13~meV (c), and +130~meV (d). We find both charge pooling on the mosaic segments as well as strong localization on the nodes of the network, see especially panel (d). We also note localization on partials with, interestingly, the same energy ordering - type 3 below the Dirac point, type 2 above the Dirac point - as may be observed in the more complex experimentally derived partial network, see Fig.~\ref{expd}.
  }
  \label{hexd}
\end{figure}
%-------------------------- FIG --------------------------

As may be observed in Fig.~\ref{hexd}, qualitatively similar features are seen to those noted in the experimental partial network - in particular the energy order in which the type 3 and 2 partials become "hot" and accumulate charge is the same; the type 1 partial also, as before, does not accumulate charge. This charge accumulation is therefore largely independent of the global details of partial network and rather is a consequence of the local partial structure. A number of features that are difficult to detect in the experimental network may be seen much more clearly in this designed network, in particular the localization of charge on the nodes of the network is much more pronounced, see panel (d) of Fig.~\ref{hexd}.

%-------------------------- FIG --------------------------
\begin{figure}
  \centering
  \includegraphics[width=0.98\linewidth]{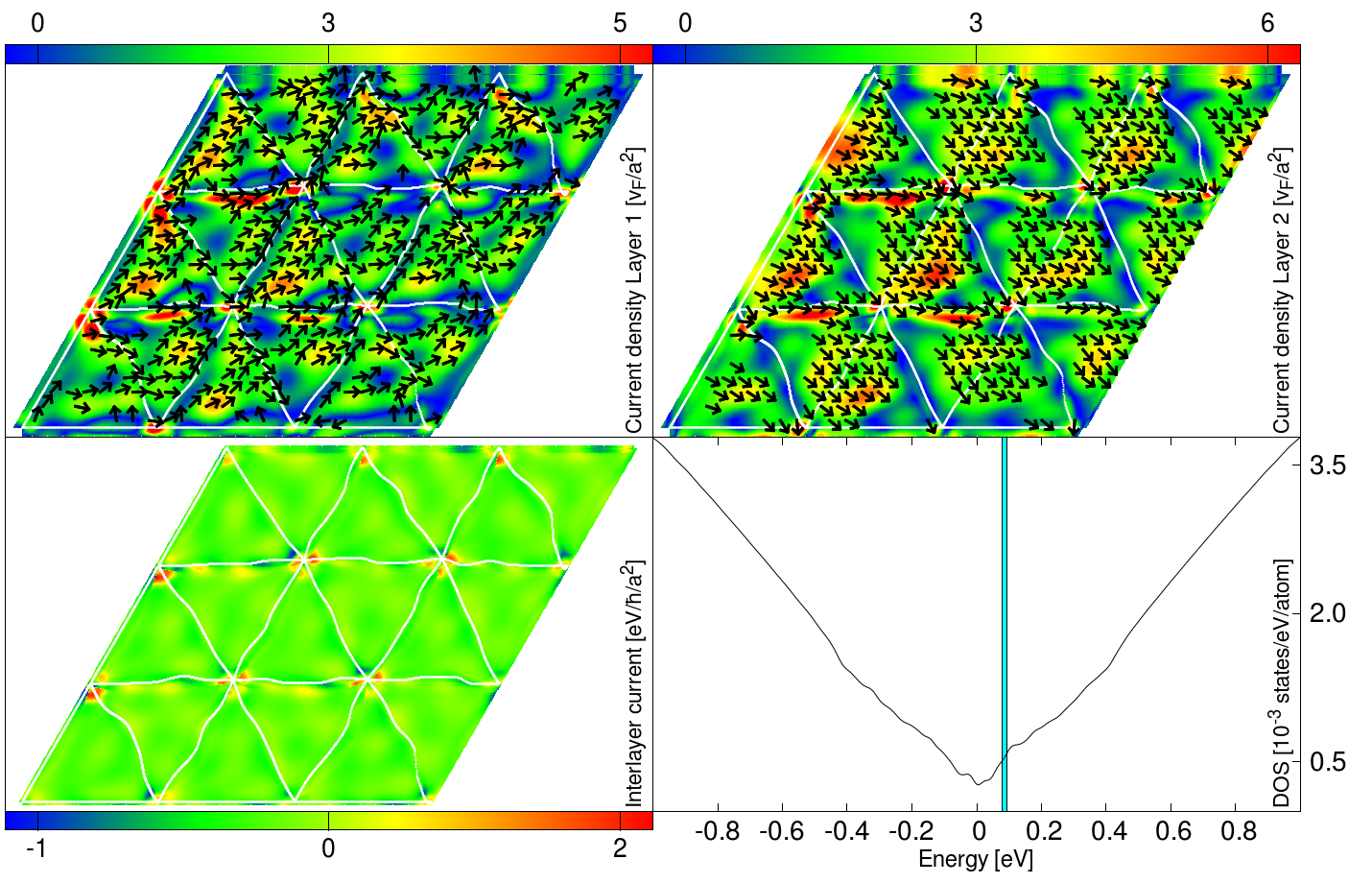}
  \caption{Interlayer and intralayer currents for the charge accumulation states on type 2 partial dislocations; compare with the charge density pictures in Fig.~\ref{hexd}, especially panel (c). Both interlayer currents, panel (c), associated with the partial as well as strong intralayer currents, see panels (a) and (b), can be observed. Note that this calculation is performed in the absence of a magnetic field and therefore integrating all states (up to the Fermi energy) will result in zero net current.
  }
  \label{hexj}
\end{figure}
%-------------------------- FIG --------------------------

Finally we examine the nature of these localized states on the partial dislocations. One might imagine that these could represent current carrying states, similar to those recently observed in experiment\cite{ju15} (although note that as we have no magnetic field the overall current will be zero, and a finite current found only by a restricted integration over eigenstates.). Shown in Fig.~\ref{hexj}(a-c) is both the intralayer as well as interlayer current density integrated over states indicated in energy window in Fig.~\ref{hexj}(d). Interestingly, one notes that the (type 2) partials are associated with both in-plane currents, which flow from left to right in Fig.~\ref{hexj}, and \emph{interlayer} currents and thus the charge accumulation on partials seen in the previous section indeed represents current carrying states. The formalism used for the current calculations (which is a simple extension of that presented in Section II) we will, due to lack of space, expound in a subsequent publication.

%%%%%%%%%%%%%%%
% CONCLUSIONS %
%%%%%%%%%%%%%%%

\section{Discussion, summary, and extensions}

\emph{Discussion}: We have presented a theory that goes substantially beyond the $\bk.\bp$ method in its capability for generating effective Hamiltonians. This enhanced applicability derives from two primary differences with $\bk.\bp$ theory. Firstly, the approach relies on local and not global closeness to a reference state, in the sense that if the pseudospin structure of the reference state provides a good local description of the system of interest, then the method will work. This allows the generation of compact and physically intuitive effective Hamiltonians even when, globally, the system of interest is dramatically different from the reference state. 
Secondly, instead of individual optical matrix elements forming the unknown constants of the theory to be fitted, it is the tight-binding hopping function that constitutes the basic unknown object. This sharply reduces the number of variables to be fitted, in particular for systems with very low (or no) symmetry, as well as for systems that require high orders in momentum or deformation tensor for an accurate description. 

The first of these differences with $\bk.\bp$ theory is highlighted by two examples presented in this work, the twist bilayer and partial dislocation networks. These materials, both structurally and electronically, are profoundly different from the reference state from which they are derived, the AB stacked bilayer, yet in both cases the method we employ here provides a compact and intuitive effective Hamiltonian description. The ability to treat this type of non-perturbative deformation, exemplified by the interlayer (i.e. stacking) degree of freedom, will be important for the emerging class of low dimensional van der Waals heterostructures in which the weakly bonded layers are likely highly susceptible to such stacking deformations\cite{ald13,butz14}. In contrast to three dimensional materials, charge carriers in two dimensions cannot avoid stacking defects that extend throughout the sample, e.g. partial dislocations, and their impact on electronic properties is, therefore, expected to be profound. Indeed, this has recently been observed in the case of bilayer graphene\cite{kiss15}, and one wonders what the impact of stacking deformations will be on the excitonic properties of the few layer dichalcogenides, for example. This represents a new materials paradigm in which extended defects, that play almost no electronic role in three dimensional materials although a crucial role in mechanical strength, represent an important ingredient in understanding the electronic structure of weakly bonded few layer materials. This makes all the more desirable a general method effective Hamiltonian method by which they may be treated, in particular as the length scales in involved in such defects render prohibitive conventional atomistic approaches.

\emph{Summary}: We have applied our theory to the case of perturbative deformations in 2d materials that are slow on the scale of the lattice constant. In this case the structure of the theory consists of a connection formula linking lattice and pseudospin spaces that, together with \emph{universal} polynomials formed from the basic variables of the theory (the momentum operator and deformation tensor), generate effective Hamiltonians for any 2d system, both for the high symmetry phase as well as providing a systematic treatment of corrections due to deformations. We deploy this method for the case of deformations in graphene, and are able to encompass all known results from the literature, as well as providing several extensions that we show together results in almost perfect agreement with tight-binding calculations for the test case of strained graphene. These extensions include both higher order fields in momentum and the deformation tensor, as well as a companion scalar field to the remarkable geometric gauge first reported in Ref.~\onlinecite{ju12} (both the scalar and vector geometric fields are pure imaginary but, as we show, preserve the Hermiticity of the effective Dirac-Weyl theory). Application to a selection of more complex all carbon 2d allotropes - we consider graphdiyne, $\gamma$-graphyne, and 6,6,12-graphyne - yields effective Hamiltonians both for the high symmetry state as well as for arbitrary (slow) deformations, a formalism ideal for treating the electron-phonon interaction in these materials. For graphdiyne, which we use as an explicit example, the effective Hamiltonian turns out simply to be the Dirac equation, with deformations entering as a complex gap field such that positive biaxial strain opens the gap, while uniaxial positive or biaxial negative strain closes the gap.

Application of the theory to the case of a bilayer geometry leads directly to the construction of a general interlayer (matrix valued) effective field that provides a continuum description of an arbitrary bilayer system subject to \emph{any stacking deformation}. This interlayer field is, therefore, the equivalent in generality of the deformation induced effective gauge field that allow one to treat arbitrary in-plane deformations in the case of single layer graphene. For bilayer graphene this interlayer field is shown to yield both the well known twist bilayer Hamiltonian, and a Hamiltonian describing partial dislocations in the AB bilayer, simply as special cases. For the twist bilayer we find (i) a generalization from pure twist to a general linear transformation (i.e., including both possible shear and strain as well as twist) and (ii) resolve a discrepancy between different effective Hamiltonians that have appeared in the literature\cite{bist11,shall16}.  For the case of partial dislocations we present calculations of both realistic and designed partial networks uncovering interesting charge localization effects on the mosaic geometry, in particular charge accumulation on partial dislocations and, near the Dirac point, charge pooling on the mosaic segments. The charge accumulation on partials appears to be Burger vector specific in a generic way: for both the experimental and designed partial networks the accumulation occurs at specific energies for specific Burgers vectors. These charge accumulation states are shown to carry current along the partials which has both an in-plane as well as an interlayer component.

To summarize, the method yields, as we have demonstrated in numerous examples, compact and physically intuitive effective Hamiltonians even for the very complex low symmetry situations that occur in low dimensional materials. As this class of materials continuous to grow apace, the theory may provide a very useful tool for investigating their electronic structure, comparable in impact to the usefulness of $\bk.\bp$ theory for three dimensional materials. A cornucopia of further applications can easily be imagined: the electronic structure and excitonic physics of the (probably unavoidable) dislocations and stacking faults in MoS$_2$ and other layered dichalcogenides, deformations fast on the scale of the lattice constant important e.g. in silicene structures and highly strained graphene-metal hybrids, and twist faults in complex carbon allotropes. Furthermore, the extension to a three dimensional and multi-orbital case will allow for the easy generation of effective Hamiltonians to describe even the most complex topological insulators.

\begin{acknowledgments}

This work was supported by the Collaborative Research Center SFB 953 of the Deutsche Forschungsgemeinschaft (DFG). We would also like to thank M. Oliva-Leyva for a useful prompt based on his reading of a draft version of this paper.

\end{acknowledgments}

%\bibliographystyle{unsrt}
%\bibliography{Mtheory}

\end{document}